\documentstyle[aas2pp4]{article}

\input{tcilatex}
\begin{document}

\title{ANALYTICAL APPROXIMATIONS FOR CALCULATING \\
THE ESCAPE AND ABSORPTION OF RADIATION \\
IN CLUMPY DUSTY ENVIRONMENTS}
\author{Frank V\'{a}rosi\altaffilmark{1} \thinspace and \thinspace Eli Dwek \\
{\em Laboratory for Astronomy and Solar Physics}\\
{\em Code 685}\\
{\em NASA Goddard Space Flight Center}\\
{\em Greenbelt MD 20771}\\
-\\
{\small submitted: February 4, 1999 -- accepted: May 5, 1999 -- to appear in ApJ: September 20, 1999}
}

\begin{abstract}
We present analytical approximations for calculating the scattering,
absorption and escape of non-ionizing photons from a spherically symmetric
two-phase clumpy medium, with either a central point source of isotropic
radiation, a uniform distribution of isotropic emitters, or uniformly
illuminated by external sources. The analytical approximations are based on
the mega-grains model of two-phase clumpy media, as proposed by Hobson \&
Padman, combined with escape and absorption probability formulae for
homogeneous media. The accuracy of the approximations is examined by
comparison with three-dimensional Monte Carlo simulations of radiative
transfer, including multiple scattering. Our studies show that the combined
mega-grains and escape/absorption probability formulae provide a good
approximation of the escaping and absorbed radiation fractions for a wide
range of parameters characterizing the clumpiness and optical properties of
the medium.

A realistic test of the analytic approximations is performed by modeling the
absorption of a starlike source of radiation by interstellar dust in a
clumpy medium, and by calculating the resulting equilibrium dust
temperatures and infrared emission spectrum of both the clumps and the
interclump medium. In particular, we find that the temperature of dust in
clumps is lower than in the interclump medium if the clumps are optically
thick at wavelengths where most of the absorption occurs. Comparison with
Monte Carlo simulations of radiative transfer in the same environment shows
that the analytic model yields a good approximation of dust temperatures and
the emerging UV to FIR spectrum of radiation for all three types of source
distributions mentioned above.

Our analytical model provides a numerically expedient way to estimate
radiative transfer in a variety of interstellar conditions and can be
applied to a wide range of astrophysical environments, from clumpy star
forming regions to starburst galaxies.
\end{abstract}

\altaffiltext{1}{\small Raytheon ITSS, \, varosi@gsfc.nasa.gov}%

\pagestyle{myheadings}
\markright{V\'{a}rosi \& Dwek: \, Escape and Absorption of Radiation in Clumpy Dusty Environments}
\clearpage
\tableofcontents
\clearpage
%

\section{INTRODUCTION}

Radiative transfer plays an important role in the spectral appearance of
almost all astrophysical systems ranging from isolated star forming regions
to protogalactic and galactic systems. At wavelengths longer than the Lyman
limit, absorption and scattering by dust in the intervening medium is the
major factor affecting the transfer of radiation. Most of the transfer of
radiation occurs in the interstellar medium (ISM) of the host system of the
emitting sources. For simplicity, models of radiative transfer often assume
a homogeneous distribution of dust and gas, although in reality the
structure of the ISM is observed to be significantly more complex.

In our Galaxy, the ISM is known to be composed of at least five phases that
are in approximate pressure equilibrium: cold dense molecular clouds, cold
diffuse clouds, warm diffuse clouds, H II regions, and hot low density
cavities created by supernova remnants (\cite{Spitzer-ISM}, \cite{Cox95}, 
\cite{McKee95}, \cite{knapp95}). The ISM is observed to have clumpy
structure even at very small scales (\cite{Marscher93}, \cite{stutzki90}).
Data from CO line emission indicate a distribution of molecular clouds
having a power-law cloud mass spectrum (\cite{SanSco85}). \cite{DG89} find
the same cloud mass spectrum based on 21 cm emission line data. Most likely
the ISM has a spectrum of densities and temperatures with correlated
multi-scale spatial structure, as evidenced by sky surveys such as the {\sl %
Infrared Astronomical Satellite (IRAS)} and 21 cm surveys. Analysis of {\sl %
IRAS} 100$\mu $m sky flux indicates that the diffuse H I clouds have a
fractal distribution (\cite{Bazell88}, \cite{Waller97}). Analysis of CO
column densities and emission line profiles further suggests that a fractal
distribution of matter applies to the molecular component of the ISM as well
(\cite{Falg95}, \cite{elmfal96}). The complex, possibly fractal,
distribution of gas and dust is supported by theoretical arguments as well (%
\cite{PfenComb94}, \cite{RosBreg95}, \cite{norman96}, \cite{elm97}). Thus
the ISM is clearly inhomogeneous and simulations of radiative transfer
should account for this fact.

The simplest model of an inhomogeneous medium is that consisting of two
phases: dense clumps embedded in a less dense interclump medium (ICM). \cite
{NatPan84} developed simple analytic approximations for the effective
optical depth of clumpy media with no scattering and an empty ICM. Radiative
transfer with isotropic scattering in a two-phase clumpy medium (non-empty
ICM) with plane-parallel geometry was investigated by \cite{boisse90}. He
used a Markov process model of the medium to develop analytical
approximations for the intensity of radiation. Comparison with 3D Monte
Carlo simulations verified the accuracy of the approximations. \cite{hs93}
developed analytic solutions of radiative transfer with isotropic scattering
in $N-$phase clumpy media using a Markov process model, which they compared
with Monte Carlo simulations for the cases of two or three phases. Analytic
approximations for radiative transfer in two-phase clumpy media were also
developed by \cite{neufeld91} and \cite{hobpad93}, and we discuss and
utilize them in later sections of this paper. \cite{wittgor96} performed
extensive Monte Carlo simulations of the transfer of radiation from a
central point source in a spherical two-phase clumpy medium when the
scattering is not isotropic, more typical of UV photons scattered by dust.
Their simulations showed that the nonlinear variation of effective optical
depth and effective albedo with respect to parameters characterizing the
clumpy medium could lead to erroneous estimates of the dust albedo and
opacity if one assumes a homogeneous medium when modeling what may actually
be a clumpy medium. \cite{gordsb97} applied the Monte Carlo model of Witt \&
Gordon to the study of dust in starburst galaxies, concluding that a shell
of clumpy dust around a starburst provides the best model of spectral data.
Radiative transfer in a clumpy environment was also investigated by \cite
{Wolf98}, with results similar to Witt \& Gordon.

All the simulations and studies mentioned above verify the general
expectation that a medium is more transparent when it is clumpy. This can be
demonstrated in the following generic example, which then leads to the
concept of effective optical depth in an inhomogeneous medium. Consider $N$
randomly chosen parallel and equal length lines of sight of through an
inhomogeneous medium acting as a foreground screen. Let $\tau _{i}$ be the
optical depth of the $i$-th line of sight, defined as the product of column
density and cross-section of the dust, so that $e^{-\tau _{i}}$ is the
transmission. If $N$ is large (e.g. the number of photons per second emitted
by a galaxy) then we can compute the average transmission of all the lines
of sight, which defines the effective optical depth: 
\begin{equation}
\tau _{eff}\,\equiv \,-\ln \left( \frac{1}{N}\sum_{i}\exp (-\tau
_{i})\right) \,\,.  \label{eff_tau}
\end{equation}
On the other hand, the average of all the optical depths equals the optical
depth of the homogeneous medium with equal dust mass: 
\begin{equation}
\tau _{hom}\equiv \frac{1}{N}\sum_{i}\tau _{i}\quad .  \label{avg_tau}
\end{equation}
Using standard calculus one can prove the inequality 
\begin{equation}
\frac{1}{N}\sum_{i}\exp (-\tau _{i})\;\;>\;\;\exp \left( -\frac{1}{N}%
\sum_{i}\tau _{i}\right) \,,  \label{eff_trans}
\end{equation}
which states that the average transmission of the inhomogeneous medium is
greater than that of the equivalent homogeneous medium, allowing relatively
more photons to escape. The expression can be an equality only when $\tau
_{i}=\tau _{hom}$ for all lines of sight, i.e. only if the medium is
homogeneous. Applying the negative natural logarithm to the inequality (\ref
{eff_trans}) and using the definitions (\ref{eff_tau}) and (\ref{avg_tau})
gives 
\begin{equation}
\tau _{eff}\,\,<\,\tau _{hom}\,\,,  \label{tau_eff<hom}
\end{equation}
so the effective optical depth of an inhomogeneous medium is less than that
of a homogeneous medium with equal mass of dust. When the dust also scatters
photons, the above inequality can be considered to apply approximately to
the scattered photons, so we again expect generically a greater transmission
through inhomogeneous media than the equivalent homogeneous medium. In this
paper we study the dependence of the nonlinear relationship between $\tau
_{eff}$ and $\tau _{hom}$ on parameters characterizing a clumpy
medium, and also give approximations for how absorption of photons is
apportioned in each phase of the medium.

One consequence of the fact that $\tau _{eff}\,\,<\,\tau _{hom}$ for an
inhomogeneous medium is that a large mass of dust could be concealed in
dense regions even though observations of low extinction infer a small mass
of dust. Such dense clumps of dust could have a lower temperature than if
the dust were distributed homogeneously, providing an alternative
explanation of observations that show an inconsistency between
dust temperature, absorbed luminosity, and known luminosity sources,
as found in recent infrared
spectral observations of the Galactic Center (\cite{Chan97}). For these
types of modeling efforts, easily computable analytic approximations of
radiative transfer in clumpy media are desirable since they would enable the
rapid exploration of the effects of varying the parameters in an
astrophysical model. The more time consuming Monte Carlo simulations can be
used to guide the development and test the accuracy of analytical
approximations.

We have developed a general Monte Carlo Radiative Transfer (MCRT) code for
simulating radiative transfer with multiple scattering in a three
dimensional inhomogeneous medium, utilizing some novel techniques. The model
is restricted to non-ionizing radiation, hence the only absorption and
scattering considered is due to dust. We repeated many of the Monte Carlo
simulations of Witt \& Gordon and our results are in good agreement. The
medium in their model is composed of cubic clumps randomly located on a body
centered cubic lattice. We instead concentrate our studies on two-phase
media composed of spherical clumps because the radiative transfer properties
of such media are more directly approximated by analytic formulas in the
mega-grains model of \cite{hobpad93}, hereafter HP93. Furthermore, we extend
the research of HP93 by computing the fraction of photons absorbed in each
phase of the medium and improving the approximations for scattering. The
resulting formulae allow for convenient and rapid exploration of the effects
of varying the parameters defining the environment on the escape and
absorption of radiation. The cases of a central isotropic point source,
uniformly distributed isotropic emitters, and uniformly illuminating
external sources are studied for homogeneous and clumpy media, with both
MCRT simulations and analytic approximations, over a wide range of dust
optical depths, scattering albedos, and from isotropic to forward
scattering. The validity and accuracy of the analytical approximations is
tested by detailed comparison with MCRT simulations.

We have also investigated the transfer of radiation in a fractal
distribution of dust density (\cite{VD97}) using MCRT, where the fractal
construction is based on methods given in Elmegreen (1997). Briefly, the
variation of the escape of radiation in a fractal medium as a function of $%
\,\tau _{hom}$ is similar to that of clumpy media, and we find that the
mega-grains model can be used to approximate the escaping fraction of
photons if some additional assumptions are imposed. However, the
distribution of absorbed radiation is more complicated than in two-phase
clumpy media since there is a spectrum of densities in the fractal case.
Studies and approximations of radiative transfer in complex and fractal
media will be presented elsewhere (\cite{VD00}).

The paper is organized as follows. Section~2 describes the MCRT code and its
application to two-phase clumpy media. In \S \ref{Esc_Abs_Homog} we present
analytical approximations for the fraction of radiation escaping from or
absorbed by homogeneous media in spherical geometry, for each of the three
types of sources mentioned above. In \S \ref{Esc_Abs_Clumpy} we present the
mega-grains model of two-phase clumpy media, and also some improvements and
extensions. The mega-grains approach is combined with the formulae for
homogeneous media to get approximations for the escaping/absorbed fractions
in clumpy media, and also the fractions absorbed in clumps and the ICM.
Section \ref{List_EAP} gives a summary of all the equations in a convenient
list. Then in \S \ref{Compare_MC}, we demonstrate the validity of the
analytic approximations by comparison with MCRT simulations. In \S \ref
{Simul_SED_IR} we present a realistic test of the analytical model by
simulating the scattering, escape, and absorption of a starlike emission
spectrum by a clumpy distribution of interstellar dust, and then comparing
the predicted equilibrium dust temperatures and emerging UV-FIR spectrum
with the results of MCRT simulations. The analytical model is also used to
explore a large region of parameter space characterizing the clumpy medium,
and we explain why the dust temperature in clumps is lower than in the ICM.
The summary and conclusions are presented in \S 8.

\section{MONTE CARLO SIMULATION OF\protect\linebreak RADIATIVE TRANSFER}

\subsection{General Simulation Methods}

The Monte Carlo code we have developed simulates coherent scattering and
absorption of photons by dust with any kind of spatial distribution and
composition. The spatial distribution of the dust is specified by a
continuous function $\rho (x,y,z)$, the mass of dust per unit volume. In
practice this quantity is defined on a discrete three-dimensional grid. For
each wavelength simulated, the number of photons absorbed by the dust in
each volume element (voxel) of the 3D grid is saved, allowing computation of
the dust temperatures and resulting infrared emission spectrum.

The grid resolution is limited only by the available computer memory:
increasing the number of grid elements does not significantly affect the
computation time. This is achieved by employing the Monte Carlo method of
rejections (\cite{Neumann51}) in selecting the random distances each photon
travels between interactions (absorption or scattering) with the dust, as
described in \cite{LuxKob95}. The method proceeds as follows. Let ${\bf r}%
\equiv (x,y,z)$ be the initial position of the emitted photon, and $\rho
_{\max }$ be the maximum dust mass density in the direction $\widehat{{\bf l}%
}$ the photon is traveling. Assume temporarily that the density is uniform
and equal to $\rho _{\max }$ along its path. In that case the probability
that the photon will travel a distance $s$ without having an interaction is $%
\exp (-s\kappa \rho _{\max })$, where $\kappa $ is the dust mass extinction
coefficient. Applying the fundamental principle of the Monte Carlo method,
one can choose a uniformly distributed random variable $0<\zeta <1$ and set $%
\zeta $ to be equal to $1-\exp (-s\kappa \rho _{\max })$, the probability
that an interaction will occur, and then solve for $s,$ the random value for
the interaction distance, to give: 
\begin{equation}
s\,=\,-\frac{\ln (1-\zeta )}{\kappa \rho _{\max }}\qquad .  \label{randist}
\end{equation}
Note that $(\kappa \rho _{\max })^{-1}$ is the worst case mean-free-path.
One then has to play a rejection game in order to determine if the supposed
interaction at the new location, ${\bf r}^{\prime }={\bf r+}s\widehat{{\bf l}%
}$, is accepted as a real interaction, because the density is of course {\it %
not} everywhere equal to $\rho _{\max }$. We choose another uniformly
distributed random number $0<\vartheta <1$ and compare it to $\rho ({\bf r}%
^{\prime })/\rho _{\max }$, the ratio of the actual density at the event
location ${\bf r}^{\prime }$ to the maximum density along $\widehat{{\bf l}}$%
. If $\vartheta \leq \rho ({\bf r}^{\prime })/\rho _{\max }$ the supposed
interaction is accepted as real. In particular, if $\rho ({\bf r}^{\prime
})/\rho _{\max }=1,$ all values of the random variable $\vartheta $ will
fall below this ratio, and the interaction will be accepted as real. Thus $%
\rho ({\bf r}^{\prime })/\rho _{\max }$ is equal to the probability that an
interaction is real. If it happens that $\vartheta >\rho ({\bf r}^{\prime
})/\rho _{\max }$ the interaction is rejected and called a virtual
interaction. After a virtual interaction the photon is allowed to travel in
the same direction another random distance selected by eq.(\ref{randist}),
and the above steps are repeated until an interaction event is accepted as
real or the photon escapes. After evolving many photons in this manner, the
method effectively integrates the density along all directions of travel
while performing the Monte Carlo simulation of photon interactions. The 3D
medium need be specified with only a point density function, and the method
is simpler than performing numerical integrations along each photon path to
determine in which voxel the photon will interact with dust. For most
simulations one can set $\rho _{\max }$ equal to the maximum density of the
entire medium, thereby further simplifying the process. However if $\rho
_{\max }\gg \rho _{\min }$ and the high density regions occupy a very small
fraction of the volume, it is worthwhile to have an algorithm for
determining the maximum density in a given direction, otherwise the method
may iterate through many virtual interactions before getting to a real
interaction or escaping, using up more CPU time.

Our Monte Carlo code follows a large group of randomly emitted photons
simultaneously, typically $10^{5}$ at a time, through multiple scatterings
and to eventual absorption or escape. Each photon is given an initial flux
weight of unity, and this weight changes as the photon interacts with the
dust. The simulation proceeds via iterations, as follows. First, the group
of photons travel randomly selected distances (as described above) depending
on the dust density encountered, and then photons either escape the medium
(reducing the number in the group) or interact with dust. Photons that
interact are considered to experience both scattering and absorption: they
are reduced in flux weight by the dust scattering albedo $\omega \equiv
\sigma _{scat}/\sigma _{ext}$ (the ratio of scattering to extinction
cross-section), and a fraction $1-\omega $ of the energy of each interacting
photon is absorbed by the dust at the locations of interaction. The path of
each scattered photon is deflected by a random angle, $\theta _{scat},$
which is distributed according to the Henyey-Greenstein (HG) phase function (%
\cite{H-G41}), fully characterized by the parameter $g=\langle \cos \theta
_{scat}\rangle $, the average value of the cosine of the deflection angles,
also called the asymmetry parameter. The random scattering deflection angles
are selected by the method given in \cite{witt77}. The remaining photons
having new directions are fed back into the beginning of the iteration
procedure. This cycle of interaction, absorption and scattering is repeated
until the net flux of photons remaining in the medium is a small fraction of
the flux that has escaped (e.g. less than 5\%).

The flux of photons remaining after the iterations are terminated is divided
into escaping and absorbed fractions according to a new scheme that
considers the history of the escaping photons during each iteration and the
dust scattering albedo. Let $N_{k}$ be the number of photons remaining after
iteration $k$, and let $m$ be the total number of iterations performed. If
the remaining fraction after each iteration, given by the ratio $\beta
_{k}=N_{k}/N_{k-1}\leq 1$, is tending toward a constant value, then the
fraction of the final $N_{m}$ photons that will escape if the iterations
were continued ad-infinitum is approximately (see Appendix \ref{App FinF}) 
\begin{equation}
f_{m}^{\,esc}=\frac{1-\beta _{m}}{1-\omega \beta _{m}}\quad .
\label{fesc_final}
\end{equation}
Thus upon termination after $m$ scatterings we assume that a fraction $%
f_{m}^{\,esc}$ of the remaining flux escapes and a fraction $1-f_{m}^{\,esc}$
is absorbed at the current location. Testing of the formula has shown that
even if the multiple scattering is terminated after a few iterations when
the flux remaining is 20\% of the escaped flux, the final escaping fraction
of flux obtained after applying eq.(\ref{fesc_final}) is as accurate as if
the iteration was continued until the remaining flux is less than 1\% of the
escaped flux, and the total computation time is reduced. Of course when the
iterations are prematurely terminated, the locations of the photon
absorptions and exits are not quite as accurate as when the iterations are
continued, but the error is always less than the ratio of the flux that is
remaining to that which has already escaped. In any case, using eq.(\ref
{fesc_final}) is more intelligent than considering the remaining flux to be
all escaped, all absorbed, or split up using just the scattering albedo.

\subsection{Radiative Transfer in Two-Phase Clumpy Media\label%
{S.Clumpy Media}}

The main parameters defining a two-phase clumpy medium are the volume
filling factor of the clumps, $f_{c}$, the ratio of the clump to interclump
medium (ICM) densities, $\alpha =\rho _{c}/\rho _{icm}$, the total dust mass 
$M$, and the volume $V$ of the medium. Typical ranges that may describe
parts of the interstellar medium are $0.01<f_{c}<0.3$ and $10<\alpha <10^{7}$
(\cite{Spitzer-ISM}, \cite{vB89}, \cite{MWH92}, \cite{GvB93}). The number of
clumps and the clump sizes are secondary parameters, inversely related,
which we shall discuss later. To obtain a formula for the ICM density, first
define the density of the equivalent homogeneous medium to be $\rho
_{hom}=M/V.$ Then since the total dust mass is simply the sum of mass in
each phase, the homogeneous density can be expressed as 
\begin{eqnarray}
\rho _{hom} &=&\,\frac{M_{c}+M_{icm}}{V}  \nonumber \\
&=&\,\frac{V_{c}}{V}\frac{M_{c}}{V_{c}}\,+\,\frac{V_{icm}}{V}\frac{M_{icm}}{%
V_{icm}}  \nonumber \\
&=&\,f_{c}\rho _{c}+(1-f_{c})\rho _{icm}  \nonumber \\
&=&\;(\rho _{c}-\rho _{icm})f_{c}+\rho _{icm}\,\,,  \label{dhom}
\end{eqnarray}
where $M_{c}$ and $M_{icm}$ are the dust masses in the clumps and ICM
respectively, $V_{c}$ and $V_{icm}$ are the respective volumes, and so $\rho
_{c}=M_{c}/V_{c}$ is the density of dust in the clumps, $\rho
_{icm}=M_{icm}/V_{icm}$ the density in the ICM, and $f_{c}=V_{c}/V$ is the
filling factor of the clumps. Equation (\ref{dhom}) can be rearranged by
substituting for $\rho _{c}$ with $\alpha \rho _{icm}$ to give an equation
for the ICM density 
\begin{equation}
\rho _{icm}=\frac{\rho _{hom}}{(\alpha -1)\,f_{c}+1}\,\,\,.  \label{den_icm}
\end{equation}
As $f_{c}\rightarrow 0$, $\rho _{icm}\rightarrow \rho _{hom}$ and $\rho
_{c}\rightarrow \alpha \rho _{hom}$ from below, whereas if $f_{c}\rightarrow
1$, $\rho _{icm}\rightarrow \rho _{hom}/\alpha $ and $\rho _{c}\rightarrow
\rho _{hom}$ from above.

One possible type of two-phase clumpy medium is that of cubic clumps
randomly located on a body centered cubic lattice. The construction is
similar to that used in percolation theory, where the probability of lattice
site occupation is equivalent to the filling factor and the clumps are then
the occupied sites, and the cubic clumps do not overlap. \cite{wittgor96}
studied radiative transfer in such a two-phase clumpy medium defined on a
cubic lattice, and we repeated some of their Monte Carlo simulations giving
a successful test of our methods. In this work we use spherical clumps
defined on a high resolution 3D grid, allowing clump overlaps and the random
locations of clumps to be chosen at a higher resolution than the clump size.
In addition, the radiative transfer properties are then readily approximated
by analytic formulas, as we demonstrate in \S \ref{Esc_Abs_Clumpy} and \S 
\ref{Compare_MC}. The stochastic media in the analytical models of \cite
{boisse90} and \cite{hs93} do not impose any restrictions on the shapes of
the clumps, however, the models are developed for the case of isotropic
scattering.

To construct a two-phase medium with randomly located clumps that may
overlap, we must calculate the number of clumps needed to achieve the
desired filling factor. This can be derived from the volume of a clump, $%
\upsilon _{c}$, the total volume of the medium, $V$, and the filling factor, 
$f_{c}$, but the possible occurrence of overlapping clumps complicates the
calculation. Assume that the clumps are all identical. Defining the volume
fraction of a single clump as 
\begin{equation}
p\equiv \frac{\upsilon _{c}}{V}\quad ,  \label{prob_clump}
\end{equation}
then the probability that a random point is not in any clump is equal to $%
(1-p)^{N_{c}},$ were $N_{c}$ is the number of clumps, and therefore 
\begin{equation}
f_{c}=1-(1-p)^{N_{c}}  \label{ff_pvc}
\end{equation}
(see Appendix \ref{App Nc} for more discussion). Equation (\ref{ff_pvc}) is
easily solved for the total number of clumps, 
\begin{equation}
N_{c}\,=\,\,\frac{\ln (1-f_{c})}{\ln (1-p)}\quad ,  \label{Nclump}
\end{equation}
in terms of the total filling factor, $f_{c}$, and the single clump filling
factor, $p$. Although the above equations apply to identical clumps of any
shape, we shall consider only spherical clumps of radius $r_{c}$ with $%
\upsilon _{c}=4\pi r_{c}^{3}/3$. Given $f_{c},$ $r_{c},$ and $V,$ the clumpy
medium is constructed by calculating $N_{c}$ using equations (\ref
{prob_clump}) and (\ref{Nclump}), then selecting $N_{c}$ random points
uniformly distributed in $V,$ placing a sphere of radius $r_{c}$ and density 
$\rho _{c}$ around each point, setting the density between the spheres to $%
\rho _{icm},$ and finally discretizing the medium on a high resolution 3D
grid. When the randomly located clumps do overlap, the densities are {\it not%
} summed, so the medium is always characterized by only two possible
densities: that of the ICM and the clumps.

\begin{figure}[tbp]
\plotone{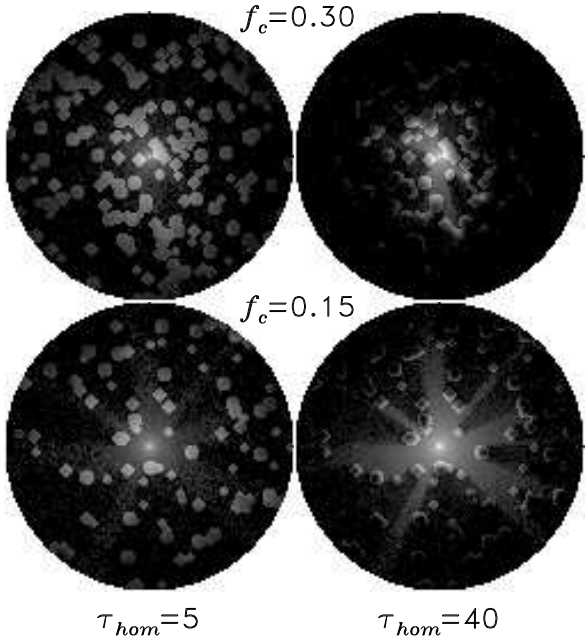}
\caption{ Spatial distribution of photons absorbed by dust in a
two-phase clumpy medium heated by a central point source.
The simulations were performed by Monte Carlo calculations,
and the images depict a 2D slice through the center of the sphere.
Scaling is logarithmic from the minimum of $0.1$ photons (black)
to the maximum of $4 \times 10^4$ photons absorbed per voxel (white).
For more details see \S2.2. }
\label{abs_map_sc}
\end{figure}

\begin{figure}[tbp]
\plotone{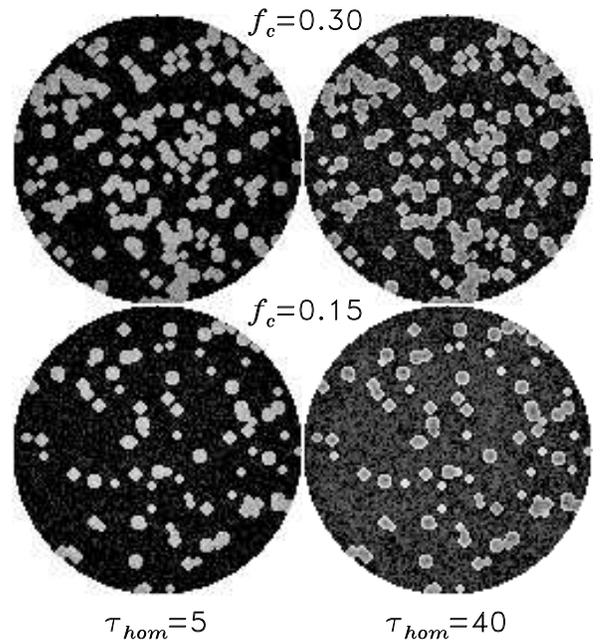}
\caption{ Same as Figure~\ref{abs_map_sc} for dust heated by
a uniform distribution of internal sources.
Scaling is logarithmic from the minimum of 0.1 photons (black)
to the maximum of 40 photons absorbed per voxel (white).
For further description see \S2.2. }
\label{abs_map_su}
\end{figure}

\begin{figure}[tbp]
\plotone{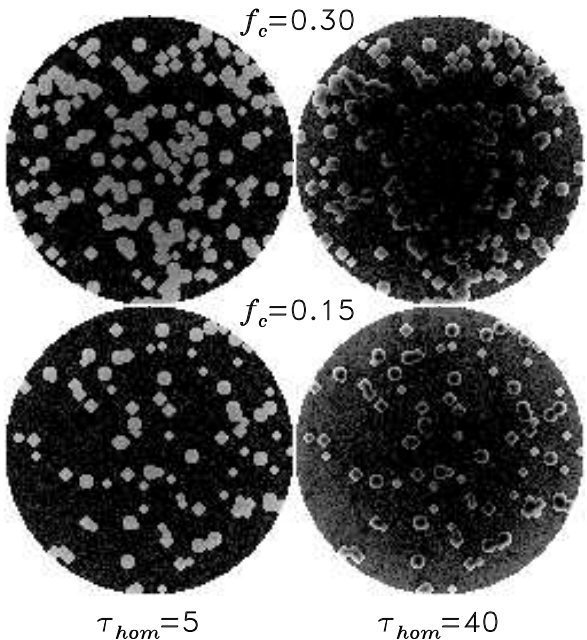}
\caption{ Same as Figure~\ref{abs_map_sc} for dust heated by
an external uniformly illuminating source. Scaling is logarithmic
from the minimum of 0.1 photons (black) to
the maximum of 80 photons absorbed per voxel (white).
For further description see \S2.2. }
\label{abs_map_sx}
\end{figure}

Monte Carlo simulations of radiative transfer were computed for three types
of photon sources in a two-phase clumpy medium contained within a sphere of
unit radius $R_{S}=1$, in which the spherical clumps have radius $r_{c}=0.05$
and are $100$ times denser than the ICM ($\alpha =100$). The 3D rectangular
grid used to represent the medium has resolution of $127^{3}$ voxels, so
that the clump diameters are about $6$ grid elements. Figures~\ref
{abs_map_sc}, \ref{abs_map_su} and \ref{abs_map_sx} show our results for an
array of clump filling factors and optical depths, keeping $r_{c}$ and $%
\alpha $ constant. Three million photons were followed in each Monte Carlo
simulation. Each of the images is the map of photons absorbed in a 2D slice (%
$127\times 127$ pixels) through the center of the sphere. The grey scale
from black to white indicates minimum to maximum absorption, on a
logarithmic scale. Figure~\ref{abs_map_sc} represents the case of a single
central isotropic luminosity source, analogous to a star or a centrally
condensed cluster of stars in an H II region. Figure~\ref{abs_map_su}
represents the case of isotropic photon emission distributed uniformly in
the sphere, analogous to a uniform distribution of stars in a galaxy. Figure~%
\ref{abs_map_sx} represents the case of an external uniformly illuminating
source of isotropic photons, analogous to cold molecular clouds illuminated
by the diffuse interstellar radiation field (IRF). In all cases the dust is
characterized by a scattering albedo of $\omega =0.6$ and asymmetry
parameter $g=0.6$, which are typical values for UV photons scattering off
dust grains (\cite{gordalb94}). Recall that $g=0$ results in isotropic
scattering and $g=1$ gives forward scattering.

Moving vertically in the array of slices in the Figures corresponds to
increasing the volume filling factor $f_{c}$ of the clumps while keeping the
total dust mass constant (in each column). Thus the number of clumps
appearing in each slice increases and $\rho _{icm}$ decreases from bottom to
top. Moving horizontally increases the equivalent homogeneous optical depth
of the sphere, 
\begin{equation}
{\tau }_{hom}\,=\,\,\kappa \rho _{hom}\,R_{S}\,\,,
\end{equation}
which is the radial optical depth of extinction (absorption plus scattering)
that would result if the dust were distributed uniformly throughout the
sphere instead of being clumpy. Increasing ${\tau }_{hom}$ can be viewed as
either increasing the dust abundance (increasing $\rho _{icm}$ and $\rho
_{c} $) or changing the wavelength of the photons, either case resulting in
more absorption.


For the case of a central source (Figure~\ref{abs_map_sc}), as ${\tau }%
_{hom} $ increases the ICM absorbs more photons and the clumps become
opaque, creating the apparent shadows behind the clumps, when $f_{c}$ is
small. However scattering by the dust causes photons to go behind the clumps
and become absorbed, thus diminishing the effect of what would otherwise be
completely dark shadows in the case of no scattering. As the clumps become
opaque (${\tau }_{hom}=40$), absorption occurs more at the clump surfaces.
When $f_{c}$ is increased (keeping the total dust mass constant) the effect
of shadows merges into the appearance of an absorption cavity. Similar
effects occur for the case of uniformly distributed sources (Figure~\ref
{abs_map_su}), however not as dramatic as when the clumps are illuminated by
a central source, since there are no shadows when the photons are emitted
everywhere in the medium. At high $f_{c}$ the clumps dominate the medium and
absorb most of the photons. In the case of a uniformly illuminating external
source (Figure~\ref{abs_map_sx}), when ${\tau }_{hom}$ and $f_{c}$ are large
we find that most of the impacting photons are absorbed in the outer layers
of the spherical region and the center is shielded from radiation. However,
for small $f_{c}$ the center of the sphere is less shielded and instead the
clump centers are shielded, most photons being absorbed on the surfaces of
the clumps. In general, the clumpy medium allows photons to penetrate
farther into the spherical region than if the medium was homogeneous, and
this fact was the motivation and a conclusion in the work of \cite{boisse90}%
, \cite{hs93}, and HP93, for the case of plane-parallel slab geometry.

\begin{figure}[tbp]
\plotone{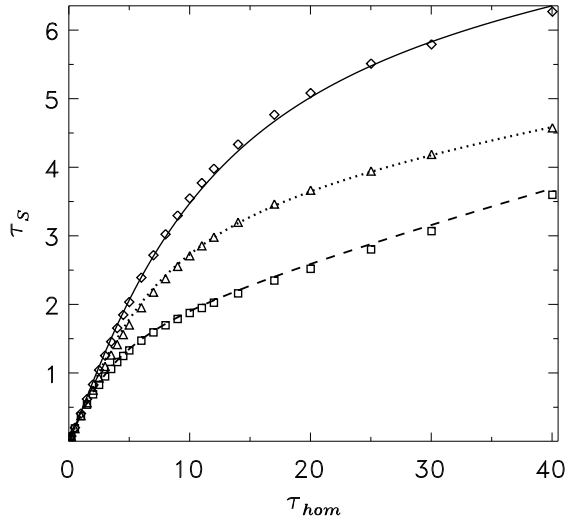}
\caption{ The effective optical radius, $\tau_S$,
of clumpy media in a sphere is shown as a function of
the equivalent homogeneous optical depth, ${\tau }_{hom}$,
for three values of $f_{c}$, the clump filling factor.
The symbols show $\tau_S$ given by eq.(\ref{tau_eff})
using the results of Monte Carlo simulations including scattering
($\omega=0.6$ and $g=0.6$),
with $f_{c}=0.1$ (squares), 0.2 (triangles), and 0.3 (diamonds).
See \S2.2 for more discussion.
The curves are produced by analytic approximations developed in this paper,
and are discussed in \S\ref{Compare_MC}. }
\label{tauS_ffs}
\end{figure}

By observing the fraction of photons that escape in the case of a central
point source we can compute the effective optical depth of the medium at a
given wavelength $\lambda $ as 
\begin{equation}
\tau _{S}(\lambda )=-\ln \left[ \frac{L_{esc}(\lambda )}{L_{0}(\lambda )}%
\right]  \label{tau_eff}
\end{equation}
where $L_{0}(\lambda )$ is the luminosity of the central source and $%
L_{esc}(\lambda )$ the escaping luminosity (for simplicity we shall
hereafter drop any explicit dependence on $\lambda $). We use the notation $%
\tau _{S}$ instead of $\tau _{eff}$ because $\tau _{eff}$ is reserved for
the case of no scattering (absorption plus scattering considered as
interaction), whereas $\tau _{S}$ includes the effects of scattering and is
therefore dependent on the geometry of the medium, which in this case is
spherical. Equation (\ref{tau_eff}) is analogous to eq.(\ref{eff_tau}), and
it follows that $\tau _{S}<\tau _{eff}<{\tau }_{hom}$ for a clumpy medium.
Figure~\ref{tauS_ffs} shows the behavior of $\tau _{S}$ as a function of ${%
\tau }_{hom}$ for three cases of $f_{c}$, with other parameters having the
values mentioned above. The diamonds indicate results of Monte Carlo
simulations when $f_{c}=0.3$ (corresponding to top row in Figure~\ref
{abs_map_sc}), the triangles show $f_{c}=0.2$, and the squares show $%
f_{c}=0.1$. There are two slopes in each curve of $\tau _{S}$ versus ${\tau }%
_{hom}$, corresponding to the two phases of the medium. The steeper slope at
low values of ${\tau }_{hom}$ is due to absorption by the increasing optical
depth of the clumps. As ${\tau }_{hom}\rightarrow \infty $ the clumps become
opaque and any further increase in absorption is due to the ICM, and because
it has a lower density the slope of the $\tau _{S}$ curves is less. The
solid, dotted and dashed curves are produced by the analytic approximations
presented in sections \ref{Esc_Abs_Homog} and \ref{Esc_Abs_Clumpy}, and the
agreement with Monte Carlo results is discussed further in \S \ref
{Compare_MC}.

\section{ESCAPE AND ABSORPTION IN\protect\linebreak HOMOGENEOUS MEDIA\label%
{Esc_Abs_Homog}}

Since Monte Carlo simulations can require a large amount of computer time,
it is useful to have analytical approximations for the basic results of
radiative transfer, such as, the fraction of photons that escape a bounded
medium. In this section we present such approximations for homogeneous
spherical media with internal or external sources. The escape and absorption
probability approximations are also compared with results of Monte Carlo
simulations to assess their accuracy. These escape probability formulae are
later applied to the case of clumpy media.

\begin{figure}[tbp]
\plotone{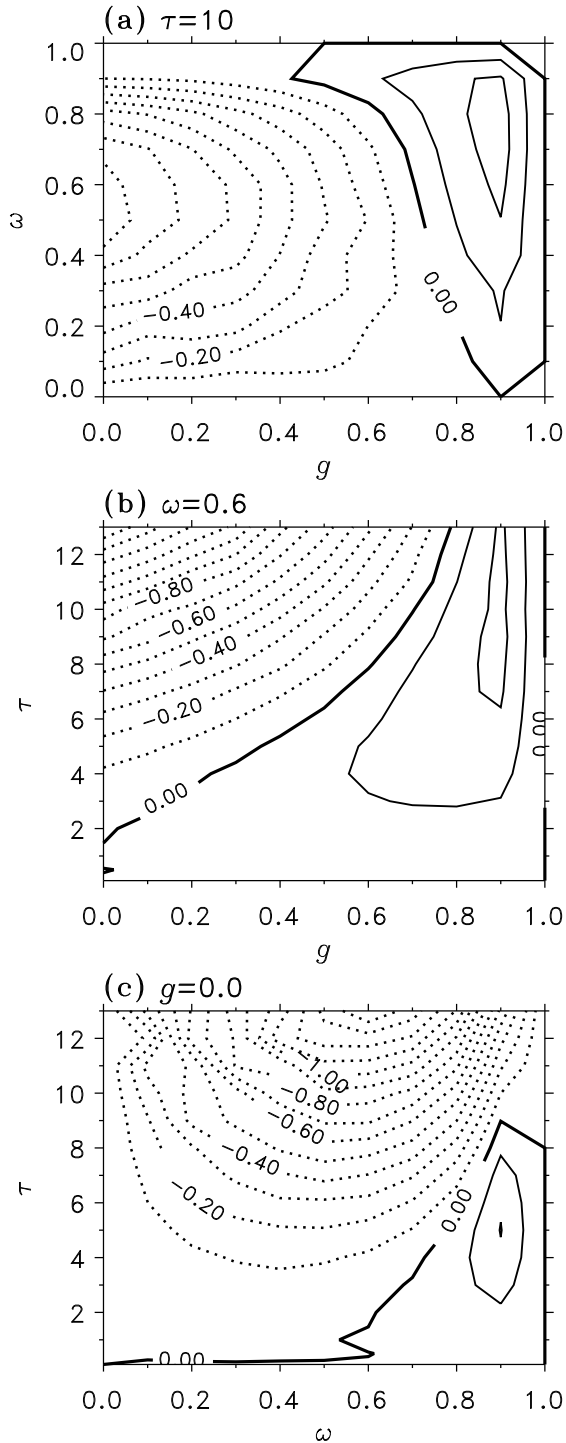}
\caption{ Contours show the difference between
the effective optical depth of a homogeneous sphere as
predicted by eq.(\ref{tau_cs_sphere}) and
that resulting from Monte Carlo simulations. }
\label{cs_ep_mc}
\end{figure}%

\subsection{Central Isotropic Point Source}

Consider an isotropic point source in the center of a spherical homogeneous
medium. When the medium does not scatter photons the escape probability is
simply $e^{-\tau }$, where $\tau =\rho \kappa R$ is the optical radius of
the sphere, characterized by a radius $R$ and a mass density $\rho ,$ and
where $\kappa $ is the absorption cross-section of the dust per unit mass.
If the medium also scatters photons, we can construct an analytical
approximation for the effective optical radius of the sphere, $\tau _{S}$,
defined as the negative natural logarithm of the escaping fraction of
photons, as in eq.(\ref{tau_eff}). The approximation formula is based on
limiting cases of the optical parameters. When the optical depth of
extinction (absorption and scattering), $\tau =\tau _{abs}+\tau _{scat}$, is
large and the scattering is isotropic ($g\sim 0$), then theoretical analysis
(\cite{Rybicky79}) suggests that $\tau _{S}$ is approximately 
\begin{equation}
\tau _{S}\,\simeq \,\tau \,(1-\omega )^{\frac{1}{2}}\,\,;\,\,(\tau \gg 1%
\text{ and }g\sim 0),  \label{taueff_g=0}
\end{equation}
where $\omega \equiv \tau _{scat}/\tau $ is the scattering albedo. When the
optical radius is small ($\tau \ll 1$) and $g$ is any value, we expect that 
\begin{equation}
\tau _{S}\,\simeq \,\tau \,(1-\omega )\,\,\,;\,\,(\tau \ll 1\text{ and any }%
g),
\end{equation}
which is also an exact formula in the case of purely forward scattering at
any optical depth. We can interpolate between these extreme cases using the
following formula: 
\begin{equation}
\tau _{S}(\tau ,\omega ,g)\,\,\equiv \,\,\tau \,(1-\omega )^{\chi (\tau
,g)}\quad ,  \label{tau_cs_sphere}
\end{equation}
where the interpolation exponent is given by 
\begin{equation}
\chi (\tau ,g)\,\,\equiv \,\,1\,-\,\frac{1}{2}\left( 1-e^{-\tau /2}\right)
(1-g)^{\frac{1}{2}}\quad .  \label{tau_cs_sphere_2}
\end{equation}
The probability of escaping from the homogeneous sphere is then defined as 
\begin{equation}
{\cal P}_{esc}^{c}(\tau ,\omega ,g)\,\,\equiv \,\,\exp \left[ -\tau
_{S}(\tau ,\omega ,g)\right] \,\quad ,  \label{EP1_cs}
\end{equation}
where the superscript ``{\it c}'' indicates that this is for a central point
source.

To check this approximation, the Monte Carlo code was used to compute a 3D
matrix of effective optical radii as a function of optical parameters in the
ranges $0<\tau \leq 13$ with $\Delta \tau \leq 1$, $0\leq \omega \leq 1$
with $\Delta \omega =0.1$, and $0\leq g\leq 1$ with $\Delta g=0.1$,
following more than $10^{6}$ photons in each simulation. Figure~\ref
{cs_ep_mc} shows contours of the difference between $\tau _{S}(\tau ,\omega
,g)$ given by eq.(\ref{tau_cs_sphere}) and $\tau _{S}$ from Monte Carlo
simulations. The differences are shown on 2D slices through the 3D parameter
space, as a function of (a): $(g,\omega )$ with $\tau =10$; (b): $(g,\tau )$
with $\omega =0.6$; and (c): $(\omega ,\tau )$ with $g=0$. The bold line in
each panel depicts the region in parameter space where the agreement is
perfect. The dotted contour lines indicate where eq.(\ref{tau_cs_sphere})
gives values less than the Monte Carlo results. The worst case of the
approximation occurs when $\omega =0.6$, $g=0$, and $\tau >4$, as panels (b)
and (c) show that $\tau _{S}(\tau ,\omega ,g)$ underestimates the Monte
Carlo results by a value exceeding unity. Also, panels (a) and (c) show that
eq.(\ref{taueff_g=0}) is not always a good approximation. However the
difference is relatively small compared to the value of $\tau _{S}\approx 10$
at $(\tau ,\omega ,g)=(13,0.6,0)$. In addition, when modeling radiative
transfer through interstellar dust, isotropic scattering occurs at
near-infrared (NIR) wavelengths for which the optical depth is smaller than
at UV wavelengths where the scattering is more forward directed, and so the
approximation is likely to be good at all wavelengths, as shown in panel (b).

\subsection{Uniformly Distributed Emission\label{EPSS}}

Consider a spherical homogeneous medium with a uniform distribution of
isotropically emitting sources. When there is no scattering, \cite{oster89}
derived an exact solution for the photon escape probability given by 
\begin{equation}
P_{e}(\tau )\,\equiv \,\frac{3}{4\tau }\left[ 1-\frac{1}{2\tau ^{2}}+\left( 
\frac{1}{\tau }+\frac{1}{2\tau ^{2}}\right) e^{-2\tau }\right]
\label{Oster_EP}
\end{equation}
where $\tau =$ $\rho \kappa R$ is the optical radius of the sphere ($R=$
radius of sphere, $\rho =$ density of absorbers, and $\kappa =$ absorption
cross-section). We find that $P_{e}(\tau )$ agrees exactly with the escaping
fraction of photons computed in Monte Carlo simulations. Using the optical
depth of extinction resulting from both absorption and scattering, $\tau
_{ext}=\tau _{abs}+\tau _{scat}$, we can get the probability, $P_{e}(\tau
_{ext})$, that photons will escape the spherical medium without any
interactions (absorption or scattering), and we call it the extinction
escape probability. Making the assumption that the scattered photons have
the same spatial and angular distribution as the emission sources, the
extinction escape probability can be applied recursively to arrive at the
scattering escape probability \cite{Lucy91} 
\begin{equation}
{\cal P}_{esc}^{u}(\tau ,\omega )\,\equiv \,\frac{P_{e}(\tau )}{1-\omega
[1-P_{e}(\tau )]}\,\,,  \label{Lucy_EP}
\end{equation}
where $\tau \equiv \tau _{ext}$ and $\omega =\tau _{scat}/\tau _{ext}$ is
the scattering albedo. This formula estimates the effects of scattering,
namely that scattered photons may eventually escape, thereby increasing the
escape probability. The superscript ``{\it u}'' indicates that ${\cal P}%
_{esc}^{u}$ is for uniformly distributed emitters in a sphere. Actually, $%
P_{e}(\tau )$ in the formula can be any extinction escape probability for
any geometry, but this paper will concentrate on the case of spherical
geometry, for which eq.(\ref{Oster_EP}) applies. The derivations of
equations (\ref{Oster_EP}) and (\ref{Lucy_EP}) are reviewed in Appendix \ref
{App Eprob}.

\begin{figure}[tbp]
\plotone{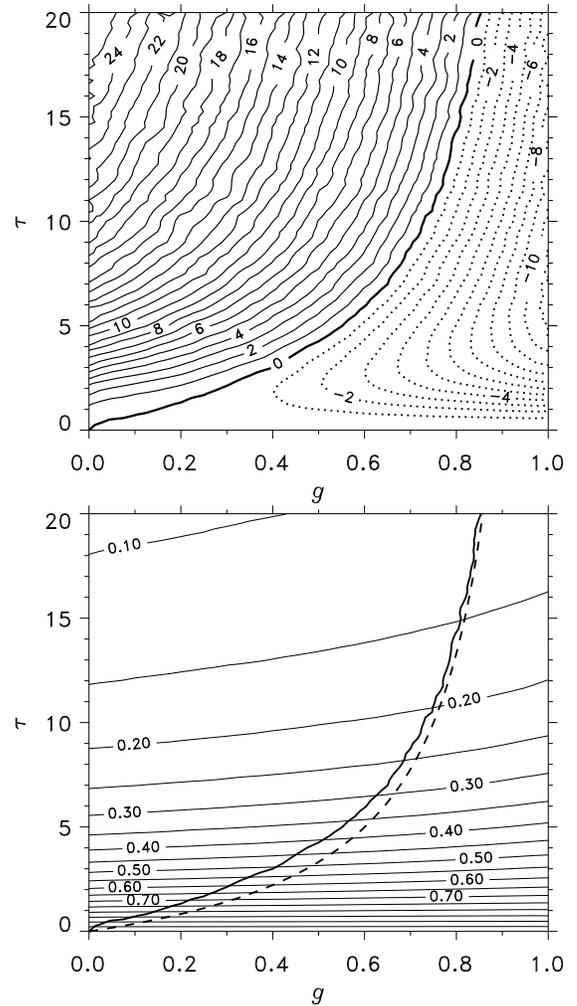}
\caption{ Analysis of the escaping fraction of photons that are emitted
by sources uniformly distributed in a homogeneous sphere,
as a function of the scattering asymmetry parameter $g$,
and $\tau$, the extinction optical radius of the sphere,
with albedo $\omega =0.7$.
The top panel shows
contours of the percent relative difference between the escaping
fraction given by the OLEP formula
[eqs.(\ref{Oster_EP}) and (\ref{Lucy_EP})]
and that predicted by Monte Carlo simulations.
The bottom panel shows
contours labeled with the escaping fraction predicted by
Monte Carlo simulations, over-plotted with the zero difference contour
(thick solid line) from the top graph, referred to as $g^{*}(\tau)$.
The dashed line is an approximation of $g^{*}(\tau)$ given by eq.(\ref{g_tau}). }
\label{ep_mc_gp}
\end{figure}%

Equations (\ref{Oster_EP}) and (\ref{Lucy_EP}), which we shall call the
Osterbrock-Lucy escape probability (OLEP) formula, were tested extensively
against Monte Carlo radiative transfer simulations with multiple scattering
and found to be a reasonable approximation of the fraction of photons
escaping from a spherical homogeneous medium. Since the scattering asymmetry
parameter, $g=\langle \cos \theta _{scat}\rangle $, does not enter into the
formula for ${\cal P}_{esc}^{u}(\tau ,\omega )$, the value of $g$ for which
the OLEP formula agrees exactly with Monte Carlo simulations is found to
depend on the extinction optical depth $\tau $. The upper panel in Figure~%
\ref{ep_mc_gp} shows contours of the percent relative difference between the
OLEP formula and escaping fraction of photons found by Monte Carlo
calculations (following $10^{6}$ photons in each simulation), as a function
of $g$ and $\tau $, when the albedo is $\omega =0.7$. The values of $g$ for
which the OLEP agrees with Monte Carlo results are indicated by the zero
contour level drawn with a thick solid line. The dotted contours indicate
where the escape probability formula underestimates the escaping fraction,
and the thin solid contours indicate where it overestimates the escaping
fraction. For optically thin situations ($\tau <1)$ the OLEP formula agrees
well with the Monte Carlo simulations for all values of $g$, with the best
agreement occurring when the scattering is nearly isotropic ($g\sim 0$). As
the optical depth increases the agreement shifts toward more forward
scattering cases ($g\rightarrow 1$) of the Monte Carlo simulations. The
maximum relative difference in the range shown is 26\% overestimation at $%
\tau =20$, $\omega =0.7$, and $g=0$, where the escaping fraction is only
0.10. The behavior of the relative difference is similar for other values of
the albedo, and always decreases with albedo, to zero as $\omega \rightarrow
0$, since then ${\cal P}_{esc}^{u}(\tau ,\omega )\rightarrow P_{e}(\tau )$.
The value of $g$ for which ${\cal P}_{esc}^{u}(\tau ,\omega )$ agrees with
Monte Carlo calculations (the zero difference contour) is found to be a
function of optical depth only, independent of the albedo $\omega $, and so
we designate this special value by $g^{*}(\tau )$.

The lower panel in Figure~\ref{ep_mc_gp} shows contours labeled with the
escaping fraction resulting from the Monte Carlo calculations, and the thick
solid line is again $g^{*}(\tau )$, the zero difference contour from the
upper panel. We find that a reasonable empirical approximation for $%
g^{*}(\tau )$ is given by 
\begin{equation}
g^{*}(\tau )\,\simeq \,1\,-\,\frac{3.3}{\tau +3.3}\quad ,  \label{g_tau}
\end{equation}
as shown by the dashed line.

There is a simple explanation for why the best agreement between ${\cal P}%
_{esc}^{u}(\tau ,\omega )$ and Monte Carlo simulations occurs at a value of $%
g$ that approaches unity as $\tau \rightarrow \infty $. To explain this fact
we rewrite eq.(\ref{Lucy_EP}) to read 
\begin{equation}
{\cal P}_{esc}^{u}(\tau ,\omega )=\frac{1}{\omega \,+\,(1-\omega
)/P_{e}(\tau )}\,,  \label{EP_scat}
\end{equation}
where $P_{e}(\tau )$ is the extinction escape probability from a sphere
given by eq.(\ref{Oster_EP}). As $\tau \rightarrow \infty $ the extinction
escape probability approaches zero as $P_{e}(\tau )\sim 3/4\tau $ . Applying
this limiting behavior to eq.(\ref{EP_scat}) gives 
\begin{equation}
{\cal P}_{esc}^{u}(\tau ,\omega )\sim \frac{1}{\omega +4\tau (1-\omega )/3}%
\sim \frac{3}{4\tau (1-\omega )}\,,  \label{EP_Lim_g=1}
\end{equation}
since the term \thinspace $4\tau \,(1-\omega )/3\,\gg \,\omega $ as $\tau
\rightarrow \infty $. Now we recognize that $\tau \,(1-\omega )$ is exactly
the optical depth in the case of purely forward scattering, when $g=1$,
since then scattering acts like a reduction in absorption. The forward
scattering escape probability is then exactly $P_{e}\left[ \tau (1-\omega
)\right] ,$ which has the same limiting behavior as eq.(\ref{EP_Lim_g=1})
when $\tau \rightarrow \infty $ : 
\begin{equation}
P_{e}\left[ \tau (1-\omega )\right] \sim \,\frac{3}{4\tau \,(1-\omega )\,}%
\quad .  \label{EP_g=1_Lim}
\end{equation}
Therefore as $\tau \rightarrow \infty $ the OLEP approximation approaches
from below the exact escape probability for purely forward scattering: 
\begin{equation}
{\cal P}_{esc}^{u}\left( \tau ,\omega \right) \,\rightarrow \,P_{e}\left[
\tau (1-\omega )\right] \,.  \label{EP_Lim_EP_g=1}
\end{equation}
Since the left side is valid at $g=g^{*}(\tau )$ and the right side is valid
at $g=1$, the limiting behavior implies that $g^{*}(\tau )\rightarrow 1$ as $%
\tau \rightarrow \infty $, which is satisfied by the approximation equation (%
\ref{g_tau}).

In summary, ${\cal P}_{esc}^{u}(\tau ,\omega )$ overestimates the actual
escape probability when $g<g^{*}(\tau )$ and underestimates it when $%
g>g^{*}(\tau )$. For values of $g$ in the range $g^{*}(\tau )<g<1$ the
escape probability approximation can be improved by linear interpolation in $%
g$ between the values ${\cal P}_{esc}^{u}(\tau ,\omega )$ $\,$at $g^{*}(\tau
)$ and $\,P_{e}\left[ \tau (1-\omega )\right] $ at $g=1$. If an analytical
approximation for the escape probability when $g=-1$ (purely backward
scattering) were known, then three point parabolic interpolation could be
used to obtain a very good approximation to the actual escape probability
for all $g,$ and hence all $(\tau ,\omega ,g)$. When modeling absorption and
scattering by dust, large optical depths usually occur at UV wavelengths for
which the scattering by dust tends to be more forward oriented, having
values of $g$ near $g^{*}(\tau )$, so the error incurred by using the OLEP
approximation may be relatively small.

\subsection{Finite Number of Uniformly Distributed Point Sources\label%
{Sec_Nstar}}

\begin{figure}[tbp]
\plotone{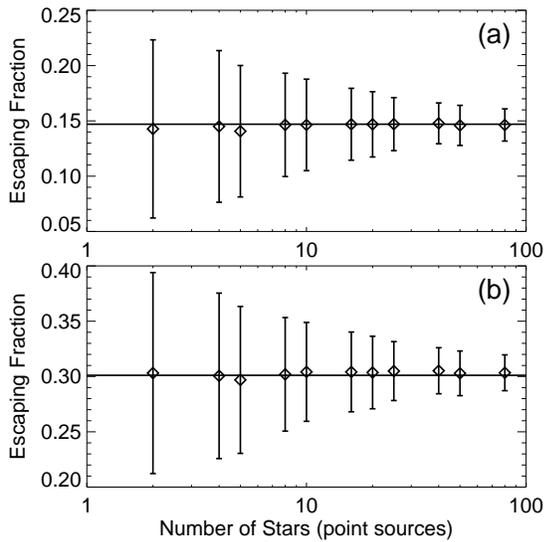}
\caption{ Monte Carlo simulations of the fraction of photons that escape
from a homogeneous sphere of dust after being emitted by a small number of
point sources (``stars''), as indicated on the horizontal axes. Case (a) is
for absorption only ($\omega=0$), case (b) is with scattering ($\omega=0.6$
and $g=0.6$), and the optical radius of the sphere is $\tau=5$ in both
cases. The diamonds and error bars are each the means and standard deviations
of 200 Monte Carlo simulations. The
solid horizontal lines are predictions by the OLEP formula.
The figure is discussed in \S\ref{Sec_Nstar}. }
\label{ep_nstar}
\end{figure}%

The question arises as to whether the escape probability formula can be
successfully applied to a discrete collection of randomly placed isotropic
point sources (``stars'') instead of a continuous uniform distribution of
emitters, and in particular what is the minimum number of stars for which
the escape probability formula gives correct results. To answer these
questions we performed Monte Carlo simulations for cases of 2 to 80 randomly
placed ``stars'' in a homogeneous sphere, and for each case we simulated 200
trials, where in each trial $10^{5}$ photons were followed to find the
fraction that eventually escapes. The results are shown in Figure~\ref
{ep_nstar} for when the optical radius of the sphere is $\tau =5$, for the
case (a) of no scattering ($\omega =0$) and case (b) with scattering ($%
\omega =0.6$ and $g=0.6$). The horizontal lines in cases (a) and (b) are the
escape probabilities $P_{e}(\tau )$ and ${\cal P}_{esc}^{u}(\tau ,\omega ) $%
, respectively, given by equations (\ref{Oster_EP}) and (\ref{Lucy_EP}).
Each diamond symbol shows the average escaping fraction of the Monte Carlo
trials for a fixed number of stars, and the error bars are plus and minus
one standard deviation from the average value. The distribution of escaping
fractions from the Monte Carlo trials is approximately Gaussian. We find
that the standard deviation of the trials goes as the inverse square root of
the number of stars $n_{s}$, as expected for sums of random variables, and
in particular $\sigma \sim 0.1\,(n_{s})^{-0.5}$ for all $\tau >1.$ Since the
escaping fraction in the case of no scattering behaves as $P_{e}(\tau )\sim
3/4\tau $ when $\tau \rightarrow \infty ,$ the expected relative deviation ($%
\sigma /P_{e}$) in the escaping fraction of photons from $n_{s}$ stars
randomly located in a sphere of optical depth $\tau \gg 1$ will be
approximately $0.133\,\tau \,n_{s}^{-0.5}$ for the case of absorption only,
and less when scattering is also involved. Thus the number of stars must
increase as the square of the optical depth to maintain the same expected
relative deviations. The analytic escape probability agrees well with the
average escaping fraction found for each group of Monte Carlo trials, it is
just a question of what relative deviation is acceptable.

\subsection{Uniformly Illuminating External Source\label{external_src}}

In this section we introduce approximations for the probability that
externally emitted isotropic photons will interact with and get absorbed by
a homogeneous medium contained in a sphere. The term interaction includes
both absorption and scattering events. Given that a distribution of photons
encounters the sphere at a random impact parameters, the probability of
interaction at any location inside the sphere is 
\begin{equation}
P_{i}(\tau )=1-\frac{1}{2\tau ^{2}}+\left( \frac{1}{\tau }+\frac{1}{2\tau
^{2}}\right) e^{-2\tau } \,\, ,  \label{Pinteract}
\end{equation}
where $\tau =$ $\rho \kappa R$ is the extinction optical radius of the
sphere. This equation is derived by averaging the transmission over all
possible impact parameters and computing the ratio of net absorption versus
impacting flux, using the same techniques as for the derivation of eq.(\ref
{Oster_EP}) (see Appendix \ref{App Eprob}). When $\tau \ll 1$ we have $%
P_{i}(\tau )\sim 4\tau /3,$ which is the optical path length through a
sphere averaged over all impact parameters. Equation (\ref{Pinteract}) was
utilized by \cite{neufeld91} and \cite{hobpad93} in their models for clumpy
media, as we shall discuss in \S \ref{MG_Approx}. Note that equations (\ref
{Pinteract}) and (\ref{Oster_EP}) are related by 
\begin{equation}
P_{i}(\tau )=\frac{4\tau }{3}\,P_{e}(\tau )\quad ,  \label{Pe_Pi}
\end{equation}
giving a duality between absorption (interaction) of an external source and
escape (non-interaction) of an internal source.

If the medium scatters photons, we would like to know what fraction of $%
P_{i}(\tau ),$ the interacting fraction, will eventually escape due to
multiple scattering events, and thus determine what fraction of photons are
actually absorbed. In the special case of purely forward scattering there is
an exact formula for the absorbed fraction of photons. When the scattering
deflection angle is always zero, it is equivalent to no scattering with a
reduced optical depth equal to $\tau (1-\omega ),$ where $\omega $ is the
scattering albedo. So the actual fraction of photons absorbed in the purely
forward scattering case is 
\begin{equation}
{\cal P}_{abs}^{x}(\tau ,\omega \,;\,g=1)\,\equiv \,P_{i}\left[ \tau
(1-\omega )\right] \quad ,  \label{Pxabs_g=1}
\end{equation}
where the superscript ``$x$'' indicates that this is for the case of an
external source and $g$ is the scattering asymmetry parameter. As expected,
equations (\ref{Pinteract}) and (\ref{Pxabs_g=1}) agree exactly with Monte
Carlo simulations.

For non-forward scattering cases, when $g<1$, we can use the methods
discussed in \S \ref{EPSS} for the case of internal uniformly distributed
emitters, namely the OLEP equations (\ref{Oster_EP}) and (\ref{Lucy_EP}).
Given that photons interact with the medium inside the sphere, the
probability that they scatter is simply $\omega $. These scattering events
can be regarded as re-emitted photons, and assuming that they are
approximately uniformly distributed in the sphere, the scattered photons
have a probability ${\cal P}_{esc}^{u}(\tau ,\omega )$ of escaping [eq.(\ref
{Lucy_EP})]. The fraction of the interacting photons that actually get
absorbed is then $1-$ $\omega {\cal P}_{esc}^{u}(\tau ,\omega )$. Thus an
approximation for the fraction of photons absorbed is 
\begin{equation}
{\cal P}_{abs}^{x}(\tau ,\omega \,)\,\equiv \,P_{i}\left( \tau \right)
\,\left[ 1-\omega {\cal P}_{esc}^{u}(\tau ,\omega )\right] .  \label{Pxabs}
\end{equation}
Recall that ${\cal P}_{esc}^{u}(\tau ,\omega )$ is valid only for $%
g=g^{*}(\tau )$ as discussed in \S \ref{EPSS}, and so we expect ${\cal P}%
_{abs}^{x}(\tau ,\omega \,)$ to follow the same pattern. When $\tau _{c}$ is
large, the penetration depth of photons decreases so they will not be
uniformly distributed in the sphere, as assumed above, but this does not
severely affect the application the approximation as we show next.

\begin{figure}[tbp]
\plotone{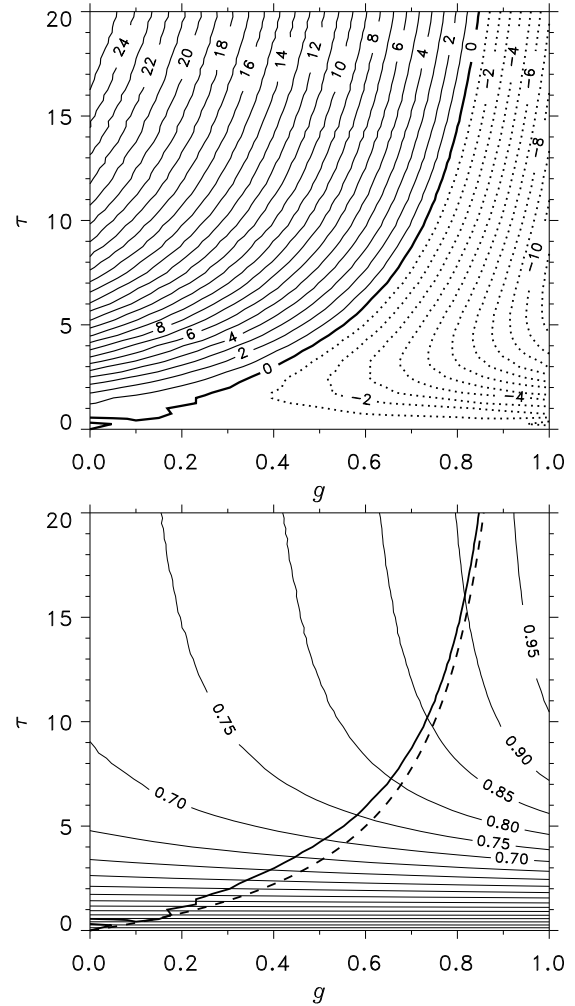}
\caption{ Analysis of the absorbed fraction of photons from
an external source which uniformly illuminates a homogeneous sphere,
when the scattering albedo is $\omega=0.7 $.
The top panel shows
the percent relative difference between the absorption probability
given by eq.(\ref{Pxabs}) and 
the absorbed fraction found by Monte Carlo
simulations, as a function of the asymmetry parameter $g$,
and $\tau$, the extinction optical radius of the sphere.
The contours in the bottom panel are labeled with the absorbed fractions
computed by Monte Carlo simulations.
The thick solid line is the locus of
agreement between ${\cal P}_{abs}^{x}$ [eq.(\ref{Pxabs})] and Monte
Carlo simulations (zero difference contour from top graph),
referred to as $g^{*}(\tau)$.
The dashed line
is an approximation of $g^{*}(\tau)$ given by eq.(\ref{g_tau}). }
\label{iap_mc_gp}
\end{figure}%

The upper panel in Figure~\ref{iap_mc_gp} shows contours of the percent
relative difference between ${\cal P}_{abs}^{x}(\tau ,\omega )$ and absorbed
fraction of photons found by Monte Carlo calculations, as a function of $g$
and $\tau $, for $\omega =0.7$. The values of $g$ for which eq.(\ref{Pxabs})
agrees with Monte Carlo results are indicated by the zero contour level
drawn with a thick solid line. The dotted contours indicate where the
absorption probability formula (\ref{Pxabs}) underestimates the absorbed
fraction, and the thin solid contours indicate where it overestimates. The
difference is small and independent of $g$ for optically thin situations ($%
\tau <1)$, and as the optical depth increases the agreement shifts toward
more forward scattering cases ($g\rightarrow 1$) of the Monte Carlo
simulations. Not surprisingly, the behavior of the relative difference
contours is identical to that in Figure~\ref{ep_mc_gp} of \S \ref{EPSS},
since we are using ${\cal P}_{esc}^{u}(\tau ,\omega )$ in the formula. The
relative difference decreases as $\omega \rightarrow 0$, since then ${\cal P}%
_{abs}^{x}(\tau ,\omega )\rightarrow P_{i}\left( \tau \right) $ and $%
P_{i}\left( \tau \right) $ is an exact formula.

The lower panel in Figure~\ref{iap_mc_gp} shows contours labeled with values
of the absorbed fraction obtained from Monte Carlo simulations of a
uniformly illuminating isotropic external source. Comparison with Figure~\ref
{ep_mc_gp} shows that the behavior of the absorbed fraction of an external
source is quite different from the absorbed fraction of an internal source
(one minus the escaping fraction shown in lower panel of Figure~\ref
{ep_mc_gp}). Note that for $g<1$, as $\tau \rightarrow \infty $ the absorbed
fraction tends asymptotically toward a constant value, since in the opaque
limit scattering occurs at the surface of the sphere. The thick solid line
is again the contour of zero difference between ${\cal P}_{abs}^{x}(\tau
,\omega )$ and Monte Carlo simulations, and the dashed line is the
approximation of $g^{*}(\tau )$ given by eq.(\ref{g_tau}). This zero
difference contour is found to be independent of $\omega $ and fit well by
eq.(\ref{g_tau}), also because ${\cal P}_{esc}^{u}(\tau ,\omega )$ is used
in the approximation formula.

Let us now examine the behavior of the absorption probability as $\tau
\rightarrow \infty $ to show in more detail why the best agreement with
Monte Carlo simulations occurs at a value of $g$ that approaches unity.
Using eqs.(\ref{Pe_Pi}) and (\ref{Lucy_EP}), and defining $P\equiv
P_{e}\left( \tau \right) $ for convenience, eq.(\ref{Pxabs}) can be
rewritten as 
\begin{eqnarray}
{\cal P}_{abs}^{x}(\tau ,\omega \,) &=&\frac{4\tau }{3}P\,\left[ 1-\frac{%
\omega P}{1-\omega \left( 1-P\right) }\right]   \nonumber \\
&=&\frac{4\tau }{3}P\,\left[ \frac{1-\omega }{1-\omega \left( 1-P\right) }%
\right]   \nonumber \\
&=&\frac{4\tau \left( 1-\omega \right) }{3}\,\left[ \frac{P}{1-\omega \left(
1-P\right) }\right]   \nonumber \\
&=&\frac{4\tau \left( 1-\omega \right) }{3}\,\,{\cal P}_{esc}^{u}(\tau
,\omega )\,\,,  \label{Pxabs_2}
\end{eqnarray}
thus extending eq.(\ref{Pe_Pi}), the duality between the cases of internal
and external sources, to the case of $\omega >0$. Equation (\ref{Pxabs_g=1})
can be rewritten as 
\begin{equation}
{\cal P}_{abs}^{x}(\tau ,\omega \,;\,g=1)\,\,=\,\,\frac{4\tau \left(
1-\omega \right) }{3}P_{e}\left[ \tau (1-\omega )\right] 
\end{equation}
where we have used eq.(\ref{Pe_Pi}) again. The analysis given by eqs.(\ref
{EP_Lim_g=1}-\ref{EP_Lim_EP_g=1}) proved that ${\cal P}_{esc}^{u}(\tau
,\omega )\rightarrow P_{e}\left[ \tau (1-\omega )\right] $ from below as $%
\tau \rightarrow \infty $, thereby proving that 
\begin{equation}
{\cal P}_{abs}^{x}(\tau ,\omega ;g=g^{*})\rightarrow {\cal P}_{abs}^{x}(\tau
,\omega ;g=1)
\end{equation}
from below, implying that $g^{*}(\tau )\rightarrow 1$ as $\tau \rightarrow
\infty $.

\section{ESCAPE AND ABSORPTION IN\protect\linebreak CLUMPY MEDIA\label%
{Esc_Abs_Clumpy}}

A viable approach to the problem of estimating the escaping and absorbed
fractions of photons in an inhomogeneous medium is to find effective values
for the absorption and scattering properties of the inhomogeneous medium and
then use these effective values in the escape/absorption probability
formulae that were developed for homogeneous media. The work of \cite
{hobpad93} (HP93) provides the means to such an approach for two-phase
clumpy media by a model they called the ``mega-grains'' approximation, which
reduces the inhomogeneous radiative transfer problem to an effectively
homogeneous one. In the following sections we review the mega-grains
approximation, and then discuss how it is utilized to give escape/absorption
probability approximations for clumpy media. We also introduce some
improvements to the mega-grains approximation of HP93. In addition, we
present new formulae giving the approximate fraction of photons absorbed in
each phase of the medium, which can then be used to estimate the dust
temperature and infrared emission from the clumps and ICM separately.

\subsection{The Mega-Grains Approximation\label{MG_Approx}}

\cite{NatPan84} developed analytic approximations for the extinction by dust
having various kinds of inhomogeneous distributions, in particular, a clumpy
distribution with an empty inter-clump medium (ICM). \cite{neufeld91}
proposed the approach of treating spherical clumps in a two-phase medium as
large grains, and applied it to the special case of the scattering,
absorption, and escape of Ly$\alpha $ radiation, involving both gas and
dust. HP93 proposed a more general set of formulae called the mega-grains
approximation, which also treats the spherical clumps as large grains but
with absorption and scattering coefficients in direct analogy with dust
grains. Their mega-grains model gives equations for the effective optical
depth and effective scattering albedo of a two-phase clumpy medium, with a
non-empty ICM. In this section we review the derivation of the mega-grains
approximation and introduce an improved equation for the effective
scattering albedo. We also develop a new approximate formula for the
effective scattering asymmetry parameter of the clumpy medium. To
distinguish between similar properties of the clumps (mega-grains) and the
dust we will assign them the subscripts ``{\it c}'' and ``{\it d}'',
respectively. So $\omega _{c}$ will be the macroscopic albedo of the clumps
and $\omega _{d}$ is the microscopic albedo of the dust. Also, an implicit
dependence on wavelength is assumed in all the following.

\subsubsection{Effective Optical Depth}

Assume that a collection of spherical clumps, each of radius $r_{c}$ and
having dust mass density $\rho _{c}\,$, are randomly distributed in an
inter-clump medium (ICM) having a constant dust mass density of $\rho
_{icm}\leq \rho _{c}$. Following HP93, define the clump optical radius 
\begin{eqnarray}
\tau _{c}\, &=&\,(\rho _{c}-\rho _{icm})\,\kappa \,r_{c}  \label{tau_clump}
\\
&=&\,(\alpha -1)\,\kappa \rho _{icm}\,r_{c}  \nonumber
\end{eqnarray}
where $\kappa $ is the total absorption plus scattering cross-section of the
dust grains per unit mass and recall that $\alpha = \rho _{c}/\rho _{icm}$.
The reason for the subtraction of densities in this definition is to
disregard the clumps in the limit of when the clump density approaches the
density of dust in the ICM, since then there are effectively no clumps. Thus
the mega-grains are considered to be just the over-density, $\rho _{c}-\rho
_{icm}$, allowing for analysis as a separate component superimposed on a
continuous medium of lower density. Now considering the clumps as large
grains, the probability that a randomly emitted photon will encounter a
clump is just $\pi r_{c}^{2}$. Once a clump is encountered, the probability
that the photon will interact (get absorbed or scatter) with dust in clump
is $P_{i}(\tau _{c}),$ given by eq.(\ref{Pinteract}), the interaction
probability for a sphere of optical radius $\tau _{c}$. Then we can define
the interaction coefficient per unit length, $\Lambda _{mg},$ in the medium
for just the mega-grains as 
\begin{equation}
\Lambda _{mg}=n_{c}\,\pi r_{c}^{2}\,P_{i}(\tau _{c})  \label{MG_xsec}
\end{equation}
where $n_{c}\ $is the number of clumps per unit volume. Here the use of the
interaction probability $P_{i}(\tau _{c})$ is analogous to the absorption
and scattering efficiency coefficient of dust grains.

Define the effective interaction coefficient per unit length, $\Lambda
_{eff},$ in the two-phase clumpy medium by including the ICM: 
\begin{equation}
\Lambda _{eff}=\Lambda _{mg}\,+\,\kappa \rho _{icm}\quad .  \label{xsec_eff}
\end{equation}
This reduces to the interaction coefficient for a homogeneous medium, $%
\kappa \rho _{hom},$ when $\alpha =\rho _{c}/\rho _{icm}=1$ or $f_{c}=0,$
i.e. when the medium is homogeneous [see eqs.(\ref{dhom}) \& (\ref{den_icm}%
)]. For a clumpy medium in a plane-parallel slab of thickness $L$, or sphere
of radius $R=L,$ the effective optical depth is then 
\begin{equation}
\tau _{eff}\,=\,L\,\Lambda _{eff}=L\Lambda _{mg}\,+\,L\kappa \rho _{icm}\,,
\label{tau_eff_mg}
\end{equation}
where by effective we mean the optical depth corresponding to the average
transmission of a large collection of randomly chosen paths through the
slab, as expressed by eq.(\ref{eff_tau}). In \S \ref{Extend_MG} we analyze
in more detail the implementation of the mega-grains equation (\ref{MG_xsec}%
), developing an improved version, and discuss the behavior of $\tau _{eff}$
as a function of all the parameters characterizing the clumpy medium.
Equation (\ref{MG_xsec}) is compared with the approximation developed by 
\cite{NatPan84} in \S \ref{Old_Theory}.

\subsubsection{Effective Scattering Albedo}

The effective scattering albedo of the two-phase clumpy medium is logically
defined as a weighted combination of the albedos for each phase in the
medium: 
\begin{equation}
\omega _{eff}\,\equiv \,\frac{\omega _{c}\,\Lambda _{mg}\,\,+\,\,\omega
_{d}\,\kappa \rho _{icm}}{\Lambda _{eff}}\quad ,  \label{albedo_eff}
\end{equation}
where $\omega _{d}$ is the albedo of the dust and $\omega _{c}$ is the
effective albedo of a clump, which we discuss below. First note that in the
limit as the medium becomes homogeneous ($f_{c}\rightarrow 0$) we have $%
\,\Lambda _{mg}\rightarrow 0$ and $\Lambda _{eff}\rightarrow \kappa \rho
_{icm},$ so then $\omega _{eff}\,\rightarrow \omega _{d}$ as expected.

Given that interactions with dust in a clump have occurred for a group of
photons, the effective albedo of the clump is the fraction of those photons
that manage to eventually escape from the clump by means of multiple
scattering. HP93 suggested an approximation for the clump albedo for which
we shall use the symbol $\omega _{c}^{HP}$: 
\begin{equation}
\omega _{c}^{HP}\equiv \frac{\omega _{d}}{1\,+\,(1-\omega _{d})\,4\tau _{c}/3%
}\quad ,  \label{HP_cl_alb}
\end{equation}
where $\tau _{c}$ is the optical radius of a clump as given by eq.(\ref
{tau_clump}). As $\tau _{c}\rightarrow 0$ we have $\omega
_{c}^{HP}\rightarrow \omega _{d}$ as required. As $\tau _{c}\rightarrow
\infty $ then $\omega _{c}^{HP}\rightarrow 0,$ but this cannot be generally
true since we expect always some backscattering from the surface layer of
the clump as long as the dust scattering asymmetry parameter $g_{d}<1.$ Note
that the approximation does not consider the distribution of scattering
angles. In addition we shall show that for small $\tau _{c}$, equation (\ref
{HP_cl_alb}) is below the minimum possible value for the clump albedo.

Let us examine the special case of purely forward scattering, when the
asymmetry parameter of the dust is $g_{d}=1$. In this case there is an exact
formula for the clump albedo. The forward scattering case for photons
randomly impacting a spherical clump was discussed previously in the
presentation of equations (\ref{Pinteract}) and (\ref{Pxabs_g=1}), and the
actual fraction of photons absorbed is $P_{i}\left[ \tau _{c}(1-\omega
_{d})\right] .$ The fraction that escapes is then $P_{i}(\tau
_{c})-P_{i}\left[ \tau _{c}(1-\omega _{d})\right] ,$ and the ratio of
escaping fraction over interacting fraction is the clump albedo: 
\begin{equation}
\omega _{c}^{1}\,=\,\frac{P_{i}(\tau _{c})-P_{i}\left[ \tau _{c}(1-\omega
_{d})\right] }{P_{i}(\tau _{c})}  \label{clump_alb_g=1}
\end{equation}
where the superscript ``1'' indicates that $g_{d}=1$. As $\tau
_{c}\rightarrow 0$, we have $P_{i}(\tau _{c})\rightarrow 4\tau _{c}/3$ and
then 
\begin{eqnarray*}
\omega _{c}^{1}\, &\rightarrow &\,\frac{4\tau _{c}/3-4\tau _{c}(1-\omega
_{d})/3}{4\tau _{c}/3}\, \\
&=&\,1-(1-\omega _{d})\,=\,\omega _{d}
\end{eqnarray*}
reaching the correct limit. On the other hand, if $\tau _{c}\rightarrow
\infty $ then $\omega _{c}^{1}\,\rightarrow 0,$ which is expected in this
case of purely forward scattering. We shall show that $\omega _{c}^{1}$ is
the minimum clump albedo over all $-1\leq g_{d}\leq 1$.

\begin{figure}[tbp]
\plotone{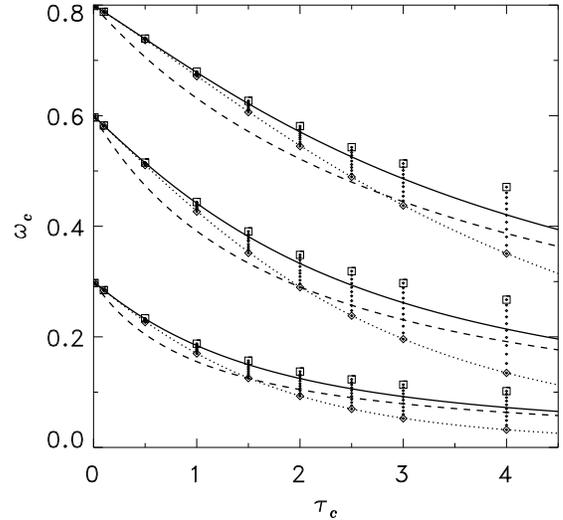}
\caption{ Comparison of theoretical approximations for the clump albedo
given by HP93, [$\omega_c^{HP}$ of eq.(\ref{HP_cl_alb}), dashed line],
and those developed in this paper,
$\omega_c^{*}$ [eq.(\ref{clump_albedo_pe}), solid line]
and $\omega_c^{1}$ [eq.(\ref{clump_alb_g=1}), dotted line],
with Monte Carlo simulations (squares for $g_{d}=0$, diamonds for $g_{d}=1$),
as a function of the clump optical depth $\tau_c$. }
\label{eff_cl_alb}
\end{figure}%

To determine the actual clump albedo and how it depends on dust scattering
parameters we performed Monte Carlo simulations of photons randomly
impacting a spherical clump for many values of the optical parameters in the
ranges of $0<\tau _{c}\leq 25,$ $0<\omega _{d}<1,$ and $0\leq g_{d}\leq 1$.
The fraction of photons that interact with the clump is found to be exactly
given by eq.(\ref{Pinteract}), as expected. Of these interacting photons,
the fraction that scatters and eventually escapes from the clump is the
clump albedo $\omega _{c}$. Figure~\ref{eff_cl_alb} shows $\omega _{c}$
versus $\tau _{c}$ for three values of the dust albedo $\omega
_{d}=0.3,\,0.6,$ and $0.8\,$ (apparent when $\tau _{c}=0$). The square
symbols indicate the case of $g_{d}=0$ (isotropic scattering), the diamonds
show $g_{d}=1$ (forward scattering), and intermediate values of $0<g_{d}<1$
at increments of 0.1 are plotted as dots vertically connecting the squares
and diamonds. The dotted line is $\omega _{c}^{1}$ given by eq.(\ref
{clump_alb_g=1}), the dashed line is $\omega _{c}^{HP}$ given by eq.(\ref
{HP_cl_alb}), and the solid line we shall introduce shortly. Clearly, the
theoretical forward scattering clump albedo, $\omega _{c}^{1}$ (dotted
line), agrees exactly with the Monte Carlo results for $g_{d}=1$ (diamond
symbols), as expected, and is a lower bound on the clump albedo. From the
dashed line it is apparent that $\omega _{c}^{HP}<\omega _{c}^{1}$ when $%
\tau _{c}<2,$ which violates the lower bound.

Another formula for the clump albedo can be derived using the OLEP formula
[eqs. (\ref{Oster_EP}) and (\ref{Lucy_EP})] in the same manner as was
discussed in the derivation of equation (\ref{Pxabs}). Given that a photon
interacts with the dust inside a clump, the probability that it scatters is
just $\omega _{d}$. These scattering events can be regarded as re-emitted
photons, and assuming that they are approximately uniformly distributed in
the clump, they have a probability ${\cal P}_{esc}^{u}(\tau _{c},\omega
_{d}) $ of escaping, thus obtaining the clump albedo 
\begin{equation}
\omega _{c}^{*}=\omega _{d}\,{\cal P}_{esc}^{u}(\tau _{c},\omega _{d})\quad ,
\label{clump_albedo_pe}
\end{equation}
which gives the solid line shown in Figure~\ref{eff_cl_alb}. The escape
probability equation (\ref{Lucy_EP}) for ${\cal P}_{esc}^{u}(\tau
_{c},\omega _{d})$ can be rearranged to give 
\begin{equation}
\omega _{c}^{*}=\frac{\omega _{d}}{\omega _{d}\,+\,(1-\omega
_{d})\,\,/\,P_{e}(\tau _{c})}  \label{clump_albedo}
\end{equation}
where $P_{e}(\tau _{c})$ is the extinction escape probability, as given by
eq.(\ref{Oster_EP}). As $\tau _{c}\rightarrow 0$, we have $P_{e}(\tau
_{c})\rightarrow 1$ and so $\omega _{c}^{*}\rightarrow \omega _{d}$, because
after a single scattering a photon will most likely escape an optically thin
clump. As $\tau _{c}\rightarrow \infty $ , we have $P_{e}(\tau _{c})$ $%
\rightarrow 3/4\tau _{c}$ and then $\omega _{c}^{*}\rightarrow 0$. However, $%
\omega _{c}^{*}$ gives the actual clump albedo for a single value of $g_{d}$
for each $\tau _{c}$, and this value, $g_{d}^{*}=g^{*}(\tau _{c}),$ is
approximately given by eq.(\ref{g_tau}), since we are using the approximate
escape probability ${\cal P}_{esc}^{u}(\tau _{c},\omega _{d})$ (see Figure~%
\ref{ep_mc_gp}). Therefore the fact that $\omega _{c}^{*}\rightarrow 0$ as $%
\tau _{c}\rightarrow \infty $ is just a consequence of the fact that $%
g^{*}(\tau _{c})\rightarrow 1$ as $\tau _{c}\rightarrow \infty $, coupled
with the fact that the exact equation (\ref{clump_alb_g=1}) for the case of $%
g_{d}=1$ gives the limit of zero for the albedo of increasingly opaque
clumps.

The clump albedo depends more strongly on $g_{d}$ as $\tau _{c}$ increases
(see Figure~\ref{eff_cl_alb}), creating a large spread in the values of $%
\omega _{c}$ between the cases of forward scattering ($g_{d}=1$, diamonds)
and isotropic scattering ($g_{d}=0$, squares). This is because as $\tau _{c}$
increases the average photon penetration depth to first scattering
decreases, and then forward scattered photons are likely to be absorbed in
the clump whereas the spherical geometry presents many opportunities for the
escape of any non-forward scattered photon.. Thus a higher probability of
backscattering, corresponding to smaller values of $g_{d}$, increases the
probability that a photon will escape, giving a larger effective clump
albedo. For any $g_{d}<1$ our Monte Carlo simulations indicate that the
albedo of increasingly opaque clumps ($\tau _{c}>20$) remains non-zero.

In summary, Figure~\ref{eff_cl_alb} compares the theoretical approximations
for the clump albedo developed in this paper, $\omega _{c}^{*}$ [solid line,
eq.(\ref{clump_albedo})], $\omega _{c}^{1}$ [dotted line, forward
scattering, eq.(\ref{clump_alb_g=1})], with the approximation given by HP93, 
$\omega _{c}^{HP}$ [dashed line, eq.(\ref{HP_cl_alb})], and with the Monte
Carlo results (symbols). From the figure it is clear that $\omega
_{c}^{*}>\omega _{c}^{HP}$ and $\omega _{c}^{*}>\omega _{c}^{1}$ for all
clump optical depths, and for $\tau _{c}<2$ our proposed eq.(\ref
{clump_albedo}) for $\omega _{c}^{*}$ does better at predicting the clump
albedo computed by Monte Carlo than $\omega _{c}^{HP}$ given by eq.(\ref
{HP_cl_alb}). We shall assume that $\omega _{c}^{*}$ given by eq.(\ref
{clump_albedo}) is a sufficiently good approximation of the actual clump
albedo, and will use it in eq.(\ref{albedo_eff}) to obtain the effective
albedo of the clumpy medium. If $g^{*}(\tau _{c})\leq g_{d}\leq 1$, one can
interpolate in the variable $g_{d}$ between values $\omega _{c}^{*}$ and $%
\omega _{c}^{1}$ corresponding to the endpoints of the range to obtain a
more accurate estimate of the clump albedo.

\subsubsection{Effective Scattering Asymmetry Parameter}

Hobson and Padman did not consider the angular scattering distribution of
photons in their paper introducing the mega-grains approximation, but a
similar approach as used for the effective albedo of the clumpy medium can
be followed. We define the effective phase function asymmetry parameter for
the two-phase clumpy medium as a weighted combination of the asymmetry
parameters of each phase in the medium: 
\begin{equation}
g_{eff}\equiv \frac{g_{c}\,\Lambda _{mg}\,\,+\,\,g_{d}\,\kappa \rho _{icm}}{%
\Lambda _{eff}}\quad \quad ,  \label{geff}
\end{equation}
where $g_{d}$ is the asymmetry parameter of the dust and $g_{c}$ is the
clump scattering asymmetry parameter, which we discuss below. Note that $%
g_{eff}\,\rightarrow g_{d}$ as the medium becomes homogeneous ($%
f_{c}\rightarrow 0$), since then $\,\Lambda _{mg}\rightarrow 0$ and $\Lambda
_{eff}\rightarrow \kappa \rho _{icm}$.

\begin{figure}[tbp]
\plotone{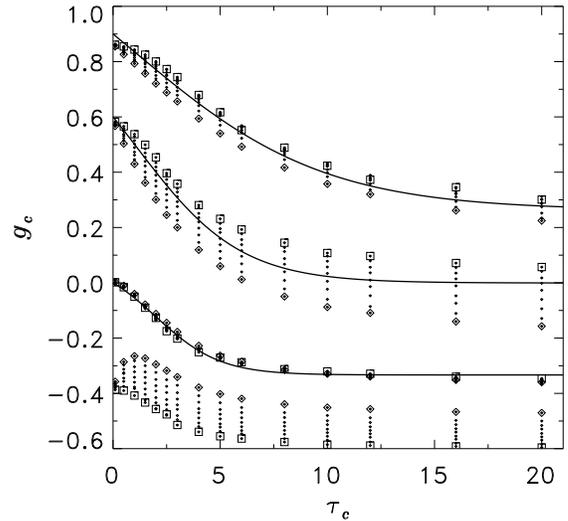}
\caption{ The effective clump scattering
asymmetry parameter $g_{c}=\langle \cos \theta _{exit}\rangle $,
where $\theta _{exit} $ is the angle between impacting and exiting directions
as simulated by Monte Carlo calculations, is plotted as a
function of the clump optical depth $\tau_c$, for four values of the dust
scattering asymmetry parameter, $g_d = -0.4$, 0.0, 0.6, and 0.9,
corresponding to the intersection of each curve with the vertical axis at $\tau_c=0$.
The square symbols are for  dust scattering albedo of $\omega _{d}=0.1$,
the diamonds for $\omega _{d}=0.9,$ and  the dots show intermediate
values of $\omega _{d}$ in increments of 0.1.
The solid lines show the empirical approximation given by eq.(\ref{g_c}) 
for $\omega _{d}=0.5$ . }
\label{eff_cl_gp}
\end{figure}

Let us assume that a collection of photons enter a clump with parallel ray
paths, are scattered by the dust in the clump, and eventually escape from
the clump. How does the exiting angular distribution depend on the dust
scattering properties and the optical depth of the clump? The results of our
Monte Carlo simulations shown in Figure~\ref{eff_cl_gp} answer this
question. We computed $g_{c}=\langle \cos \theta _{exit}\rangle $ for all
simulations, where $\theta _{exit}$ is the exiting angle respect to the
entering parallel rays, and plot $g_{c}$ versus the clump optical radius $%
\tau _{c},$ for values of $g_{d}=\{-0.4,\,0.0,\,0.6,\,0.9\}$ (evident when $%
\tau _{c}=0$), and albedos $0.1\leq \omega _{d}\leq 0.9$. The square symbols
are for when $\omega _{d}=0.1$, the diamonds for $\omega _{d}=0.9,$ and
intermediate values of $\omega _{d}$ in increments of 0.1 are plotted as
dots connecting the squares and diamonds. The distribution of exit angles
does not actually follow the HG phase function with $g=g_{c}$, since the
scattering is complicated by the spherical geometry, however, the HG phase
function is a reasonable approximation. In the case of forward scattering ($%
g_{d}>0$), the clump scattering distribution becomes more isotropic as the
optical radius $\tau _{c}$ of the clump increases and as the dust albedo $%
\omega _{d}$ increases. If $g_{d}=0$ (isotropic scattering by dust), then as
the clump becomes opaque the clump asymmetry parameter approaches a
backscattering distribution of $g_{c}=-1/3$ independent of the dust albedo.
In addition, when $\left| g_{d}\right| $ is near zero and $\tau _{c}$ is
large, it is apparent that $g_{c}$ approaches $-1/3$ as $\omega _{d} $
increases. Our computations of $g_{c}$ agree with the three cases computed
and mentioned by \cite{codeblob95}.

The solid lines in Figure~\ref{eff_cl_gp} are produced by the following
empirical formula: 
\begin{equation}
g_{c}(\tau _{c},\omega _{d},g_{d})=g_{d}\,-\,\,C\left( 1\,-\,\frac{1+e^{-B/A}%
}{1+e^{\left( \tau _{c}-B\right) /A}}\,\right)  \label{g_c}
\end{equation}
where 
\begin{eqnarray*}
A &\equiv &\,1.5+4g_{d}^{3}+2\omega _{d}\sqrt{g_{d}}\exp (-5g_{d}) \\
B &\equiv &2-g_{d}(1-g_{d})-2\omega _{d}g_{d} \\
C &\equiv &\frac{1}{3-\sqrt{2g_{d}}-2\omega _{d}g_{d}(1-g_{d})}\;\;.
\end{eqnarray*}
The formula is a good approximation of the Monte Carlo results when $%
g_{d}\geq 0$. For $g_{d}$ slightly negative one can shift the $g_{d}=0$
curve downward to get an approximation of $g_{c}$, or just use $g_{c}=g_{d}$
when $g_{d}<-0.2$. The exact value of $g_{c}$ is of importance mainly for
the case of a central point source, since the escaping fraction is then very
sensitive to the distribution of angular scattering.

\subsection{The Extended Mega-Grains Approximation\label{Extend_MG}}

The mega-grains approximation as presented by Hobson \& Padman is limited to
small values of the clump filling factor. In addition, we found that for
optically thin situations it predicts an effective optical depth that is
slightly greater than the equivalent homogeneous optical depth, in violation
of eq.(\ref{eff_trans}). In this section we develop an improved version of
the mega-grains approximation that resolves these problems, and extends the
approximation to all values of the clump filling factor.

An important term in eq.(\ref{MG_xsec}) is $n_{c}$, the density of clumps,
which we shall now discuss in detail since it is the key to our extension of
the mega-grains approximation. Define the porosity, $Q_{c}$, of a randomly
located collection of identical clumps as the ratio of the total volume of
clumps, {\em including} possible overlaps, to the volume $V$ of the medium: 
\begin{equation}
Q_{c}=\frac{N_{c}\upsilon _{c}}{V}  \label{porosity}
\end{equation}
where $N_{c}$ is the total number of clumps and $\upsilon _{c}$ is the
volume of just one clump. Equation (\ref{porosity}) is easily solved for the
density of clumps 
\begin{equation}
n_{c}\,=\frac{N_{c}}{V}=\frac{Q_{c}}{\upsilon _{c}}=\,\frac{3Q_{c}}{4\pi
r_{c}^{3}}\quad ,\,  \label{denclumps}
\end{equation}
as a function of the porosity and the radius of a spherical clump $r_{c}$.
Substituting for $n_{c}$ in eq.(\ref{MG_xsec}) gives 
\begin{equation}
\Lambda _{mg}=\frac{3Q_{c}}{4r_{c}}\,P_{i}(\tau _{c})  \label{MG_xq}
\end{equation}
for the interaction coefficient of the mega-grains, where $\tau _{c}$ is the
optical radius of a clump.

The filling factor $f_{c}$ is related to the porosity, $Q_{c},$ of the
clumps by 
\begin{equation}
f_{c}=1-e^{-Q_{c}}  \label{ff_q}
\end{equation}
because the probability that a random point is not in a clump goes as $%
e^{-Q_{c}}$ (this is a very good approximation to eq.(\ref{ff_pvc}) when $%
\upsilon _{c}\ll V$, see Appendix \ref{App Nc}). When $f_{c}\ll 1$ then $%
f_{c}\simeq Q_{c}$\thinspace , however as $f_{c}\rightarrow 1$ we have $%
Q_{c}=\,-\ln (1-f_{c})\,\rightarrow \infty ,$ since then the clumps tend to
overlap. Thus an obvious problem with equation (\ref{MG_xq}) is that as $%
f_{c}\rightarrow 1$ we have $Q_{c}\rightarrow \infty $ causing $\Lambda
_{mg}\rightarrow \infty .$ Indeed, Hobson \& Padman found that $f_{c}<0.3$
is the useful range for the mega-grains approximation. The clump overlapping
fraction is calculated as $(Q_{c}-f_{c})/Q_{c}$, and for $f_{c}<0.3$ it is
less than 16\% of the volume of clumps, but as $f_{c}>0.3$ the overlapping
volume fraction increases rapidly toward 100\% so that the clumps can no
longer be treated as separate mega-grains.

There is another problem with the previous equation, which is more subtle
but still important. Consider the behavior of eq.(\ref{MG_xq}) as $\tau
_{c}\rightarrow 0,$ which occurs when either $r_{c}\rightarrow 0,$ $\kappa
\rightarrow 0$ or $\,\rho _{hom}\rightarrow 0$ (holding $f_{c}$ and $\alpha =
$ $\rho _{c}/\rho _{icm}$ constant). In that case $P_{i}(\tau _{c})\sim
4\tau _{c}/3$, which simplifies eq.(\ref{MG_xq}), and upon substituting for $%
\tau _{c}$ with eq.(\ref{tau_clump}), the clump radius cancels giving 
\begin{equation}
\Lambda _{mg}\;\approx \;\frac{Q_{c}}{r_{c}}\,\tau _{c}\,=\,(\rho _{c}-\rho
_{icm})\,\kappa \,Q_{c}  \label{tau_zero}
\end{equation}
when the clump optical radii are very small. Substituting this $\,$into eq.(%
\ref{xsec_eff}) gives 
\begin{eqnarray}
\Lambda _{eff} &=&\Lambda _{mg}\,+\,\kappa \rho _{icm}  \nonumber \\
&\approx &\kappa \left[ (\rho _{c}-\rho _{icm})Q_{c}\,+\,\rho _{icm}\right] 
\label{tau_zero2}
\end{eqnarray}
for the behavior of the effective interaction coefficient of the two-phase
clumpy medium as $\tau _{c}\rightarrow 0$. The problem with this behavior is
that since $Q_{c}>f_{c}$ we get that 
\begin{equation}
\Lambda _{eff}\,>\,\kappa \left[ (\rho _{c}-\rho _{icm})f_{c}\,+\,\rho
_{icm}\right] =\,\kappa \rho _{hom}  \label{eff_overshoot}
\end{equation}
using eq.(\ref{dhom}) for the final equality. Thus as the clump optical
depths become small, $\Lambda _{eff}$ exceeds the interaction coefficient of
the equivalent homogeneous medium. If $L$ is the geometrical thickness of
the medium then eq.(\ref{eff_overshoot}) implies that 
\begin{equation}
\tau _{eff}\,=\,L\,\Lambda _{eff}\,\,>\,L\,\kappa \rho _{hom}\,=\,\tau
_{hom}\,\,,  \label{eff_tau_over}
\end{equation}
contradicting equations (\ref{eff_trans}) and (\ref{tau_eff<hom}), which
state that a clumpy medium is more transparent than the equivalent
homogeneous medium. The inequality in eqs.(\ref{eff_overshoot}) and (\ref
{eff_tau_over}) becomes greater in error as $f_{c}$ increases since then $%
Q_{c}\rightarrow \infty .$

\begin{figure}[tbp]
\plotone{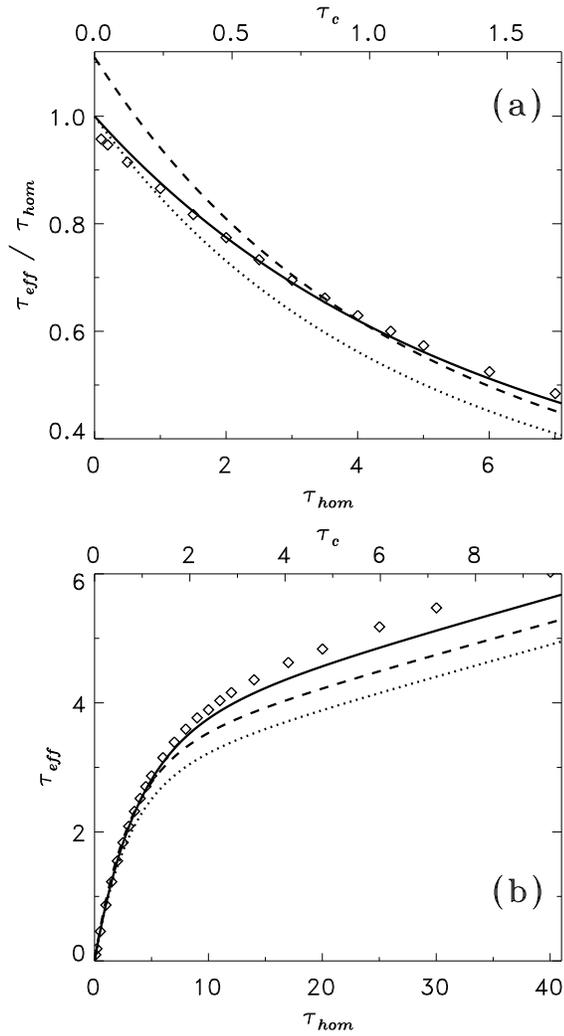}
\caption{ Comparison of three versions of the mega-grains
approximations (MGA) with Monte Carlo results,
for the case of no scattering,
in a clumpy medium having $f_{c}=0.2,\,$ $\alpha =100$, and $r_{c}=0.05$.
In both panels (a) and (b)
the dashed line results from the original version of MGA
using eq.(\ref{MG_xq}),
the dotted line from the modified version using eq.(\ref{MG_xf}),
and the solid line is our new extended version
of MGA using eq.(\ref{MG_xfv}).
The diamonds are results from Monte Carlo simulations.
Panel (a) shows the ratio of the effective optical depth of the clumpy medium,
$\tau_{eff}$, to the equivalent homogeneous optical depth, $\tau_{hom}$,
as a function of $\tau_{hom}$ and  the clump optical radii
$\tau_c$ (upper horizontal axis).
Panel (b) shows $\tau_{eff}$ as a function of larger values
of $\tau_{hom}$ and $\tau_c$. }
\label{mg_comp_tau}
\end{figure}

An immediate solution to the problem is to substitute $f_{c}$ in place of $%
Q_{c}$ in eq.(\ref{MG_xq}), then obtaining 
\begin{equation}
\Lambda _{mg}\,=\,\frac{3f_{c}}{4r_{c}}\,P_{i}(\tau _{c})  \label{MG_xf}
\end{equation}
which gives the correct behavior of $\Lambda _{eff}<\kappa \rho _{hom}$ as $%
\tau _{c}\rightarrow 0$ for values of the clump filling factor. However, now
consider the behavior as $\tau _{c}\rightarrow \infty $, occurring when $%
\rho _{hom}\rightarrow \infty $ or $\kappa \rightarrow \infty $ while
holding $f_{c}$, $\alpha ,$ and $r_{c}$ constant. Then $P_{i}(\tau
_{c})\rightarrow 1$ and $\Lambda _{mg}\rightarrow 3f_{c}/4r_{c}$ as the
clumps become opaque. For values of $f_{c}>0.1$ the resulting effective
interaction coefficient $\Lambda _{eff}=3f_{c}/4r_{c}\,+\,\kappa \rho _{icm}$
underestimates the actual value found from Monte Carlo simulations (see
Figure~\ref{mg_comp_tau}(b) and discussion below). In addition, as $%
f_{c}\rightarrow 1$ we should have $\Lambda _{eff}\rightarrow \kappa \rho
_{hom}$, but instead eq.(\ref{MG_xf}) gives $\Lambda _{eff}\rightarrow
3/4r_{c}+\kappa \rho _{icm},$ which because of the dependence on $r_{c}$
will in general not reach the correct limit.

To extend the mega-grains approximation to all filling factors and retain
the correct behavior as the clumps become optically thin or opaque, we
propose to renormalize the clump radii by the factor $(1-f_{c})^{\gamma }$,
where $0<\gamma \leq 1$ is a parameter that can fine tune the behavior as
the clumps become opaque. In Appendix \ref{App Cover} we discuss the
motivation and a possible interpretation of this renormalization of the
clump radii. Substituting $r_{c}(1-f_{c})^{\gamma }$ for all instances of $%
r_{c}$ in eq.(\ref{MG_xf}) we get 
\begin{equation}
\Lambda _{mg}=\frac{3f_{c}}{4r_{c}}\,\frac{P_{i}\left[ \tau
_{c}(1-f_{c})^{\gamma }\right] }{(1-f_{c})^{\gamma }}\quad ,  \label{MG_xfv}
\end{equation}
our new definition of the mega-grains interaction coefficient. Now as $\tau
_{c}\rightarrow 0$ or $f_{c}\rightarrow 1$ we have that $P_{i}\left[ \tau
_{c}(1-f_{c})^{\gamma }\right] \sim \frac{4}{3}\tau _{c}(1-f_{c})^{\gamma }$
and so the behavior of eq.(\ref{MG_xfv}) is 
\begin{equation}
\Lambda _{mg}\;\sim \;\frac{f_{c}}{r_{c}}\,\tau _{c}\;=\;(\rho _{c}-\rho
_{icm})\,\kappa \,f_{c}  \label{tau_0_ff_1}
\end{equation}
since the $(1-f_{c})^{\gamma }$ factor cancels out. This gives the desired
result of $\Lambda _{eff}\rightarrow \,\,\kappa \rho _{hom}$ as $\tau
_{c}\rightarrow 0$ or $f_{c}\rightarrow 1$, and keeps $\Lambda _{eff}<\kappa
\,\rho _{hom}$ for a clumpy medium, thereby extending the mega-grains
approximation to the full range of $0\leq f_{c}\leq 1.$

In the other extreme, when $\tau _{c}\gg (1-f_{c})^{-\gamma }$ then $%
P_{i}\left[ \tau _{c}(1-f_{c})^{\gamma }\right] \sim 1$ resulting in 
\begin{equation}
\Lambda _{mg}\,\sim \,\frac{3f_{c}}{4r_{c}(1-f_{c})^{\gamma }}\,\,\,.
\end{equation}
Comparing this with the previously discussed versions of the mega-grains
approximation gives the following sequence of inequalities in the limit as
the clumps become opaque: 
\begin{equation}
\frac{3f_{c}}{4r_{c}(1-f_{c})}\,>\,\,\frac{3Q_{c}}{4r_{c}}\,>\,\frac{3f_{c}}{%
4r_{c}}\,\,,  \label{MGA_ineq_seq}
\end{equation}
corresponding to equations (\ref{MG_xfv}), (\ref{MG_xq}), and (\ref{MG_xf})
respectively, with $\gamma =1$ and $f_{c}<1$. Varying the parameter $\gamma $
allows for adjustment of the opaque clump limit behavior of eq.(\ref{MG_xfv}%
) over the complete range in the above sequence of inequalities: as $\gamma
\rightarrow 0$ the left side of the inequality approaches the right side. We
find that the optimal value of $\gamma $ in our extended mega-grains
approximation depends on the arrangement of the source of photons with
respect to the geometry of the clumpy medium: $\gamma \sim 0.5$ gives a good
approximation for the situation of photons impacting a slab of clumps with a
fixed angle with respect to the surface normal, whereas $\gamma \sim 0.75$
works better for a central source within a sphere, and $\gamma =1$ works
best for uniformly distributed sources. When $f_{c}\ll 1$ the variation of
the extended mega-grains approximation is small with respect to the
parameter $\gamma $, and of course vanishes as the filling factor goes to
zero. We shall use $\gamma =1$ unless indicated otherwise.

Figure~\ref{mg_comp_tau} compares the three versions of the mega-grains
approximation (MGA) to Monte Carlo simulations. Figure~\ref{mg_comp_tau}(a)
shows the ratio $\tau _{eff}\,/\,\tau _{hom}$ versus $\tau _{hom}$ for the
case of $f_{c}=0.2,\,$ $\alpha =100,$ and with $r_{c}=0.05R$, where $R$ is
the extent of the medium ( $f_{c}$ and $\alpha $ are dimensionless). By
definition, $\tau _{eff}=R\,\Lambda _{eff}$ and $\tau _{hom}=R\,\kappa \rho
_{hom}$. The upper horizontal axis is the clump optical depth, $\tau _{c}$,
which is proportional to $\tau _{hom}$ via 
\begin{equation}
\tau _{c}\,=\,\left( \frac{r_{c}}{R}\right) \tau _{hom}\left[ \frac{%
\,(\alpha -1)\,}{(\alpha -1)\,f_{c}+1}\right] \,\,,  \label{tau_clump_f}
\end{equation}
derived by combining equations (\ref{den_icm}) and (\ref{tau_clump}). The
dashed line in Figure~\ref{mg_comp_tau}(a) results from the use of equation (%
\ref{MG_xq}), the straightforward implementation of HP93, and it clearly
exceeds unity for $\tau _{c}<0.2$ and $\tau _{hom}<1$, violating the
requirement that $\tau _{eff}<\tau _{hom}$. The overshoot of unity becomes
worse for larger filling factors. The diamonds are results from Monte Carlo
simulations of a central source in a sphere of radius $R=1$ containing a
two-phase clumpy medium characterized by the same parameters but with no
scattering ($\omega _{d}=0$). Each diamond plotted is the ratio $\tau
_{eff}\,/\tau _{hom}$, where $\tau _{eff}$ is computed by eq.(\ref{tau_eff}%
). The dotted line results from the use of equation (\ref{MG_xf}), giving $%
\tau _{eff}<\tau _{hom}$ as required, however the disagreement with Monte
Carlo results becomes worse as $\tau _{hom}$ increases. The solid line
results from the use of equation (\ref{MG_xfv}), our new definition of the
mega-grains interaction coefficient, also giving $\tau _{eff}<\tau _{hom}$
and agreeing well with the Monte Carlo results.

Figure~\ref{mg_comp_tau}(b) compares $\tau _{eff}$ resulting from the three
version of the MGA at large $\tau _{hom}$ (equivalent to large $\tau _{c}$)
for the same clumpy medium. As before, the diamonds are results from the
same Monte Carlo simulations. The dotted line is produced by using equation (%
\ref{MG_xf}), giving 
\begin{equation}
\tau _{eff}\rightarrow R\left( 3f_{c}/4r_{c}\,+\,\kappa \rho _{icm}\right)
\label{MGA_xf_Lim}
\end{equation}
as $\tau _{c}\rightarrow \infty $, which clearly disagrees with the Monte
Carlo results. The dashed line is produced by using equation (\ref{MG_xq}),
giving 
\begin{equation}
\tau _{eff}\rightarrow R\left( 3Q_{c}/4r_{c}\,+\,\kappa \rho _{icm}\right)
\label{MGA_xq_Lim}
\end{equation}
as $\tau _{c}\rightarrow \infty $, which is in better agreement with the
Monte Carlo results. The solid line is produced by our extended MGA, using
eq.(\ref{MG_xfv}), and as $\tau _{c}\rightarrow \infty $ 
\begin{equation}
\tau _{eff}\rightarrow R\left[ \frac{3f_{c}}{4r_{c}(1-f_{c})}+\kappa \rho
_{icm}\right] \,,  \label{MGA_xfv_Lim}
\end{equation}
which provides the best approximation of the Monte Carlo results. There is a
smooth transition in the slope of the $\tau _{eff}$ curves as $\tau _{c}$
increases, and when the clumps become optically thick ($\tau _{c}>4$) their
contribution to the absorption becomes a fixed quantity so the only further
change is due to the ICM density. Thus all versions of the MGA become linear
as $\tau _{c}\gg 1$ and $\tau _{hom}\rightarrow \infty $, basically having
the slope 
\begin{equation}
\frac{\partial \tau _{eff}}{\partial \tau _{hom}}\,\rightarrow \,\frac{%
\partial \rho _{icm}}{\partial \rho _{hom}}\,=\,\frac{1}{(\alpha -1)\,f_{c}+1%
}\,,  \label{tau_eff/tau_hom}
\end{equation}
which for the parameters used for Figure~\ref{mg_comp_tau}(b) is a value of
about $\frac{1}{21}$. The intercepts of the asymptotic lines with the
vertical axis are given by $R$ times the first term in each of the above
eqs.(\ref{MGA_xf_Lim}-\ref{MGA_xfv_Lim}), and these values are approximately
the average number of clumps encountered along a random line of sight (see
Appendix \ref{App Nc}).

\begin{figure}[tbp]
\plotone{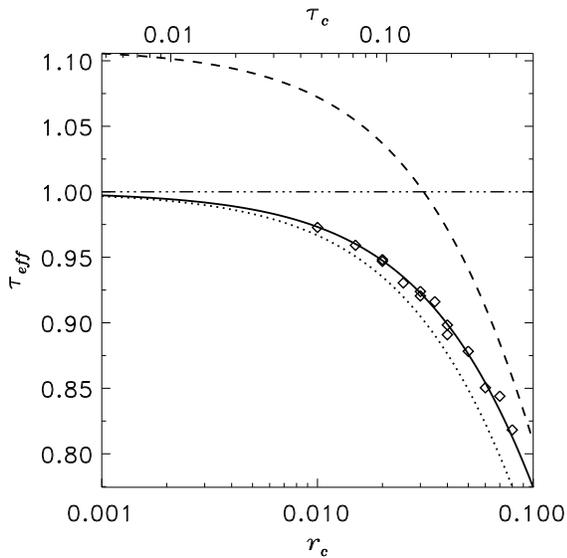}
\caption{ Comparison of HP93 version [eq.(\ref{MG_xq}), dashed line],
modified version [eq.(\ref{MG_xf}), dotted line], and new extended version
[eq.(\ref{MG_xfv}), solid line] of the MGA as a
function of the clump radius $r_c$. Results of Monte Carlo simulations are
plotted as diamonds, and the clumpy medium is defined by
parameters $f_{c}=0.2$, $\alpha =100$, and $\tau _{hom}=1$. }
\label{mg_comp_rc}
\end{figure}%

Now consider the behavior of the MGA as the clump radii vanish and other
parameters are held fixed. This is shown in Figure~\ref{mg_comp_rc} for the
case of $f_{c}=0.2,\,$ $\alpha =100,$ and $\tau _{hom}=1,$ where $\tau
_{eff} $ is plotted versus $r_{c}$ and $\tau _{c}$. The dashed line produced
by using eq.(\ref{MG_xq}) exceeds the value of $\tau _{hom}=1$ (thin long
dashed line) when $r_{c}<0.03$ ($\tau _{c}<0.2$), because as $%
r_{c}\rightarrow 0$ we have $\tau _{c}\rightarrow 0$, and by eqs.(\ref
{tau_zero}) and (\ref{tau_zero2}) 
\[
\tau _{eff}\,\;\rightarrow \;R\kappa \left[ (\rho _{c}-\rho
_{icm})Q_{c}\,+\,\rho _{icm}\right] \,>\,\tau _{hom}\,\,.\, 
\]
Thus the same problematic behavior of $\tau _{eff}>\tau _{hom}$ that was
shown in Figure~\ref{mg_comp_tau}(a) for $\tau _{hom}\rightarrow 0$ occurs
as $r_{c}\rightarrow 0$ and $\tau _{hom}$ is held constant. The solid line
is produced by our extended MGA, using eq.(\ref{MG_xfv}), and clearly yields
the expected behavior $\tau _{eff}<\tau _{hom}$ for the full range of clump
radii, since then as $r_{c}\rightarrow 0$ we have the exact limit of 
\[
\tau _{eff}\,\;\rightarrow \;R\kappa \left[ (\rho _{c}-\rho
_{icm})f_{c}\,+\,\rho _{icm}\right] =\,\tau _{hom} 
\]
[see eq.(\ref{tau_0_ff_1})]. The diamonds are results from Monte Carlo
simulations, in good agreement with our extended MGA. The dotted line is
produced by using eq.(\ref{MG_xf}), which has the correct limit as $%
r_{c}\rightarrow 0$ but tends to underestimate $\tau _{eff}$ at larger clump
radii.

\begin{figure}[tbp]
\plotone{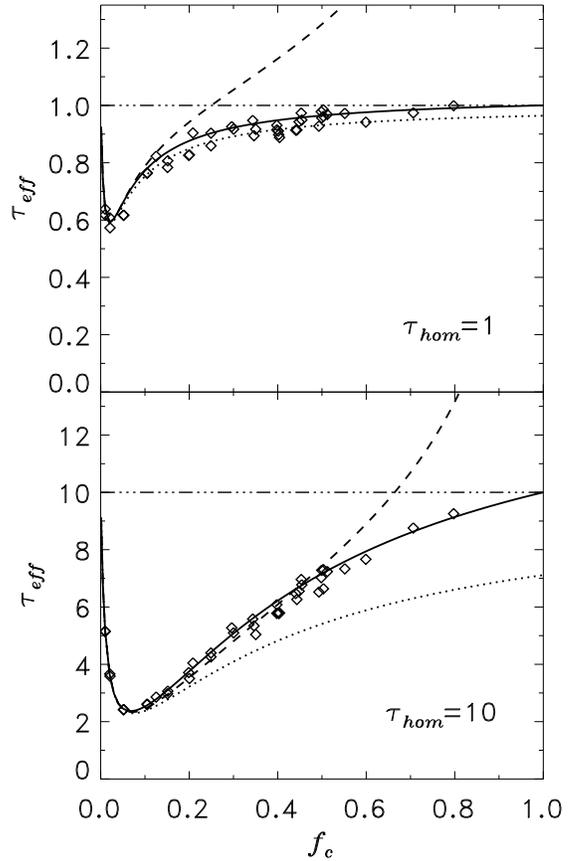}
\caption{ Comparison of $\tau _{eff}$ (vertical axes) predicted by the new
extended version [eq.(\ref{MG_xfv}) with $\gamma=1$, solid line], modified version
[eq.(\ref{MG_xf}), dotted line],
and original version [eq.(\ref{MG_xq}), dashed line] of
the MGA, with Monte Carlo results (diamonds), over
the full range of clump filling factors ($0 < f_c < 1$ on horizontal axis),
for cases $\tau _{hom}=1$ and $\tau _{hom}=10$, as indicated.
The medium consists of clumps with radii $r_{c}=0.05$
and density ratio $\alpha=100$. }
\label{mg_comp_fc}
\end{figure}%

Figure~\ref{mg_comp_fc} compares the extended MGA to the original version
over the full range of clump filling factors, $0\leq f_{c}\leq 1,$ with $%
\alpha =100$ and $r_{c}=0.05$ held constant, for cases $\tau _{hom}=1$ and $%
\tau _{hom}=10$, as indicated. In each case the horizontal axis is $f_{c}$
and the vertical axis is $\tau _{eff}$, the effective optical depth. The
diamonds are the $\tau _{eff}$ resulting from Monte Carlo simulations of a
central source in a sphere of unit radius containing a two-phase clumpy
medium with clump filling factors in the range $0.01\leq f_{c}\leq 0.8$,
also with $\alpha =100$ and $r_{c}=0.05$ held constant. Clearly the extended
MGA [using eq.(\ref{MG_xfv}) in eq.(\ref{xsec_eff})], shown as the solid
lines, gives the best agreement with the Monte Carlo results, and satisfies
the condition $\tau _{eff}<\tau _{hom}$ for the full range of clump filling
factors. The original version of the MGA [using eq.(\ref{MG_xq})], shown as
the dashed lines, starts diverging to infinity for $f_{c}>0.1$ in the case
of $\tau _{hom}=1,$ and diverges to infinity for $f_{c}>0.5$ in the case of $%
\tau _{hom}=10.$ The dotted line is produced using eq.(\ref{MG_xf}), which
underestimates the value of $\tau _{eff}$ and fails to reach the correct
limit of $\tau _{eff}\rightarrow \tau _{hom}$ as $f_{c}\rightarrow 1$,
especially when $\tau _{hom}$ is large. The fluctuations in the Monte Carlo
results are due to variations in particular realizations of clumps in a
finite volume, and these fluctuations grow larger as $f_{c}\rightarrow 1$.

\begin{figure}[tbp]
\plotone{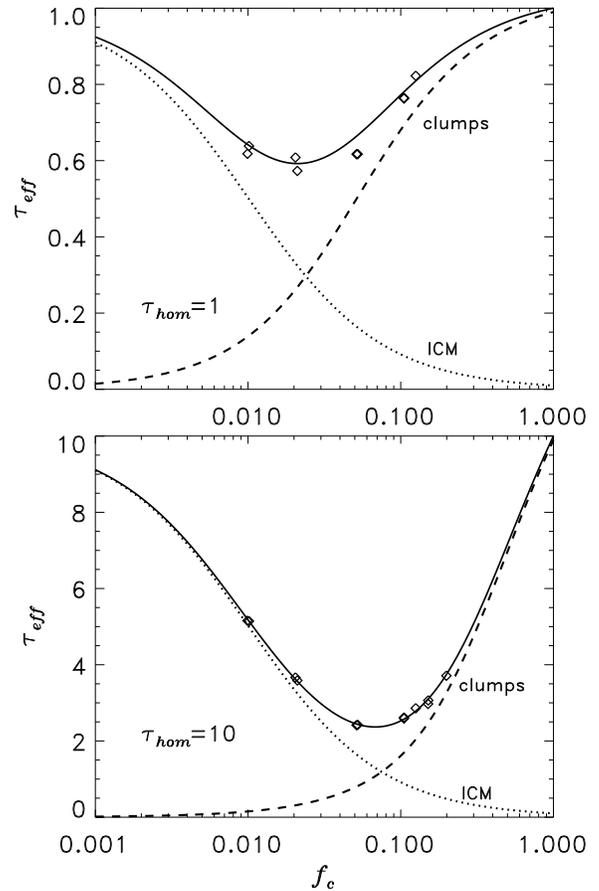}
\caption{ Analysis of $\tau _{eff}$ versus $f_{c}$ predicted by our new
extended version of the MGA
for the same cases shown in Figure~\ref{mg_comp_fc}.
The dashed lines are $\tau _{mg}=R \Lambda_{mg}$
[eq.(\ref{MG_xfv})],
the dotted lines are $\tau _{icm}=R \kappa \rho_{icm}$, and
the solid lines are the sum $\tau _{eff} = \tau _{mg} + \tau _{icm}$. }
\label{mg_ff_min}
\end{figure}%

The behavior of $\tau _{eff}$ versus $f_{c}$ always exhibits a minimum at a
single value of $f_{c}$ that has a complicated dependence on the other
parameters characterizing the clumpy medium. To study this behavior, Figure~%
\ref{mg_ff_min} shows the prediction of $\tau _{eff}$ by the extended MGA
(solid lines) and the contribution from each phase on the clumpy medium: the
dashed lines are ICM component, $\tau _{icm}=R\kappa \rho _{icm}$, and the
dotted lines are the component of the effective optical depth due to clumps, 
$\tau _{mg}=R\Lambda _{mg}$, so that $\tau _{eff}=\tau _{mg}+\tau _{icm}$.
The horizontal axis is now $\log _{10}f_{c}$ and the cases of $\tau
_{hom}=R\kappa \rho _{hom}=1$ and $10$ are shown, with $\alpha =100$, and $%
r_{c}=0.05$ (same as Figure~\ref{mg_comp_fc}). Some of the same Monte Carlo
results are shown (diamonds) to verify the behavior. Starting at $f_{c}\sim 0
$, $\tau _{icm}$ decreases rapidly as $f_{c}$ increases since $\rho _{icm}$,
given by eq.(\ref{den_icm}), is inversely related to $f_{c}$ and the total
mass is held constant. In fact the shape of the $\tau _{icm}=R\kappa \rho
_{icm}$ curve depends only on $\alpha $ and $f_{c}$ so only the magnitude
changes with the value of $\tau _{hom}$: the shape of the variation of $\tau
_{icm}$ versus $f_{c}$ is exactly the same for the two cases of $\tau _{hom}$
in Figure~\ref{mg_ff_min}. The opacity of the clumps is also inversely
related to their filling factor, as seen in eq.(\ref{tau_clump_f}), and so $%
\tau _{c}$ is near maximum when $f_{c}\sim 0$, and then $\tau _{mg}\sim
R(4f_{c}/3r_{c})$. Thus $\tau _{mg}$ increases linearly with $f_{c}$ until
the clumps become optically thin and they fill up the medium, causing $\tau
_{mg}\rightarrow \tau _{hom}$. The linear increase of $\tau _{mg}$ does not
compensate for the rapid decrease of $\tau _{icm}$ with $f_{c}$, the net
effect being that $\tau _{eff}$ decreases with increasing $f_{c}$ until the
clumps start to fill the medium. Therefore $\tau _{eff}$ goes through a
minimum as a function of $f_{c}$ when the number of lines of sight that pass
through the very low density ICM greatly exceeds the number of paths
intersecting optically thick clumps. For $f_{c}\ll 1$, the magnitude of $%
\tau _{mg}$ is independent of $\tau _{hom}$ whereas the magnitude of $\tau
_{icm}$ is directly proportional to $\tau _{hom}$, so the minimum of $\tau
_{eff}$ occurs at a larger value of $f_{c}$ for the case of $\tau _{hom}=10$
than for the case of $\tau _{hom}=1$.

In general, the minimum of $\tau _{eff}$ occurs at nearly the same value of $%
f_{c}$ for which $\tau _{mg}=\tau _{icm}$, and analysis of the equations
shows why this is true. Assuming that $f_{c}\ll 1$ then 
\begin{equation}
\frac{\partial \tau _{mg}}{\partial f_{c}}\approx R\left( \frac{4}{3r_{c}}%
\right) \approx \frac{\tau _{mg}}{f_{c}}\,.
\end{equation}
Assuming that $\alpha \gg 1$ then 
\begin{eqnarray}
\frac{\partial \tau _{icm}}{\partial f_{c}} &=&-R\kappa \rho _{icm}\left[ 
\frac{\alpha -1}{\left( \alpha -1\right) f_{c}+1}\right]  \nonumber \\
&\approx &-\frac{\tau _{icm}}{f_{c}+\frac{1}{\alpha }}\,.
\end{eqnarray}
Now since $\tau _{eff}=\tau _{mg}+\tau _{icm}$ we have that 
\begin{equation}
\frac{\partial \tau _{eff}}{\partial f_{c}}\approx \frac{\tau _{mg}}{f_{c}}%
\,-\,\frac{\tau _{icm}}{f_{c}+\frac{1}{\alpha }}\,.
\end{equation}
The local minimum occurs when the partial derivative is zero so that 
\begin{equation}
\frac{\partial \tau _{eff}}{\partial f_{c}}=0\quad \Longleftrightarrow \quad
\left( 1+\frac{1}{\alpha f_{c}}\right) \tau _{mg}\approx \tau _{icm} \,\,,
\label{tau_eff_min_f}
\end{equation}
which is approximately the relationship between the minimum of $\tau _{eff}$
and the intersection of the $\tau _{mg}$ and $\tau _{icm}$ curves in Figure~%
\ref{mg_ff_min}. Consequently, as a function of $f_{c}$ (with constant $%
\alpha \gg 1$), the total luminosity absorbed by a clumpy medium attains a
minimum when the luminosities absorbed in the clumps and in the ICM are
equal, and this is demonstrated over a large region of parameter space in \S 
\ref{MGEP_explore}.

The previous discussion of $\tau _{eff}$ has been for the case of no
scattering, or for scattering and absorption together considered as
interaction with dust. To model scattering in a clumpy medium we need the
effective scattering albedo, $\omega _{eff}$, given by eq.(\ref{albedo_eff}%
), which depends on $\omega _{c}$, the clump albedo given by eq.(\ref
{clump_albedo}). We find that the clump radius renormalization technique
introduced above also needs to be applied to the clump albedo formula in
order to produce the correct approximation for scattering in the clumpy
medium as $f_{c}\rightarrow 1.$ The new equation is then 
\begin{equation}
\omega _{c}\,=\,\omega _{d}\,\,{\cal P}_{esc}^{u}\left[ \tau
_{c}(1-f_{c})^{\gamma },\,\omega _{d}\right] \,  \label{clump_albeff}
\end{equation}
so that $\omega _{c}\,\rightarrow \,\omega _{d}$ as $f_{c}\rightarrow 1$,
since then the escape probability becomes unity. This leads to the correct
behavior of $\omega _{eff}\rightarrow \,\omega _{d}$ for as $%
f_{c}\rightarrow 1$. For the same reasons, $\tau _{c}$ in eq.(\ref{g_c}) is
replaced by $\tau _{c}(1-f_{c})^{\gamma }$ to give the correct limits as $%
f_{c}\rightarrow 1$ for $g_{c}$, the clump asymmetry parameter, and $g_{eff}$%
, the effective scattering asymmetry parameter.

In summary, we have extended and improved the formulas for the effective
optical depth and albedo of a clumpy medium, and introduced an approximation
for the effective scattering asymmetry parameter. The validity of the
extended mega-grains formulas, with scattering and for other source
distributions, is further demonstrated by comparison with Monte Carlo
simulations in \S \ref{Compare_MC} and \S \ref{Simul_SED_IR}.

\subsection{Comparison with Earlier Theory\label{Old_Theory}}

\begin{figure}[tbp]
\plotone{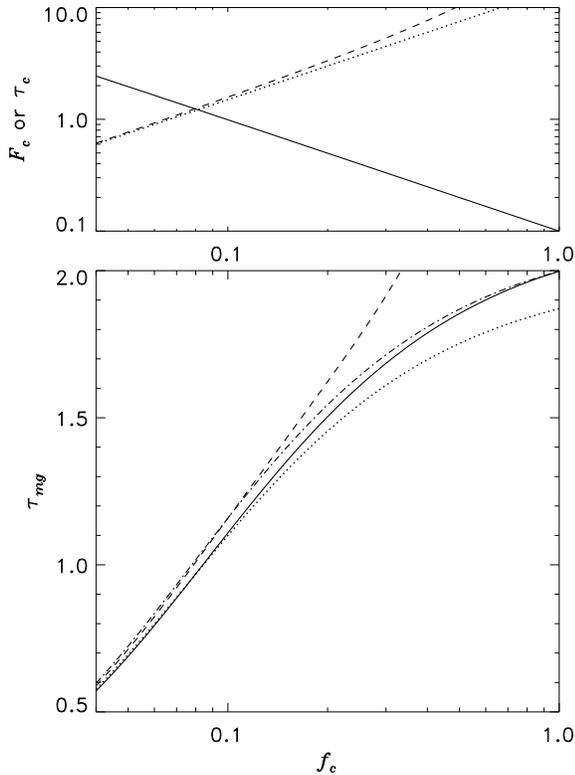}
\caption{
Comparison our extended MGA
with the effective optical depth of a clumpy medium derived by \cite{NatPan84},
for the case of $\tau _{hom}=2$, $\alpha=1000$, and $r_c=0.05$.
The top panel shows $\tau_c$ (solid line), the clump optical radii,
or upper and lower bounds for $F_c$,
the covering factor of clumps (dashed and dotted lines).
The solid line in bottom graph is $\tau _{mg}=R \Lambda_{mg}$
produced by eq.(\ref{MG_xfv}).
The dashed and dotted lines are produced by eq.(\ref{tau_mg_NP}),
using $\tau_c$ and bounds on $F_c$ from the top panel.
The dash-dotted line is produced by eq(\ref{tau_mg_NP_renorm}). \label{mg_np} }
\end{figure}%

Here we compare the extended mega-grains approximation (MGA) for the
effective optical depth of just clumps with the theory of \cite{NatPan84}
(hereafter NP84), when $\alpha $ is large so the ICM can be ignored. Figure 
\ref{mg_np} compares the approximations and quantities involved as function
of $\log _{10}f_{c}$, for the case of $\tau _{hom}=2$, $\alpha =1000$, and $%
r_{c}=0.05$. The upper panel shows $\tau _{c}$ (solid line), the clump
optical radii, and two extreme values bounding $F_{c}$, the average number
of clumps encountered along a random line of sight: 
\begin{equation}
F_{c}^{q}=R\left( \frac{3Q_{c}}{4r_{c}}\right)   \label{covering_factor}
\end{equation}
is the dashed line that eventually diverges to infinity, where $Q_{c}$ is
the porosity given by eq.(\ref{porosity}), and the dotted line is 
\begin{equation}
F_{c}^{f}=R\left( \frac{3f_{c}}{4r_{c}}\right) \,,
\end{equation}
which stays finite, so that $F_{c}^{f}\leq F_{c}\leq F_{c}^{q}$ (see
Appendix \ref{App Nc}). $F_{c}$ is also called the covering factor of the
clumps, and the random lines of sight are perpendicular to the plane of a
slab of thickness $R$, or radial in the case of a spherical medium of radius 
$R$. In the lower panel the solid line is $\tau _{mg}=R\Lambda _{mg}$ using
eq.(\ref{MG_xfv}), the extended MGA. Assuming that the number of clumps
encountered along a random line of sight is Poisson distributed with mean
number $F_{c}$, NP84 derived an equation for the effective optical depth: 
\begin{equation}
\tilde{\tau}=F_{c}\left( 1-e^{-4\tau _{c}/3}\right) \,\,,  \label{tau_mg_NP}
\end{equation}
where $\frac{4}{3}\tau _{c}$ is the expected optical depth of a random path
through a spherical clump. The result of using $F_{c}=F_{c}^{q}$ in eq.(\ref
{tau_mg_NP}) is shown by the dashed line in the lower panel of Figure~\ref
{mg_np}, and $\tilde{\tau}$ agrees with $\tau _{mg}$ (solid line) for $%
f_{c}<0.1$, corresponding to $F_{c}\leq 1$, but $\tilde{\tau}$ exceeds $\tau
_{hom}$ when $f_{c}>0.3$ and $F_{c}>5$, diverging to infinity.. If instead $%
F_{c}=F_{c}^{f}$ is used in eq.(\ref{tau_mg_NP}) then $\tilde{\tau}$ remains
finite as shown by the dotted line. Applying the clump radii renormalization
gives 
\begin{equation}
\tilde{\tau}=F_{c}^{f}\left( \frac{1-\exp \left[ -4\tau
_{c}(1-f_{c})/3\right] }{1-f_{c}}\right) \,,  \label{tau_mg_NP_renorm}
\end{equation}
shown as the dash-dotted line, which is almost identical to $\tau _{mg}$. So
the approach used by NP84 gives almost the same result for $\tau _{mg}$ as
the extended MGA when the average optical depth of a sphere, $\frac{4}{3}%
\tau _{c}$, is used. The quantity $1-e^{-4\tau /3}$ happens to be a good
approximation of $P_{i}(\tau )$, with a maximum relative difference of 5\%
overestimation occurring at $\tau \approx 1.5$ when $P_{i}(1.5)\approx 0.8$.

\subsection{Escape and Absorption Probabilities for Clumpy Media}

The effective optical depth, $\tau _{eff},$ effective scattering albedo, $%
\omega _{eff},$ and effective asymmetry parameter, $\,g_{eff},$ given by the
mega-grains approximation reduces the basic radiative transfer properties of
a two-phase clumpy medium to an effectively homogeneous medium. Therefore we
propose use $\tau _{eff}$, $\omega _{eff}$, and $g_{eff}$, given by
equations (\ref{MG_xfv}, \ref{xsec_eff}, \ref{tau_eff_mg}), (\ref{albedo_eff}%
, \ref{clump_albedo_pe}), and (\ref{geff}, \ref{g_c}) respectively, in the
escape and absorption probability formulae that were developed for
homogeneous media to estimate the escaping or absorbed fractions of photons
in clumpy media. In particular, for isotropic emission uniformly distributed
within a sphere we use ${\cal P}_{esc}^{u}$ given by eqs.(\ref{Oster_EP})
and (\ref{Lucy_EP}), for a central point source we use ${\cal P}_{esc}^{c}$
given by eqs.(\ref{tau_cs_sphere}-\ref{EP1_cs}), and for a uniformly
illuminating external source we use ${\cal P}_{abs}^{x}$ given by eq.(\ref
{Pxabs}). The full set of equations is summarized as a convenient list in \S 
\ref{List_EAP}.

We find these analytic approximations give results in reasonable agreement
with Monte Carlo simulations, as is demonstrated in \S \ref{Compare_MC}. As
suggested by HP93, we also conjecture that given an escape probability
function for a homogeneous medium having any geometry and any distribution
of sources, the escape probability for a two-phase clumpy medium of same
geometry and source distribution can be reasonably approximated by
substituting the effective optical parameters given by the mega-grains
equations into the homogeneous escape probability function.

\subsection{The Fractions Absorbed in Each Phase of a Clumpy Medium\label%
{AbsFrac}}

The energy of the radiation absorbed by the dust is converted into heat and
then re-radiated as infrared (IR) photons. The spectrum of the IR emission
from the dust is dependent on the dust temperature, which is most cases is
determined by the equilibrium between the energy absorbed and emitted. In
this paper, we shall not deal with very small dust grains, which can undergo
temperature fluctuations, an effect that will cause a wider distribution in
dust temperature and enhanced IR emission in the $\sim 3-20\mu $m range
(e.g. \cite{Dwek97}). For the case of a homogeneous medium with uniformly
distributed emitters, most of the dust will be at the same temperature
(accept at the boundary of the medium where the temperature will be lower).
When the uniformly distributed emission occurs in an inhomogeneous medium
there will be a distribution of dust temperatures. In a two-phase clumpy
medium, the clumps will most likely attain a different radiative equilibrium
temperature than the ICM, since the clumps will absorb/radiate a different
amount of energy. Given the total amount of energy absorbed by the dust we
need to know what fraction is absorbed in the clumps and what fraction in
the ICM, to determine the approximate equilibrium temperature of each phase
of the clumpy medium.

In the mega-grains approximation it is the over-density of the clumps that
enters into the formulation, in this way separating the clumps and the ICM
while keeping the ICM continuous. However, to determine the fraction of
photons absorbed by the clumps we must consider the clumps as being
completely separate from the ICM, therefore considering the full density of
dust in the clumps, and so the ICM is not viewed as uniform and continuous
but as having holes occupied by clumps. This approach will give the correct
absorbed fractions as $\alpha =\rho _{c}/\rho _{icm}\rightarrow 1.$ Thus we
define the full clump optical depth as: 
\begin{equation}
\tau _{c}^{a}\,\equiv \,\kappa \rho _{c}r_{c}\,=\,\alpha \kappa \rho
_{icm}r_{c}\,,  \label{tau_cl_full}
\end{equation}
where $\kappa $ is the total absorption plus scattering coefficient of the
dust grains per unit mass, and $r_{c}$ is the clump radius. We then replace $%
\tau _{c}$ with $\tau _{c}^{a}$ in all the mega-grains approximation
equations, and the superscript ``$a$'' shall indicate that all of the clump
optical depth is being used in the following quantities, in particular, $%
\Lambda _{mg}^{a}$ is the result of using $\tau _{c}^{a}$ in eq.(\ref
{MG_xsec}).

First we deal with the case of a point source in the center of a spherical
clumpy medium, and assume that the point source is not in any clump. Let $R$
be the radius of the sphere and define optical depths 
\begin{eqnarray}
\tau _{mg}^{0} &=&R\,\Lambda _{mg}^{a}\,\,, \\
\tau _{icm}^{0} &=&R\,\kappa \rho _{icm}\,,
\end{eqnarray}
due to the mega-grains (subscript ``{\it mg}'') and ICM, respectively, where
the superscript ``0'' indicates that this is for no scattering. To model the
effects of scattering we employ the analytical approximation for the
effective optical radius of a homogeneous sphere with scattering, as given
by equations (\ref{tau_cs_sphere}) and (\ref{tau_cs_sphere_2}), defining new
optical depths as 
\begin{eqnarray}
\tau _{mg} &=&\tau _{mg}^{0}\,(1-\omega _{c}^{a})^{\chi (\tau
_{mg}^{0},g_{c}^{a})}\,,  \label{tau_mg_scat} \\
\tau _{icm} &=&\tau _{icm}^{0}\,(1-\omega _{d})^{\chi (\tau
_{icm}^{0},g_{d})}\,,  \label{tau_icm_scat}
\end{eqnarray}
for the mega-grains and ICM, respectively, where the function $\chi (\tau ,g)
$ is given by eq.(\ref{tau_cs_sphere_2}). Note that we use clump scattering
properties $\omega _{c}^{a}$ and $g_{c}^{a}$ for $\tau _{mg}$, and the usual
dust scattering albedo and asymmetry parameter, $\omega _{d}$ and $g_{d}$,
for the ICM optical depth. Of the total photons absorbed by the medium, the
fraction, $A_{icm}^{c}$, that gets absorbed by the ICM is then estimated to
be 
\begin{equation}
A_{icm}^{c}\,\,\equiv \,\,\frac{(1-f_{c})\,\tau _{icm}}{\tau
_{mg}+(1-f_{c})\,\tau _{icm}}\quad ,  \label{frac_abs_icm}
\end{equation}
where the superscript ``{\it c}'' indicates a central point source. The
factor $1-f_{c}$ is introduced to get an approximation that gives the
correct behavior as $\alpha \rightarrow 1$ or $f_{c}\rightarrow 1$, because
the volume occupied by the clumps should not be included in the ICM. Now if $%
{\cal P}_{esc}^{c}$ is the fraction of photons that escapes the sphere then $%
A_{icm}^{c}(1-{\cal P}_{esc}^{c})$ is the fraction absorbed in the ICM and $%
(1-A_{icm}^{c})(1-{\cal P}_{esc}^{c})$ is the fraction absorbed by clumps.

In the case of an internal uniform distribution of emitters we consider the
emission occurring in the clumps and ICM separately, as follows. For the
fraction $1-f_{c}$ of the photons that are emitted in the ICM we predict
that $A_{icm}^{c}$ will be absorbed by the ICM. For the fraction $f_{c}$
that are emitted in the clumps we predict that a fraction ${\cal P}%
_{esc}^{u}(\tau _{c}^{a},\omega _{d})$ [given by eqs.(\ref{Oster_EP}) and (%
\ref{Lucy_EP})] of those photons will escape the clumps, and then a fraction 
$A_{icm}^{c}$ of them will be absorbed by the ICM. The total fraction of
emitted photons that will be absorbed by the ICM is then 
\begin{equation}
A_{icm}^{u}\equiv A_{icm}^{c}\left[ (1-f_{c})\,+\,f_{c}\,{\cal P}%
_{esc}^{u}(\tau _{c}^{a},\omega _{d})\right] ,  \label{frac_U_abs_icm}
\end{equation}
where the superscript ``{\it u}'' indicates that this is for the case of
uniformly distributed emission of photons. This gives $A_{icm}^{u}\,\,<%
\,A_{icm}^{c}$ because part of the emission occurs inside clumps that are
more dense than the ICM. Defining $F_{abs}\equiv 1-{\cal P}_{esc}^{u}(\tau
_{eff},\omega _{eff})$ to be the total fraction of photons absorbed in the
medium then $A_{icm}^{u}F_{abs}$ is the fraction absorbed in the ICM and $%
(1-A_{icm}^{u})F_{abs}$ is the fraction absorbed by clumps.

To estimate absorbed fractions for the case of a uniformly illuminating
external source we just take the average of the values for central source
and uniform source: 
\begin{equation}
A_{icm}^{x}\,\,\equiv \frac{A_{icm}^{u}\,\,+\,A_{icm}^{c}}{2}\quad .
\label{frac_X_abs_icm}
\end{equation}
The estimate is motivated by Monte Carlo simulations which indicate that $%
A_{icm}^{u}<A_{icm}^{x}<A_{icm}^{c}$. If ${\cal P}_{abs}^{x}(\tau
_{eff},\omega _{eff})$ is the total fraction absorbed in the sphere then $%
A_{icm}^{x}{\cal P}_{abs}^{x}$ is the fraction absorbed in the ICM and $%
(1-A_{icm}^{x}){\cal P}_{abs}^{x}$ is the fraction absorbed by clumps. The
absorbed fractions predicted by the above equations are compared with the
results of Monte Carlo simulations in \S \ref{Compare_MC}.

\section{SUMMARY OF EQUATIONS\label{List_EAP}}

\subsection{Escape and Absorption Probabilities}

In the following formulae, use $(\tau _{hom},\omega _{d},g_{d})$ for all
occurrences of $(\tau ,\omega ,g)$ if the medium is homogeneous, or if the
medium is clumpy use $(\tau _{eff},\omega _{eff},g_{eff})$ given by the
mega-grains equations below.\medskip

\noindent \underline{Central Point Source:}

\noindent 
\begin{eqnarray}
{\cal P}_{esc}^{c}(\tau ,\omega ,g) &\equiv &\exp \left[ -\,\tau _{S}(\tau
,\omega ,g)\right]  \nonumber \\
\tau _{S}(\tau ,\omega ,g) &\equiv &\tau \,(1-\omega )^{\chi (\tau ,g)}
\label{List_Pesc_1} \\
\chi (\tau ,g) &\equiv &1\,-\,\frac{1}{2}\left( 1-e^{-\tau /2}\right) (1-g)^{%
\frac{1}{2}}  \nonumber
\end{eqnarray}

\noindent \underline{Uniform Distribution of Emitters:}

\noindent 
\begin{eqnarray}
{\cal P}_{esc}^{u}(\tau ,\omega ) &\equiv &\frac{P_{e}(\tau )}{1-\omega
[1-P_{e}(\tau )]}  \label{List_Pesc} \\
P_{e}(\tau ) &=&\frac{3}{4\tau }\,P_{i}(\tau )  \label{List_Pe} \\
P_{i}(\tau ) &=&1-\frac{1}{2\tau ^{2}}+\left( \frac{1}{\tau }+\frac{1}{2\tau
^{2}}\right) e^{-2\tau }  \label{List_Pi}
\end{eqnarray}

\noindent \underline{Uniformly Illuminating External Source:}

\noindent 
\begin{eqnarray}
{\cal P}_{abs}^{x}(\tau ,\omega ) &\equiv &P_{i}\left( \tau \right) \,\left[
1-\omega {\cal P}_{esc}^{u}(\tau ,\omega )\right]  \nonumber \\
&=&\frac{4\tau \left( 1-\omega \right) }{3}\,{\cal P}_{esc}^{u}(\tau ,\omega
)  \label{List_Pabs_x}
\end{eqnarray}
Recall that ${\cal P}_{esc}^{u}$ and ${\cal P}_{abs}^{x}$ are valid (most
accurate) when $g=g^{*}(\tau )$ [see eq.(\ref{g_tau})], whereas $P_{e}(\tau
) $ and $P_{i}(\tau )$ are exact formulae.

\subsection{Mega-Grains Approximation}

\noindent \underline{Input Parameters:}

\noindent 
\begin{eqnarray*}
\kappa &=&\text{dust mass extinction coefficient} \\
\omega _{d} &=&\text{scattering albedo of dust grains} \\
g_{d} &=&\text{scattering asymmetry parameter} \\
\rho _{hom} &=&\text{average mass density of the dust} \\
f_{c} &=&\text{filling factor of the clumps} \\
\gamma &=&\text{MGA tuning parameter (}\sim \text{1)} \\
\alpha &=&\rho _{c}/\rho _{icm} \\
r_{c} &=&\text{radius of each spherical clump} \\
R_{S} &=&\text{radius of spherical medium.}
\end{eqnarray*}

\noindent \underline{Effective Optical Depth, $\tau _{eff}$ :}

\noindent 
\begin{eqnarray}
\rho _{icm} &=&\frac{\rho _{hom}}{(\alpha -1)\,f_{c}+1}\qquad \text{(ICM
density)}  \nonumber \\
&&  \nonumber \\
\tau _{c} &=&(\alpha -1)\,\rho _{icm}\,r_{c}\,\kappa \qquad \text{(clump
optical radius)}  \nonumber \\
&&  \nonumber \\
\Lambda _{mg} &=&\frac{3f_{c}}{4r_{c}}\,\frac{P_{i}\left[ \tau
_{c}(1-f_{c})^{\gamma }\right] }{(1-f_{c})^{\gamma }}  \label{List_MG_xint}
\\
&&  \nonumber \\
\Lambda _{eff} &=&\Lambda _{mg}\,+\,\kappa \rho _{icm}  \nonumber \\
\tau _{eff} &=&R_{S}\,\Lambda _{eff}  \nonumber
\end{eqnarray}
where the formula for $\Lambda _{mg}$ utilizes $P_{i}(\tau )$ given by eq.(%
\ref{List_Pi}) and $0<\gamma \leq 1$. Use $\gamma =1$ for uniform
internal/external sources, $\gamma =0.75$ for central source, and $\gamma
=0.5$ for photons impacting a slab.\medskip

\noindent \underline{Effective Albedo, $\omega _{eff}$ :}

\[
\omega _{eff}\,=\,\frac{\omega _{c}\,\Lambda _{mg}\,\,+\,\,\omega
_{d}\,\kappa \rho _{icm}}{\Lambda _{eff}}\quad , 
\]
where the clump albedo is 
\[
\omega _{c}\,=\,\omega _{d}\,\,{\cal P}_{esc}^{u}\left[ \tau
_{c}(1-f_{c})^{\gamma },\,\omega _{d}\right] \quad , 
\]
and ${\cal P}_{esc}^{u}$ is the OLEP formula stated in eq.(\ref{List_Pesc})
above.\medskip

\noindent \underline{Effective Scattering Asymmetry Parameter, $g_{eff}$ :}

\[
g_{eff}\,=\,\frac{g_{c}\,\Lambda _{mg}\,\,+\,\,g_{d}\,\kappa \rho _{icm}}{%
\Lambda _{eff}}\quad ,
\]
where the clump asymmetry parameter is 
\begin{eqnarray*}
g_{c} &=&g_{d}\,-\,C\left( 1-\frac{1+\exp \left( -B/A\right) }{1+\exp \left(
\left[ \tau _{c}(1-f_{c})^{\gamma }-B\right] /A\right) } \right)  \\
A &\equiv &1.5+4g_{d}^{3}+2\omega _{d}\sqrt{g_{d}}\exp (-5g_{d}) \\
B &\equiv &2-g_{d}(1-g_{d})-2\omega _{d}g_{d} \\
C &\equiv &\left[ 3-\sqrt{2g_{d}}-2\omega _{d}g_{d}(1-g_{d})\right] ^{-1}
\end{eqnarray*}

\subsection{Fractions Absorbed in Each Phase}

Let $P_{abs}=1-P_{esc}$ be the generic absorption probability. Then the
fraction of photons absorbed in the ICM is $A_{icm}P_{abs}$, and the
fraction absorbed by clumps is $\left( 1-A_{icm}\right) P_{abs}$, where $%
A_{icm}$ is given by one of the following formulas, corresponding to the
type of source. See below for definitions of $\tau _{mg}$ and $\tau _{icm}$.
\medskip

\noindent \underline{Central Point Source:} \noindent 
\begin{equation}
A_{icm}^{c}\equiv \frac{(1-f_{c})\tau _{icm}}{\tau _{mg}+(1-f_{c})\tau _{icm}%
}
\end{equation}

\noindent \underline{Uniform Distribution of Emitters:} \noindent 
\begin{equation}
A_{icm}^{u}\equiv A_{icm}^{c}\left[ (1-f_{c})\,\,+\,f_{c}\,{\cal P}%
_{esc}^{u}(\tau _{c}^{a},\omega _{d})\right]
\end{equation}

\noindent \underline{Uniformly Illuminating External Source:} 
\begin{equation}
A_{icm}^{x}\equiv \frac{A_{icm}^{u}\,\,+\,A_{icm}^{c}}{2}
\end{equation}

\noindent \underline{Component Optical Depths:} 
\begin{eqnarray*}
\tau _{c}^{a}\,&\equiv& \,\kappa \rho _{c}r_{c}\,=\,\alpha \kappa \rho
_{icm}r_{c} \\
\tau _{mg}^{0}\, &=&\,R_{S}\,\Lambda _{mg}^{a} \\
\tau _{icm}^{0}\, &=&\,R_{S}\,\kappa \rho _{icm}
\end{eqnarray*}
The superscript ``$a$'' indicates that the full clump optical depth is being
used, i.e. $\Lambda _{mg}^{a}$ is the result of using $\tau _{c}^{a}$ in eq.(%
\ref{List_MG_xint}). The superscript ``0'' indicates that the quantities are
for no scattering. To include the effects of scattering in spherical
geometry, apply equations (\ref{List_Pesc_1}): 
\begin{eqnarray*}
\tau _{mg}\, &=&\,\tau _{mg}^{0}\,(1-\omega _{c}^{a})^{\chi (\tau
_{mg}^{0},g_{c}^{a})} \\
\tau _{icm}\, &=&\,\tau _{icm}^{0}\,(1-\omega _{d})^{\chi (\tau
_{icm}^{0},g_{d})}\,.
\end{eqnarray*}

\subsection{The Case of $\alpha <1$}

If one desires to model randomly distributed spherical cavities (e.g. SNRs
instead of clumps) that have a lower density than the ICM (so that $\alpha <1
$), the mega-grains equations can be applied by redefining 
\begin{eqnarray}
f_{c}^{\prime } &=&1-f_{c}\quad ,  \nonumber \\
\alpha ^{\prime } &=&\frac{1}{\alpha }\quad \quad ,  \label{cavities_alpha<1}
\end{eqnarray}
and using these new inverted values in the above equations to obtain $(\tau
_{eff},\omega _{eff},g_{eff})$ for the medium with cavities. However, when
computing the fractions absorbed in each phase of the medium we do not apply
the inversion transform of eq.(\ref{cavities_alpha<1}). The inverted
mega-grains approximation is demonstrated in \S \ref{Sec_Depend-on-alpha}.

\section{COMPARISON OF ANALYTIC APPROXIMATIONS WITH MONTE CARLO SIMULATIONS%
\label{Compare_MC}}

In \S \ref{Extend_MG} we developed the extended mega-grains approximation
(MGA) and compared it with Monte Carlo simulations, which verified the
effective optical depth predicted by the approximations for the case of no
scattering and when emission is from a central point source in a spherical
clumpy medium. We shall use the acronym MCRT when referring to Monte Carlo
simulations of radiative transfer, and the acronym MGEP when referring to
the analytic approximations consisting of the mega-grains (MG)
approximations combined with escape/absorption probability (EP) formulae,
which were summarized \S \ref{List_EAP}. In this section we present more
detailed comparisons of the MGEP model with MCRT simulations when scattering
is also involved, and for all three types of source distributions discussed
in \S \ref{Esc_Abs_Homog}. In the following we designate the cases of a
central isotropic source by ``C'', uniformly distributed internal sources by
``U'', and uniformly illuminating external source by ``X''.

We shall compare the MGEP approximations with MCRT results over a wide range
of parameters that define a two-phase clumpy medium in spherical geometry,
except that the clump radii will be held fixed at $r_{c}=0.05$ relative to
the radius of the sphere $R_{S}=1$. A larger value would be less realistic
and cause the number of clumps to be less than needed for good statistics in
the MCRT model, whereas spherical clumps with radii smaller than $r_{c}=0.05$
become difficult to represent accurately on a grid. All the Monte Carlo
simulations were performed using a grid of $100^{3}$ or $127^{3}$ voxels to
represent the clumpy medium, and followed at least $10^{6}$ photons for each
result. The quantities in the comparisons will be shown as a function of
either equivalent homogeneous optical depth, $\tau _{hom}$, clump filling
factor, $f_{c}$, or clump to ICM density ratio, $\alpha $, in the following
three subsections.

\subsection{Dependence on Optical Depth}

Recall that the equivalent homogeneous optical depth of extinction, $\tau
_{hom}=\kappa \rho _{hom}R_{S}$, can vary due to changing wavelength or dust
mass ($R_{S}$ is held constant), and so is the major parameter involved in
modeling the transfer of a spectrum of radiation in a clumpy medium. Figure 
\ref{tauS_ffs} showed the behavior of the effective optical depth $\tau _{S}$
[eq.(\ref{tau_eff})] versus ${\tau }_{hom}$ resulting from Monte Carlo
simulations with scattering ($\omega _{d}=0.6$ and $g_{d}=0.6$) for three
cases of the filling factor, $f_{c}=0.1,\,0.2,$ and $0.3,$ (squares,
triangles, and diamonds, respectively) with $\alpha =100$. The dashed,
dotted, and solid lines are produced by the extended mega-grains theory
combined with eq.(\ref{List_Pesc_1}) to compute $\tau _{S}(\tau
_{eff},\omega _{eff},g_{eff})$, and the curves agree well with Monte Carlo
results. As discussed in \S \ref{Extend_MG}, the transition in the variation
of $\tau _{S}$ as a function ${\tau }_{hom}$ goes from clump dominated
(steeper slope at low values of ${\tau }_{hom}$) to ICM dominated (less
slope at large ${\tau }_{hom}$) as the clumps become opaque, and the
occurrence of this transition can be estimated by eq.(\ref{tau_clump_f}).
Unless otherwise indicated, $\alpha =100$ in this section.

\begin{figure}[tbp]
\plotone{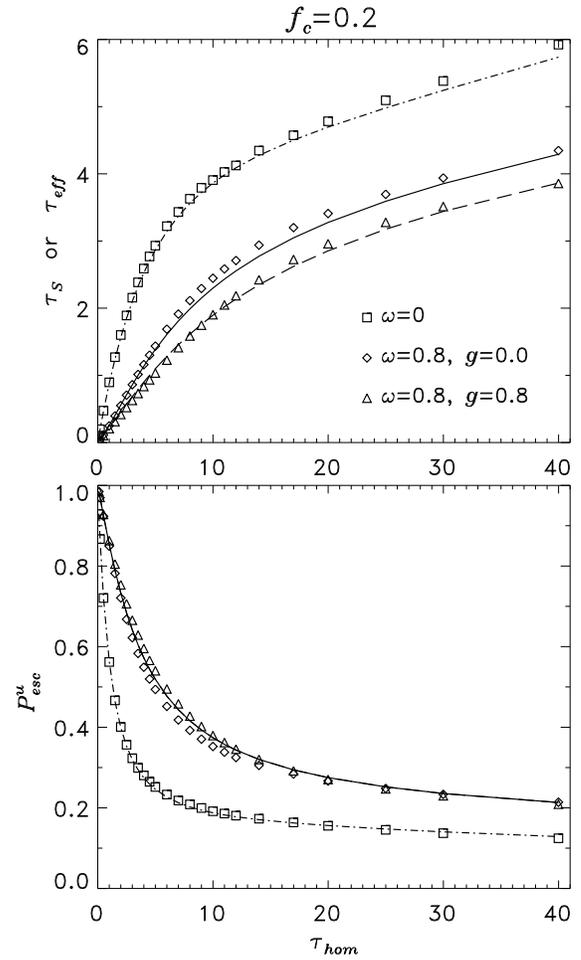}
\caption{ Comparison of MGEP theory (curves)
to MCRT results (symbols) as function of $\tau _{hom}$
with $f_{c}=0.2$ and $\alpha =100$,
for 3 different sets of dust scattering parameters
as indicated in the plot legend. }
\label{teff_vtag}
\end{figure}
%

In Figure~\ref{teff_vtag} we study the effect of changing the dust
scattering parameters for the case of $f_{c}=0.2$, with the upper panel
showing the effective optical depth with or without scattering ($\tau _{S}$
or $\tau _{eff}$ respectively) for source type C, and lower panel showing
the escaping fraction of source type U, all versus $\tau _{hom}$. The case
of no scattering is indicated with squares for MCRT results and the dash-dot
lines are the MGEP model. The diamonds and solid lines show MCRT and MGEP
results, respectively, for $\omega _{d}=0.8$ and $g_{d}=0$ (isotropic
scattering); the triangles and long-dashed lines show MCRT and MGEP results,
respectively, for $\omega _{d}=0.8$ and $g_{d}=0.8$ (almost forward
scattering). The values of $\tau _{eff}$ given by MGA (dash-dotted line) for
no scattering are used in eq.(\ref{List_Pesc_1}), along with values of $%
\omega _{eff}$ and $g_{eff}$ (not shown), to compute $\tau _{S}(\tau
_{eff},\omega _{eff},g_{eff})$, the solid and long-dashed lines in the upper
panel. The solid line in the lower panel is computed by using $\tau _{eff}$
and $\omega _{eff}$ in eq.(\ref{List_Pesc}) for source type U. The MCRT
results show that, as expected, forward scattering dust lowers the effective
optical depth and increases the escape probability as compared to isotropic
scattering dust, especially for a central source of photons, and the
analytical approximations of MGEP correctly predict this effect. In the case
of a uniformly distributed source the dependence of the escaping fraction on
the dust scattering asymmetry parameter $g_{d}$ is weak, and so it is not a
problem that $g_{d}$ does not enter into the MGEP equations for source type
U. The analytical approximations of MGEP agree well with the numerical
results of MCRT at low optical depths, whereas at high optical depths there
are some differences, especially for source type C, which are due to
specific realizations of the clumpy medium in MCRT simulations, but the
agreement is still acceptable.

\begin{figure}[tbp]
\plotone{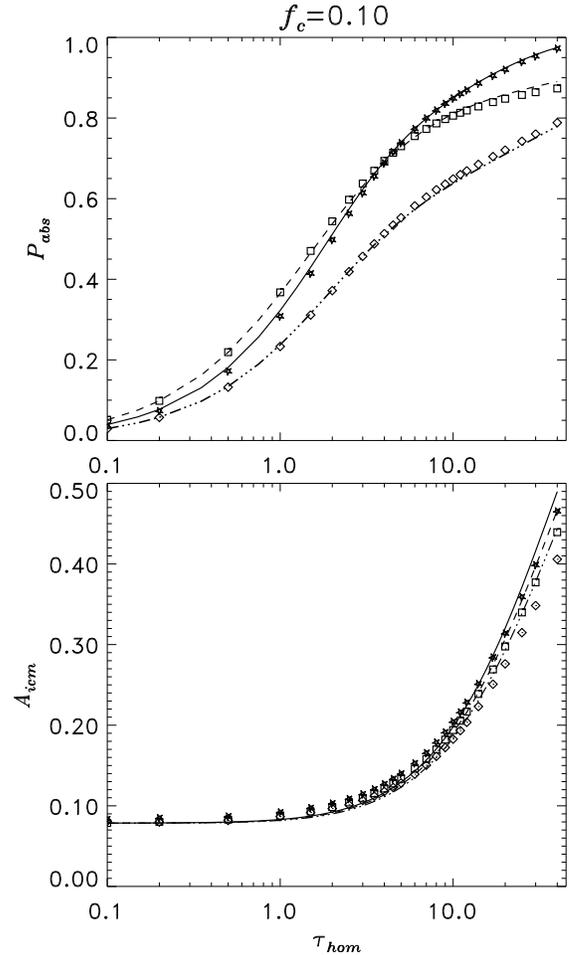}
\caption{ Comparison of MGEP absorption probabilities
to MCRT absorbed fractions as function of $\tau _{hom}$,
with $f_{c}=0.1$, $\alpha =100$, and $r_c=0.05$.
The top panel shows the total absorbed fractions whereas the bottom
panel shows the fraction of the total which is absorbed by the ICM.
Scattering parameters are $\omega _{d}=0.6$ and $g_{d}=0.6$.
The star, diamond, and square symbols represent MCRT results
for source types C, U, and X, respectively.
The solid, dot-dashed, and dashed lines
represent MGEP results for source types C, U, and X.}
\label{pabs_f1}
\end{figure}
%

\begin{figure}[tbp]
\plotone{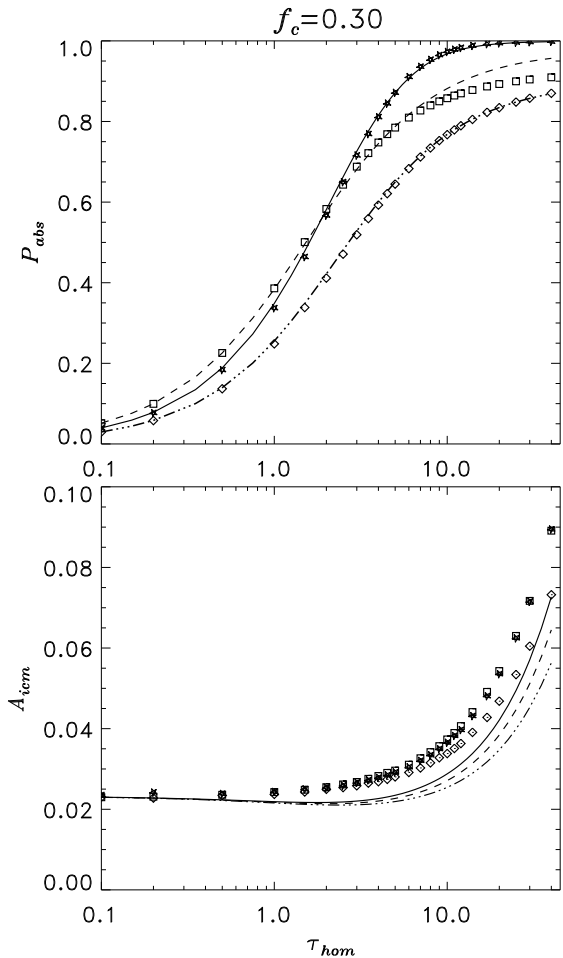}
\caption{ Same as Figure~\ref{pabs_f1}, but for $f_{c}=0.3$. }
\label{pabs_f3}
\end{figure}
%

Figure~\ref{pabs_f1} compares in the upper panel the absorbed fraction of
photons from each type of source, and in the lower panel the fraction of the
total absorbed photons that are absorbed in just the ICM, as function of $%
\tau _{hom}$, with $f_{c}=0.1$, and with scattering parameters $\omega
_{d}=0.6$ and $g_{d}=0.6$. The star, diamond, and square symbols represent
MCRT numerical results for source types C, U, and X respectively. The solid,
dot-dashed, and dashed lines represent MGEP predictions for source types C,
U, and X respectively. The absorbed fractions are computed as 
\[
P_{abs}=\left\{ 
\begin{array}{l}
1-{\cal P}_{esc}^{c}(\tau _{eff},\omega _{eff},g_{eff})\quad \text{[C, eq.(%
\ref{List_Pesc_1})]} \\ 
1-{\cal P}_{esc}^{u}(\tau _{eff},\omega _{eff})\quad \text{[U, eq.(\ref
{List_Pesc})]} \\ 
{\cal P}_{abs}^{x}(\tau _{eff},\omega _{eff})\quad \quad \text{[X, eq.(\ref
{List_Pabs_x})]}
\end{array}
\right\} 
\]
for each indicated source type. The fractions absorbed by the ICM, $A_{icm}$%
, are computed by equations (\ref{frac_abs_icm}), (\ref{frac_U_abs_icm}),
and (\ref{frac_X_abs_icm}) for source types C, U, and X, respectively. The
MGEP model agrees with the MCRT results, showing the same behavior for each
source type. Figure~\ref{pabs_f3} compares the same quantities for $f_{c}=0.3
$. The MGEP model predicts a total absorbed fraction (upper panel) for
source type X greater than the MCRT results when $\tau _{hom}>10$, probably
due to limitations of the assumptions used in deriving eq.(\ref{Pxabs}) for $%
{\cal P}_{abs}^{x}$. For this case of $f_{c}=0.3$ (and actually for $%
f_{c}>0.1$) the MGEP model predicts values for $A_{icm}$ (lower panel) that
are lower than the MCRT results. When $f_{c}<0.1$ the absorbed ICM fractions
predicted by MGEP are slightly higher than the MCRT results. Overall, the
agreement is acceptable and the MGEP model exhibits the same behavior as the
MCRT simulations.

\subsection{Dependence on Filling Factor}

\begin{figure}[tbp]
\plotone{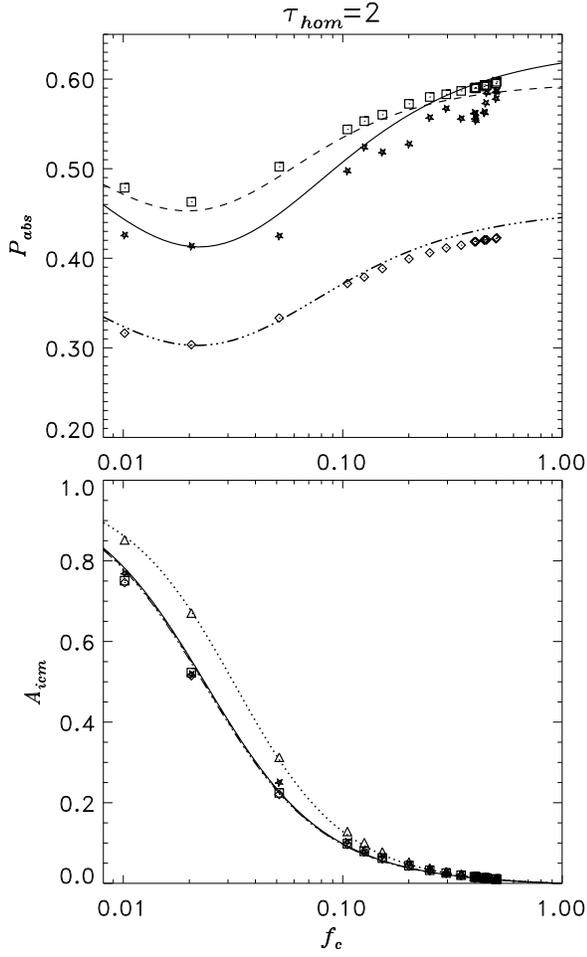}
\caption{ Comparison of MGEP absorption probabilities
to MCRT absorbed fractions as function of $f_{c}$,
with $\tau _{hom}=2$, $\alpha =100$, and $r_c=0.05$.
Scattering parameters are $\omega _{d}=0.6$ and $g_{d}=0.6$.
The star, diamond, and square symbols represent MCRT results
for source types C, U, and X, respectively.
The solid, dot-dashed, and dashed lines
represent MGEP results for source types C, U, and X.
The dotted line and the triangles
represent MGEP and MCRT results,
respectively, for the case of no scattering. }
\label{pabs_t2}
\end{figure}
%

\begin{figure}[tbp]
\plotone{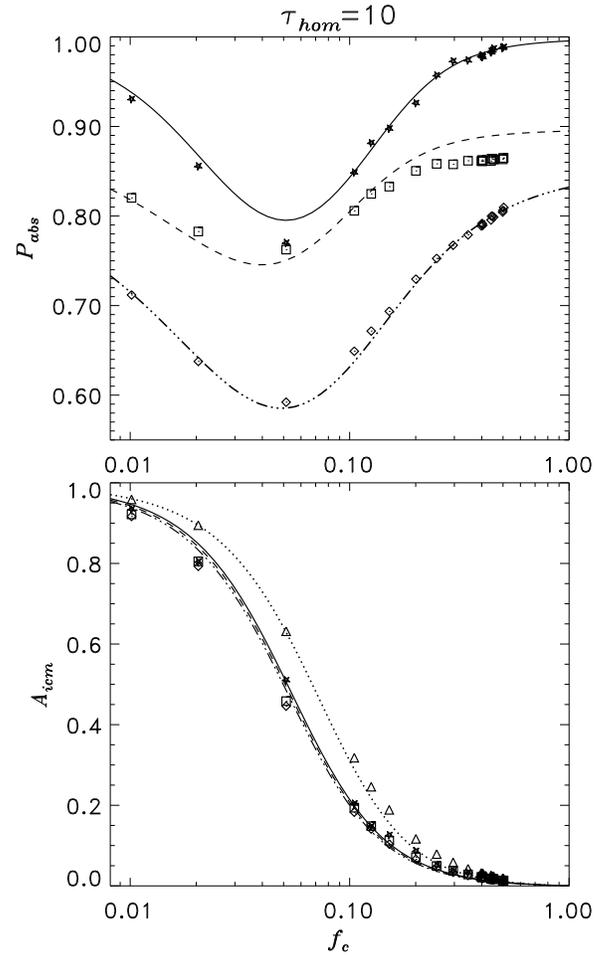}
\caption{ Same as Figure~\ref{pabs_t2}, but for $\tau_{hom}=10$. }
\label{pabs_t10}
\end{figure}
%

We next compare MGEP and MCRT results as a function of the clump filling
factor $f_{c}$, holding other parameters fixed. Figures~\ref{pabs_t2} and 
\ref{pabs_t10} compare MCRT and MGEP predictions of the fraction of emitted
photons that are absorbed in the medium, $P_{abs}$, in the upper panels, and
the fraction of the total absorbed photons that are absorbed in the ICM, $%
A_{icm}$, in the lower panels, as function of $f_{c}$, with scattering
parameters $\omega _{d}=0.6$ and $g_{d}=0.6$. Figure~\ref{pabs_t2} is for
the case of $\tau _{hom}=2$ and Figure~\ref{pabs_t10} is for $\tau _{hom}=10$
(note the different ranges for $P_{abs}$ axis). The star, diamond, and
square symbols represent MCRT numerical results for source types C, U, and
X, respectively. The solid, dot-dashed, and dashed lines represent MGEP
analytical results for source types C, U, and X, respectively. The
analytical MGEP theory agrees well with the MCRT numerical results, with the
exception of the absorbed fraction of a uniformly illuminating external
source with scattering. For that case of source type X, when $\tau _{hom}$
is large and $f_{c}\rightarrow 1$ the MGEP theory overestimates the absorbed
fraction with a difference that increases with $f_{c}$. This is probably
because the assumption of uniformly distributed photons after first
scattering (see \S \ref{external_src}) becomes invalid as $\tau
_{hom}\rightarrow \infty $ since then the externally impacting photons do
not penetrate the sphere.

The dotted line and the triangles in the lower panels represent MGEP and
MCRT results, respectively, for the case of no scattering ($\omega _{d}=0$).
The effect of turning on scattering is to decrease the fraction of photons
absorbed in the ICM, since photons scattered by the ICM then have more of a
chance of getting absorbed in the denser clumps. The MGEP model tends to
overestimate $A_{icm}$ at low clump filling factors and underestimate $%
A_{icm}$ at high filling factors, with best agreement at about $f_{c}=0.1.$
However, the MGEP theory gives the same variation of $A_{icm}$ with $f_{c}$
as the MCRT simulations, and overall the agreement is acceptable.

\subsection{Dependence on Density Ratio \label{Sec_Depend-on-alpha}}

\begin{figure}[tbp]
\epsfxsize=7in \centerline{\vbox{\epsfbox{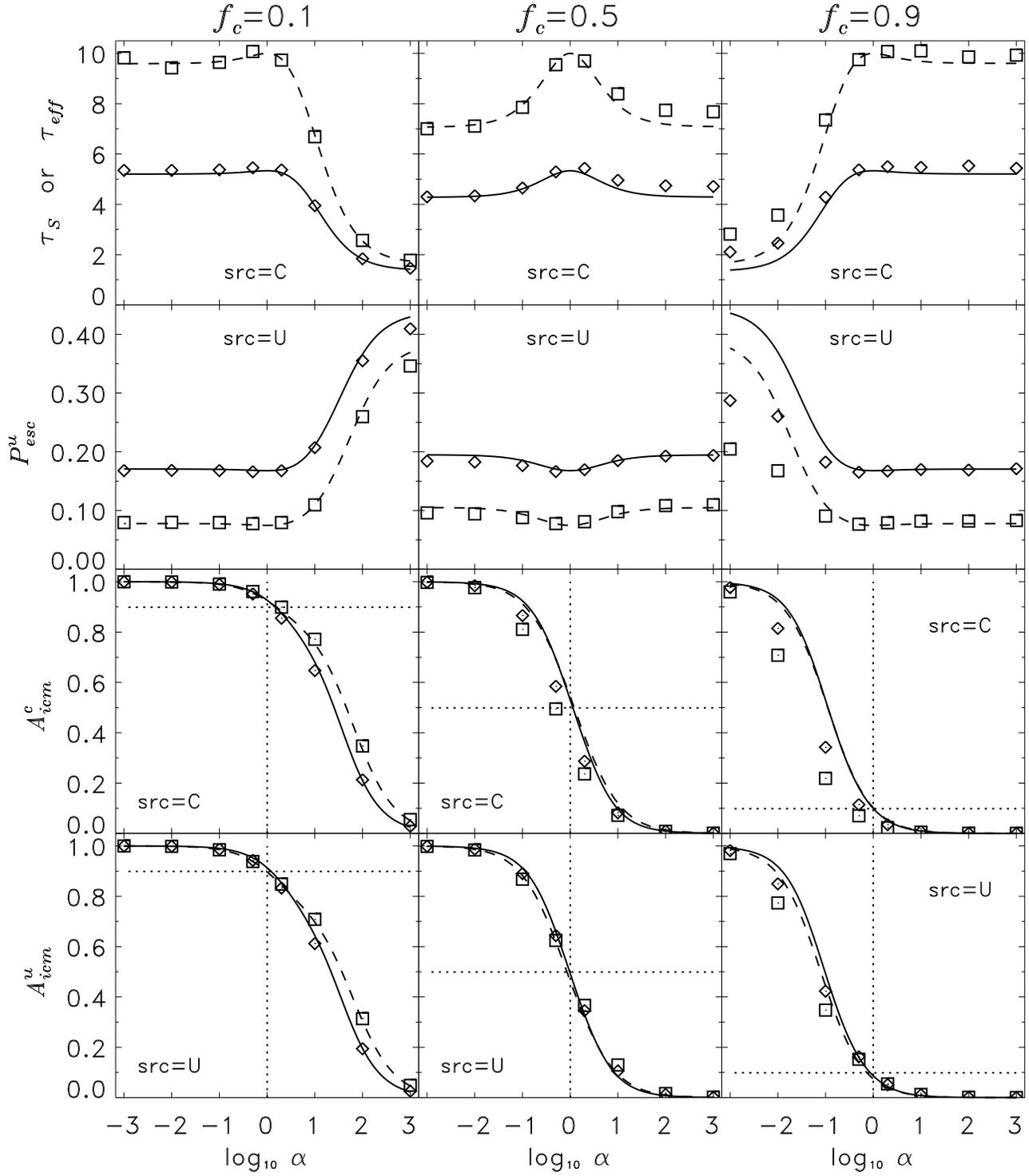}}}
\caption{ Comparison of MGEP and MCRT results as function
of $\alpha =\rho _{c}/\rho _{icm}$
for three values of $f_c=0.1$, 0.5, and 0.9, as indicated,
with $\tau _{hom}=10$ and $r_c=0.05$.
See \S \ref{Sec_Depend-on-alpha} for more information. }
\label{pesc_aicm_drat}
\end{figure}

Finally, we compare MGEP and MCRT as a function of the clump to ICM density
ratio, $\alpha =\rho _{c}/\rho _{icm}$, holding other parameters fixed.
Figure~\ref{pesc_aicm_drat} shows the effective optical depth ($\tau _{eff}$
or $\tau _{S}$) for source type C (top row), the escaping fraction of
photons (${\cal P}_{esc}^{u}$) from source type U (second row), the
fractions absorbed by the ICM for source type C ($A_{icm}^{c}$ in third
row), and for source type U ($A_{icm}^{u}$ in fourth row), all as a function
of $\alpha $, for the cases of $f_{c}=0.1$, 0.5, and 0.9 as indicated, with $%
\tau _{hom}=10$. The square and diamond symbols indicate MCRT results for no
scattering ($\omega _{d}=0$) and with scattering ( for $\omega _{d}=0.6$ and 
$g_{d}=0.6$), respectively. The solid and dashed lines indicate MGEP
predictions with and without scattering, respectively. The value $\alpha =1$
corresponds to a homogeneous medium, and then $\tau _{eff}=\tau _{hom}$ when 
$\omega _{d}=0$. Values of $\alpha <1$ show the case of spherical cavities
in a denser ICM. For the case of denser clumps ($\alpha >1$), most of the
variation in $\tau _{eff}$, $\tau _{S}$, and ${\cal P}_{esc}^{u}$ occurs for 
$1<\alpha <100$ because in this range the clumps go from being optically
thin to thick, and for $\alpha >100$ the clumps are essentially opaque.
Turning on scattering decreases the effective optical depth and allows more
photons to escape, but this effect is diminished as $\alpha \rightarrow
\infty $, and is correctly predicted by the MGEP model. The reason for the
diminished effects of scattering is that $\omega _{eff}$ and $g_{eff}$ (not
shown) decrease by more than a factor of two as $\alpha \rightarrow \infty $.

Applying the MGEP model to the case of spherical cavities ($\alpha <1$) is
accomplished by the inversion transform of eq.(\ref{cavities_alpha<1}),
which then regards the cavities as the ICM and the rest of the medium as
clumps. This is not the intended application of the mega-grains
approximation, since there are then actually no isolated clumps, but there
is some indication that the MGEP model could be tailored to give a
reasonable approximation of the escaping radiation. When $f_{c}=0.9$ we see
that the MGEP model does not give correct predictions for $\alpha <1$, the
case of spherical cavities. This is understandable, because when the
cavities occupy 90\% of the volume, the high density regions are just shells
between the cavities, which are not well approximated by clumps having the
same radius as the cavities (recall $r_{c}=0.05$). Choosing a smaller value
for $r_{c}$ would improve the MGEP approximation of the MCRT results. The
MGEP model provides an acceptable approximation of the MCRT results for $%
\alpha \geq 1$. When $f_{c}=0.5$ and as $\alpha \rightarrow \infty $ there
is an increasing difference between MGEP and MCRT results in the case of a
central source, but this can be attributed to the specific realization of
the clump locations relative to the source: there is a 50\% chance that the
point source will be inside a clump (as in this case), causing an increase
of the optical depth in the MCRT simulation. Other simulations for the same
filling factor give MCRT results that are less than the MGEP predictions, so
we expect that on average the MGEP predictions will be close to MCRT results.

The fractions absorbed by the ICM are shown in the third and bottom rows of
Figure~\ref{pesc_aicm_drat} for source types C and U, respectively. The
intersection of the horizontal and vertical dotted lines indicate the
nominal value of $A_{icm}^{u}$ expected as the medium becomes homogeneous,
since then the fraction photons absorbed by the ICM should be $1-$ $f_{c}$
of the total absorbed photons. For the case of when the clumps are denser
than the ICM ($\alpha >1$) the MGEP model agrees well with the MCRT results,
when the source is uniformly distributed. When the source is at a central
point and $f_{c}=0.5$ the MGEP predictions are less than the MCRT results,
and this can be attributed to the fact that the point source happens to be
in a clump, as discussed above. Note that the MGEP model correctly predicts
that turning on scattering causes less photons to be absorbed by the ICM and
more to be absorbed in clumps. When $\alpha <1$ the MGEP model predicts
values lower than MCRT for $f_{c}>0.3$, since the MGA was not actually
formulated for this application, but the MGEP model does exhibit the correct
behavior.

\subsection{Body Centered Cubic Lattice of Clumps}

The mega-grains approximation can be applied to the case of cubic clumps
randomly located on a body centered cubic (BCC) lattice with partial
success. The BCC lattice was used by \cite{wittgor96} in their Monte Carlo
simulations. For source type U, the MGEP model does predict the same
escaping fraction computed by MCRT when $r_{c}=1/N=0.05$, where $N=20$ is
the number of grid element along each axis of the lattice. However, this
value of $r_{c}$ in the MGEP model gives erroneous predictions for the
effective optical depth seen by a central source. The problem is related to
the fact that the apparent projected area of a cubic clump depends on the
viewing angle. To match the MGEP model with MCRT results for source type C,
we need to use $r_{c}\simeq \sqrt{2}/N=0.071$ in the mega-grains equations,
but then the predictions for source type U are wrong.

\onecolumn


%

\twocolumn

\section{MODELING THE ABSORPTION OF\protect\linebreak STELLAR RADIATION BY
DUST AND THE INFRARED EMISSION \label{Simul_SED_IR}}

In this section we model the transfer of a spectrum of radiation in a
two-phase clumpy medium, from emission by starlike sources to the
absorption and scattering by dust and the resulting infrared emission from
dust heated by the absorbed radiation. Both detailed Monte Carlo simulations
and the analytical approximations developed in previous sections are used to
model the transfer of radiation, and the emerging spectral energy
distribution (SED) from the two methods of modeling are compared. We shall
again use the acronyms MCRT when referring to the Monte Carlo radiative
transfer model, and MGEP when referring to the mega-grains approximations
combined with escape/absorption probability formulae.

The dust is assumed to be composed of 40\% graphite and 60\% silicates by
mass, and to have the grain size distribution 
\begin{equation}
\zeta (a)\,\propto \,a^{-3.5}\,\,,
\end{equation}
normalized over the following range of grain sizes: 
\begin{equation}
0.001{\mu }{\rm m}<a<0.25{\mu }{\rm m\,\,.}
\end{equation}
Using the optical constants from \cite{Draine85}, we applied Mie theory to
calculate the absorption and scattering efficiencies of dust grains, $%
Q_{abs}(a,\lambda )$ and $Q_{scat}(a,\lambda )$ respectively, as a function
of grain size, $a$, and photon wavelength $\lambda $. The scattering
asymmetry parameters, $g(a,\lambda )\,=\,\left\langle \cos \theta
_{scat}(a,\lambda )\,\right\rangle $, were also calculated by averaging with
respect to the distribution of scattering angles. Then the cross-sections
and asymmetry parameters were averaged with respect to the grain size
distribution to get the mass absorption and scattering coefficients and
average asymmetry parameters used in the models: 
\begin{eqnarray}
\kappa _{abs}(\lambda )\, &\equiv &\,\frac{\left\langle \pi
a^{2}Q_{abs}(a,\lambda )\right\rangle _{a}}{\left\langle m_{g}\right\rangle
_{a}}\, \\
\kappa _{scat}(\lambda )\, &\equiv &\,\frac{\left\langle \pi
a^{2}Q_{scat}(a,\lambda )\right\rangle _{a}}{\left\langle m_{g}\right\rangle
_{a}}  \nonumber \\
g_{d}(\lambda )\, &\equiv &\,\frac{\left\langle g(a,\lambda )\,\pi
a^{2}Q_{scat}(a,\lambda )\right\rangle _{a}}{\left\langle \pi
a^{2}Q_{scat}(a,\lambda )\right\rangle _{a}}\quad ,  \nonumber
\end{eqnarray}
where the averaging operator is defined as 
\begin{equation}
\left\langle \cdot \right\rangle _{a}\,\equiv \,\int_{a_{\min }}^{a_{\max
}}\left( \cdot \right) \,\zeta (a)\,da\quad ,
\end{equation}
and the average grain mass is 
\begin{equation}
\left\langle m_{g}\right\rangle _{a}\,\equiv \left\langle \,\frac{4}{3}\pi
a^{3}\rho \right\rangle _{a}\quad ,
\end{equation}
where $\rho $ is the mass density of a graphite or silicate grain. All the
following simulations have the same total dust mass of $1.01\,M_{\odot }$
contained in spherical region $1$pc in radius. This corresponds to a dust
mass density of $\rho _{hom}=1.63\times 10^{-23}\,{\rm gm/cm}{{^{3}}}$ in
the homogeneous case, equivalent to a gas density of $1000\,{\rm cm}{{^{-3}}}
$ with a dust to gas mass ratio of $0.007$, giving a homogeneous optical
depth of $\tau _{V}=1.67$ in the V-band.

\begin{figure}[tbp]
\plotone{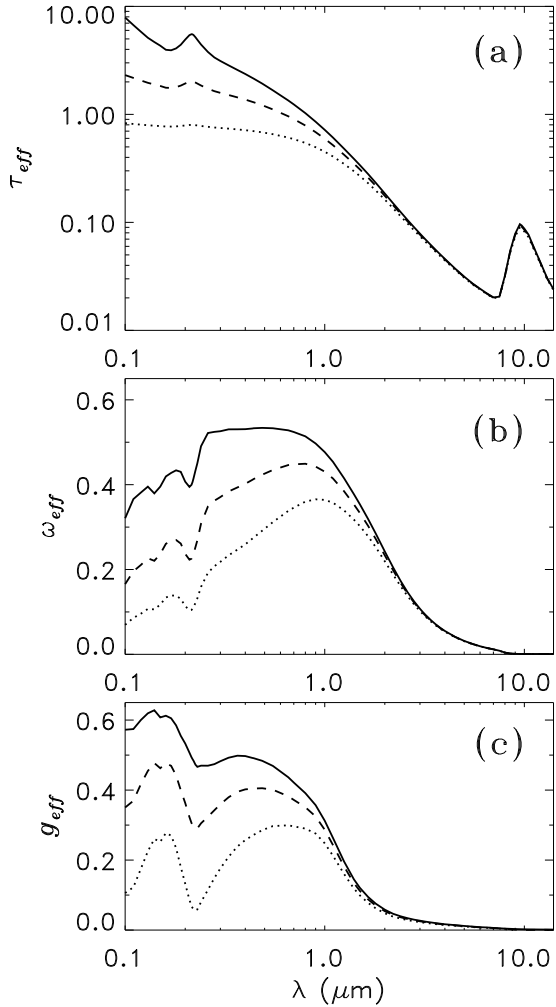}
\caption{ Comparison of radiative transfer properties of dust
distributed homogeneously or in clumps.
Panels show effective values of (a) optical depth, (b) albedo,
and (c) scattering asymmetry parameter, versus wavelength,
computed using the mega-grains approximation.
For comparison, the solid curves show the dust optical parameters for
the equivalent homogeneous medium.
The dashed curves show effective optical parameters for a
two-phase clumpy medium with filling factor $f_c=0.1$,
clump radii $r_c=0.05$pc (843 clumps),
and clump to ICM density ratio of $\alpha = \rho_c /\rho_{icm} = 10^2$.
The dotted curves show the case of
$f_c=0.01$, $r_c=0.01$pc (9918 clumps), and $\alpha = 10^4$.
See \S7 for more information. }
\label{tau_wavelen}
\end{figure}%

Before discussing the results of the MCRT and MGEP simulations, we
illustrate how the degree of clumpiness in the distribution of dust is as
important as the total dust mass and scattering albedo in affecting the
transfer of radiation through the medium. Figure~\ref{tau_wavelen} compares
as a function of wavelength, (a) the effective optical depth, $\tau
_{eff}(\lambda )$, (b) the effective albedo, $\omega _{eff}(\lambda )$, and
(c) the effective asymmetry parameter, $g_{eff}(\lambda )$, of a spherical
region of dust with different degrees of clumpiness, computed using the
mega-grains equations listed in \S \ref{List_EAP}, with $R_{S}=1$pc, $\rho
_{hom}=1.63\times 10^{-23}\,{\rm gm/cm}{{^{3}}}$, and 
\begin{eqnarray}
\kappa (\lambda ) &=&\kappa _{abs}(\lambda )+\kappa _{scat}(\lambda ) 
\nonumber \\
\omega _{d}(\lambda ) &=&\kappa _{scat}(\lambda )\,/\kappa (\lambda )\quad
\quad .
\end{eqnarray}
The solid curves in Figure~\ref{tau_wavelen} show the homogeneous case,
giving $\tau _{V}=1.67$. The dashed curves are for the case $f_{c}=0.1$ with 
$\alpha =\rho _{c}/\rho _{icm}=100$, and clump radii $r_{c}=0.05$pc, giving $%
\tau _{V}=1.11$. The dotted curves are for the extreme case of $f_{c}=0.01$
with $\alpha =10^{4}$, and $r_{c}=0.01$pc, giving $\tau _{V}=0.66$. It is
evident that the effective radiative transfer properties of the dusty medium
can be radically affected by the degree of clumpiness. The effect is
greatest at shorter wavelengths where the dust absorption and scattering
coefficient is sufficiently large to make the clumps opaque, causing $\tau
_{eff}(\lambda )$ to become almost flat and featureless. The small features
remaining are due to the ICM, but the opaque clumps dominate the effective
optical depth. An explanation for this ``gray'' behavior when $\tau
_{c}(\lambda )\gg 1$ is obtained by using eq.(\ref{tau_eff/tau_hom}): 
\begin{eqnarray}
\left| \frac{\partial \tau _{eff}}{\partial \lambda }\right|  &=&\frac{%
\partial \tau _{eff}}{\partial \tau _{hom}}\left| \frac{\partial \tau _{hom}%
}{\partial \lambda }\right|  \\
&\approx &\left( \frac{1}{\alpha f_{c}}\right) \left| \frac{\partial \tau
_{hom}}{\partial \lambda }\right| \ll \left| \frac{\partial \tau _{hom}}{%
\partial \lambda }\right| \,.  \nonumber
\end{eqnarray}
Thus, as $\alpha f_{c}\rightarrow \infty $ the variation of $\tau
_{eff}(\lambda )$ for the clumpy medium becomes a small fraction of the
variation of $\tau _{hom}(\lambda )$ in the equivalent homogeneous medium.
For longer wavelengths (infrared) there is no difference between clumpy and
homogeneous media because the clumps are optically thin.

To test the MGEP model using MCRT, we simulated a two-phase clumpy medium
with $f_{c}=0.1$, $\alpha =100$, and $r_{c}=0.05$pc, so that the effective
optical depths, albedos, and asymmetry parameters used in all the MGEP
calculations are given by the dashed curves in Figure~\ref{tau_wavelen}. The
MCRT computations always use the optical parameters given by the homogeneous
case (solid curves) in Figure~\ref{tau_wavelen}, simulating the details of
the clumpy medium on a 3D grid of $127^{3}$ voxels. The radiation source in
all cases is a black-body spectrum with $T_{s}=15000$K and $%
L_{s}=33000L_{\odot }$, so the spectrum of the emitted flux is 
\begin{equation}
S_{\lambda }=\,\frac{L_{s}}{\sigma T_{s}^{4}}\,B_{\lambda }(T_{s})\,\,.
\end{equation}
The transfer of radiation is computed by MCRT at 40 wavelengths from $%
\lambda _{\min }=0.1{\mu }{\rm m}$ to $\lambda _{\max }=14{\mu }{\rm m}$,
following $10^{7}$ photons at each wavelength. The standard three types of
source geometries are studied: uniformly distributed internal emission (U),
uniformly illuminating external source (X), and a central isotropic point
source (C).

\subsection{Absorbed Luminosities and the\protect\linebreak Distribution of
Dust Temperatures \label{Sec_Labs_Tdist}}

Since we model only non-ionizing photons experiencing coherent scattering,
for a given type of source the radiative transfer of a unit emitted flux can
be simulated at each wavelength separately, giving an escape and absorption
response function for the particular choice of source, geometry, and
clumpiness. Then we multiply the chosen source spectrum with the
escape/absorption response function to get the actual escaping SED and the
luminosity absorbed by each component of the dust. In the case of Monte
Carlo simulations (MCRT) we obtain the 3D spatial distribution of the
absorbed luminosity, whereas using the analytical approximations (MGEP)
gives the luminosity absorbed by all the clumps and luminosity absorbed by
the ICM. The equilibrium dust temperatures, $T_{d}$ for graphite or
silicate, are computed by equating the absorbed and emitted luminosities: 
\begin{equation}
\int_{\lambda _{\min }}^{\lambda _{\max }}S_{\lambda }P_{abs}(\lambda
)d\lambda \,=4\pi m_{d}\int_{0}^{\infty }\kappa _{\lambda }B_{\lambda
}(T_{d})d\lambda \,,  \label{Tdust_equilib}
\end{equation}
where $P_{abs}(\lambda )$ generically represents the fraction of flux at
wavelength $\lambda $ absorbed by graphite or silicate in a particular 3D
voxel of mass $m_{d}$ in the case of MCRT, or it is the fraction absorbed by
graphite or silicate in either the ICM or the clumps of mass $m_{d}$ in the
case of MGEP. The energy conservation eq.(\ref{Tdust_equilib}) is solved
iteratively for $T_{d}$ using Brent's method of finding the zeros of a
function \cite{NumRecip}, giving a 3D distribution of dust temperatures for
MCRT, or the average temperatures of the dust in clumps and the ICM when
using the MGEP model.

\begin{figure}[tbp]
\epsfxsize=7in \centerline{\vbox{\epsfbox{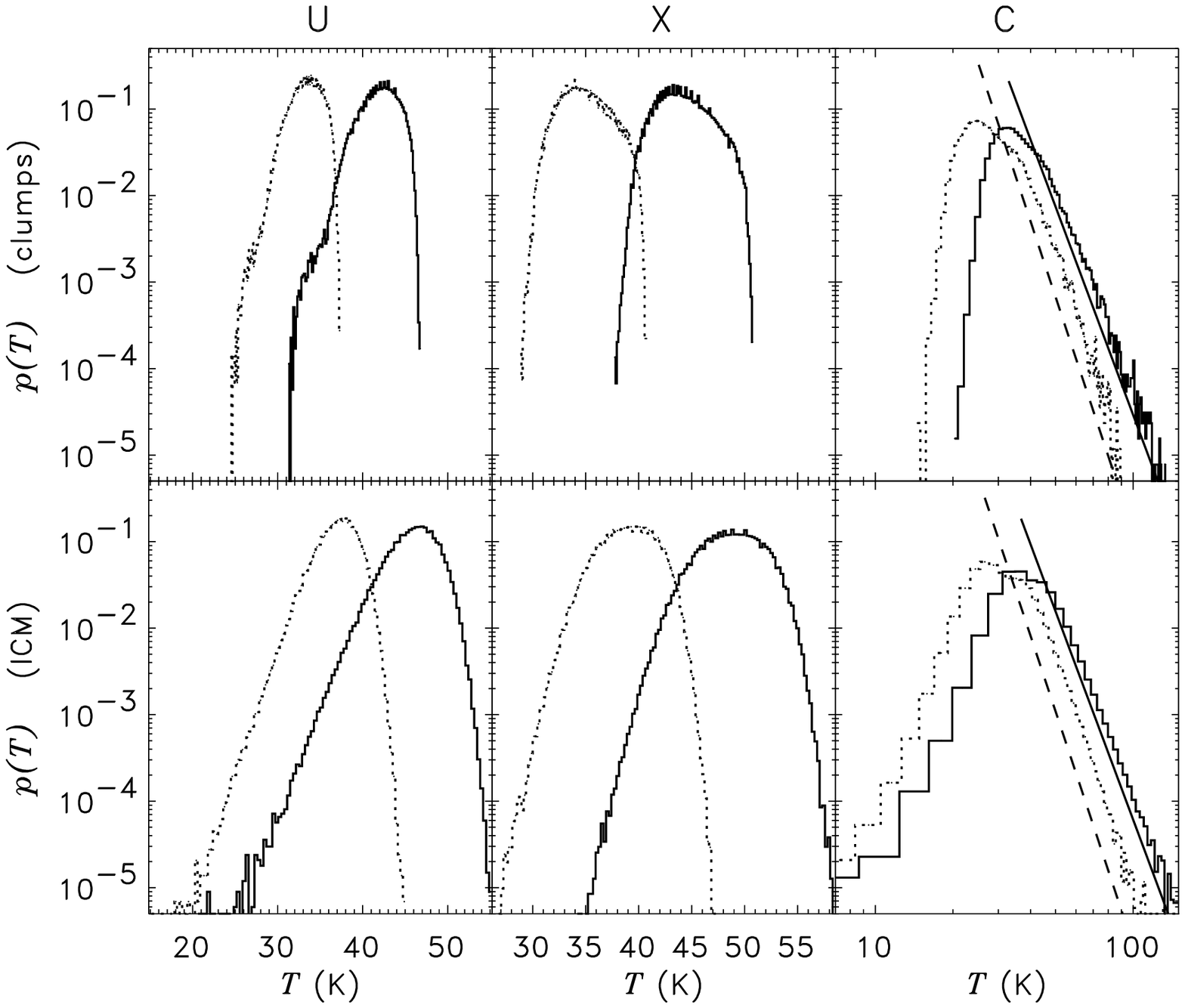}}}
\caption{ Probability distribution of equilibrium dust temperatures
from MCRT simulations for each type of source (indicated at top),
in a clumpy medium with $f_c=0.1$, $\alpha =100$, and $r_c=0.05$.
The top row represents the temperature distribution of dust in clumps
and the bottom row that for the ICM.
The solid lines are for graphite and
the dotted lines are for silicates.
See \S \ref{Sec_Labs_Tdist} for more information. }
\label{dust_T_dist}
\end{figure}

Figure~\ref{dust_T_dist} shows the distribution of equilibrium dust
temperatures computed by the MCRT model for clumps (top row) and in the ICM
(bottom row), for source types U, X, and C in each column from left to
right. The solid lines are the probability densities of graphite dust
temperatures and the dotted lines represent silicate dust. For the cases of
uniformly distributed internal and external sources the probability
densities are Gaussian on the high temperature side and of exponential form
on the low temperature side (note slightly different temperature scales). In
contrast, the central source creates a power law distribution of
temperatures (third column is a log-log plot) since dust next to the source
is heated much more than dust farther from the center. In fact, the straight
lines in Figure~\ref{dust_T_dist} are power laws: the solid line is $%
T_{d}^{-8}$ and appears to be parallel to the graphite temperature
distribution whereas $T_{d}^{-9}$ (dashed line) is parallel to the
temperature distribution of silicates.

\begin{deluxetable}{cllrrrr}
\tablewidth{0pt}
\tablecaption{Comparison of Dust Energetics and Temperatures}
\tablehead{\colhead{} & \colhead{} & \colhead{} &
           \multicolumn{2}{c}{$L_{abs}$ ($10^3 L\sun$)} &
	\multicolumn{2}{c}{$T_{dust}$ (K)} \\
           \colhead{source} & \colhead{dust} & \colhead{phase} &
           \colhead{MCRT} & \colhead{MGEP} & \colhead{MCRT} & \colhead{MGEP}}
\startdata
U & graphite  & ICM &      1.66 &     1.56 &     45.8 &     45.7 \nl
  &         & Clumps &    11.55 &    11.54 &     41.9 &     42.2 \nl
  & silicates & ICM &      0.44 &     0.42 &     37.0 &     37.0 \nl
  &         & Clumps &     2.75 &     2.74 &     33.4 &     33.6 \nl
& total & $L_{abs}/L_s=$ & 49\% &       49\% \nl
\nl
X & graphite  & ICM &      2.33 &     2.19 &     48.7 &     48.5 \nl
  &         & Clumps &    15.70 &    15.86 &     44.2 &     44.6 \nl
  & silicates & ICM &      0.62 &     0.58 &     39.0 &     39.0 \nl
  &         & Clumps &     3.66 &     3.69 &     34.9 &     35.3 \nl
& total & $L_{abs}/L_s=$ & 67\% &       67\% \nl
\nl
C & graphite  & ICM &      2.40 &     2.21 &     41.3 &     42.8 \nl
  &         & Clumps &    15.16 &    15.48 &     37.9 &     39.9 \nl
  & silicates & ICM &      0.64 &     0.59 &     32.4 &     34.8 \nl
  &         & Clumps &     3.57 &     3.63 &     29.3 &     31.9 \nl
& total & $L_{abs}/L_s=$ & 65\% &       66\% \nl
\enddata
\label{Tdist_tab}
\end{deluxetable}

Table~\ref{Tdist_tab} compares the luminosities absorbed by each component
(graphite or silicates) and phase (ICM or clumps) of the medium, as computed
by the MCRT and MGEP models, again for the cases of uniform internal,
uniform external, and central source. The MGEP results were obtained by
computing the escaping and absorbed SED using the mega-grains equations and
the escape/absorption probability formulae, combined with formulae for the
fractions absorbed by the ICM and clumps, given in \S \ref{AbsFrac}. Also
compared in Table~\ref{Tdist_tab} are the resulting dust temperatures:
averages of the distributions shown in Figure~\ref{dust_T_dist} for the case
of MCRT simulations, and in the case of the MGEP model when the source is
not a central point source, a single temperature for each component and
phase computed using eq.(\ref{Tdust_equilib}).

To compute the average temperature of the dust in the case of a central
source we use a theoretical power law distribution of dust temperatures as
described in Appendix \ref{App Tdist}, along with two assumptions: that the
absorbed luminosity decays with radial distance from the source like an
inverse power law with index $\eta =2.5$ ($\eta =2$ is normal for optically
thin case), and we use the maximum dust temperatures found by MCRT in the
voxel containing the source. Normally the maximum temperature would be the
dust sublimation temperature, however, the MCRT simulation is not set up to
fully resolve the volume next to the source so the sublimation temperature
is not reached in the simulation. Since we are comparing with MCRT results
we chose to use those maximum temperatures. We then solve for minimum
temperature of the power law distribution that matches the emitted and
absorbed luminosities, and this also yields the average dust temperature.

Table~\ref{Tdist_tab} shows that the MCRT and MGEP results are in close
agreement. The uniformly illuminating external source experiences more
absorption (67\%) than the uniformly distributed internal emission (49\%)
and therefore heats the dust to slightly higher temperatures. Approximately
the same fractions of luminosity are absorbed in the central and external
source cases, but the average dust temperature is lower in the central
source case since most of the dust is far from the source. In all cases the
clumps absorb about seven times more energy than the ICM, and this is in
part due to the fact that the clumps contain more than ten times the mass of
the ICM: 
\begin{equation}
\frac{M_{c}}{M_{icm}}\,=\,\frac{\rho _{c}\,f_{c}\,V}{\rho _{icm}(1-f_{c})V}%
\,=\,\frac{\alpha f_{c}}{1-f_{c}}\quad ,  \label{Mc/Micm}
\end{equation}
and this ratio is greater than ten for $f_{c}=0.1$ and $\alpha =100$.
However, the larger mass can easily radiate the absorbed energy in the IR so
that the temperature of dust in clumps is generally lower than in the ICM.
In \S \ref{MGEP_explore} we study in more detail the reasons for the lower
temperatures of dust in clumps.

\onecolumn

%

\begin{figure}[tbp]
\epsfxsize=7in \centerline{\vbox{\epsfbox{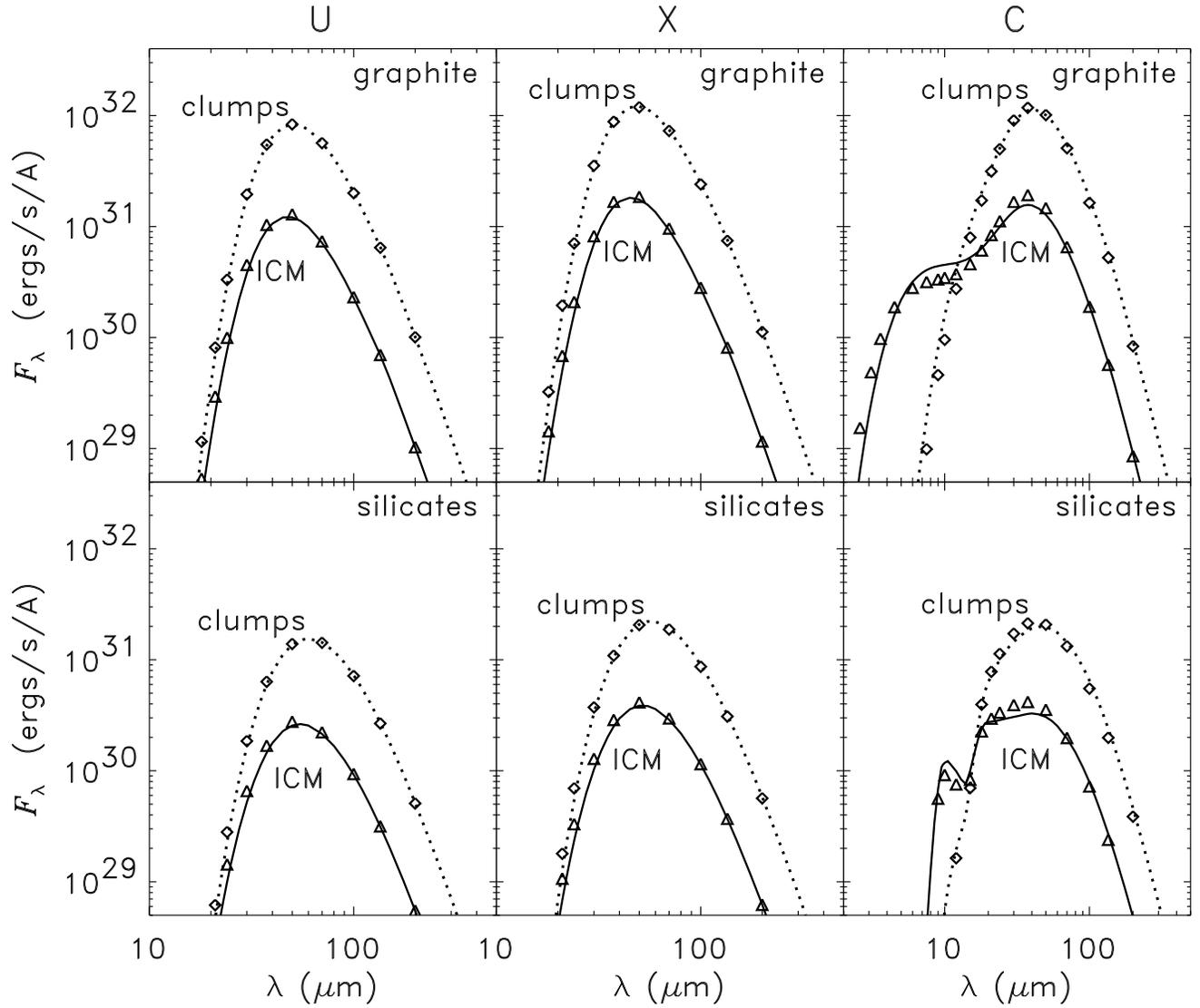}}}
\caption{ Comparison of dust emission spectra 
from MCRT and MGEP models for each type of source:
U, X, C, from left to right (indicated at top).
The clumpy medium has $f_c=0.1$, $\alpha =100$, and $r_c=0.05$.
The top row is for graphite,
bottom row is for silicates.
The triangles are for the MCRT model of dust in ICM and
the diamonds are for dust in clumps.
The solid lines are for the MGEP model of dust in ICM and
the dotted lines are for dust in clumps.
See \S \ref{Sec_UV-FIR_SED} for more information. }
\label{dust_spectrum}
\end{figure}
%

\twocolumn

\subsection{The Emerging UV-FIR SED \label{Sec_UV-FIR_SED}}

Computation of the dust temperatures also gives the IR emission from dust in
each voxel of the 3D simulation by the MCRT model, or from each phase of the
medium in the MGEP model: 
\begin{equation}
F_{\lambda }\,=\,4\pi \,m_{d}\,\kappa (\lambda )\,B_{\lambda }(T_{d})\,\,,
\label{IR_Flux}
\end{equation}
where $T_{d}$ is the temperature of dust in a voxel of mass $m_{d}$ in the
MCRT model, or where $T_{d}$ is the temperature of a dust in clumps/ICM of
mass $m_{d}$ in the MGEP model. Equation (\ref{IR_Flux}) is used by the MGEP
model only for uniformly distributed internal or external sources. In the
case of a central source with a power law distribution, $p(T_{d})$, of dust
temperatures, the MGEP model essentially computes 
\begin{equation}
F_{\lambda }\,=\,4\pi \,m_{d}\,\kappa (\lambda )\int_{T_{\min }}^{T_{\max
}}B_{\lambda }(T)\,p(T)\,dT\,,
\end{equation}
an approximation that is described in more detail in Appendix \ref{App Tdist}%
. Integrating the IR emission over the 3D volume in the case of the MCRT
approach, or just adding the IR emission from the clumps and ICM in the case
of the MGEP approach, gives the total IR emission spectrum. Figure~\ref
{dust_spectrum} shows the IR emission spectra from graphite (top row) and
silicates (lower row), with the diamonds and triangles representing emission
from clumps and the ICM, respectively, as computed by MCRT, and the solid
and dotted lines representing emission from the ICM and clumps as computed
by MGEP. The source types U, X, and C, are presented in columns from left to
right. The MGEP model is in close agreement with the MCRT results. The
emission from the clumps is in general greater than that from the ICM
because the absorbed luminosity is larger. An exception is the case of a
central source where the heating of dust adjacent to the source to much
higher temperatures causes more emission from the ICM at short IR
wavelengths.

\begin{figure}[tbp]
\plotone{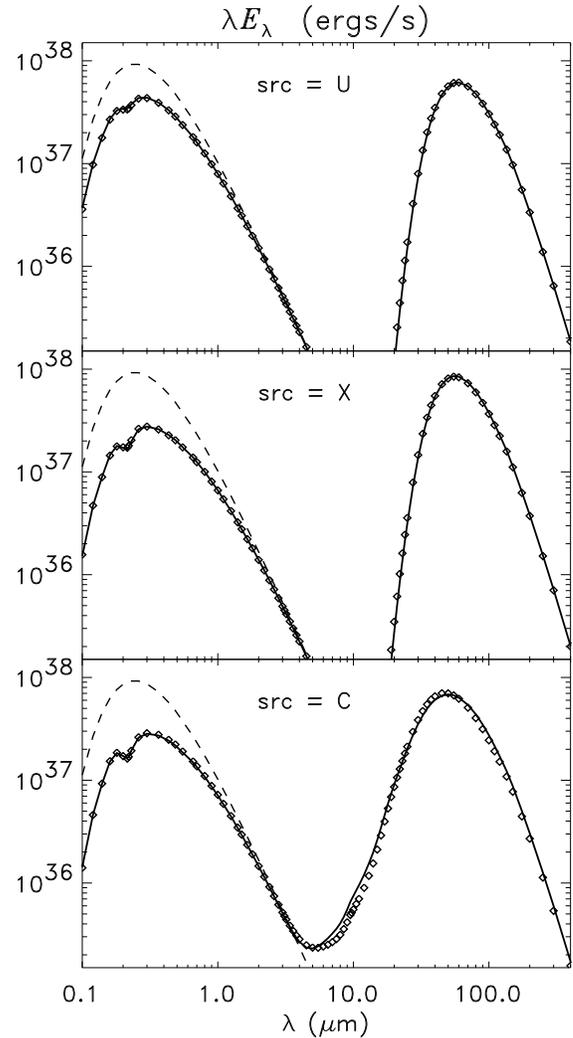}
\caption{ Comparison of the emerging UV-FIR SED resulting from
the MGEP model (solid lines) and the MCRT model (diamonds).
The dashed line is the SED of the source types U, X, and C,
from top to bottom rows. }
\label{UV-FIR_SED}
\end{figure}
%

The emerging SED is the sum of the escaping radiation and the IR\ emission
from heated dust: 
\begin{equation}
\lambda E_{\lambda }\,=\,\lambda S_{\lambda }P_{esc}(\lambda )\,+\,\lambda
F_{\lambda }\quad ,
\end{equation}
where $P_{esc}(\lambda )$ generically designates an escape probability that
is computed either numerically (MCRT) or analytically (MGEP) and depends on
the model and source type. Figure~\ref{UV-FIR_SED} compares the SED computed
by the MCRT and MGEP models for each type of source. The diamond symbols are
the MCRT results, the solid lines are the MGEP theory, and the dashed line
is the original source spectrum $\lambda S_{\lambda }$. The source types U,
X, and C, are presented in panels from top to bottom. The SED at short
wavelengths is the source radiation that escapes from the clumpy dusty
environment, whereas the SED at long wavelengths is emission from dust
heated by the absorbed radiation. The agreement between the MGEP and MCRT
models is excellent. The SEDs for the uniform internal and external sources
have very similar appearances, except that in the case of an external source
more of the source energy is absorbed (68\% versus 49\%). The central and
external source cases have the same fraction of energy absorbed by the dust,
so that the short wavelength portion of the SEDs are the same. However, in
the case of a central source the emission from hot dust near the source
increases the SED around 10$\mu $m. In all cases, the relative paucity of
emission in the $\sim 3-20\mu $m region is caused by the omission of
stochastic heating in the calculations.

\subsection{Exploring Parameter Space with the\protect\linebreak MGEP Model 
\label{MGEP_explore}}

\begin{figure}[tbp]
\plotone{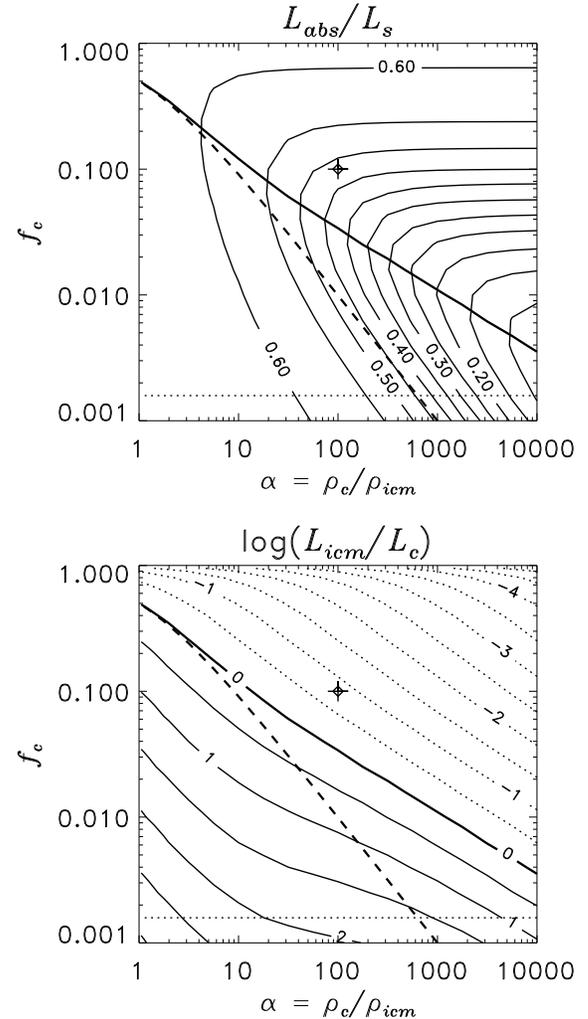}
\caption{ Results of the MGEP model over a wide range of clumpy medium
parameters $(\alpha,f_{c})$ for a uniformly distributed black-body source 
having $T_s=15,000$K and $L_s=33,000 L_{\odot}$.
(a) Upper panel shows absorbed luminosity fraction;
(b) and the lower panel shows the ratio of ICM/clump luminosities.
The diamond and cross marks the parameters for which
detailed comparisons with MCRT were presented.
In both panels the dashed line is the locus of parameters where $M_{icm}=M_c$,
and the thick solid line is the locus of parameters where $L_{icm}=L_c$.
More details are given in \S \ref{MGEP_explore} of the text. }
\label{MGEP_Lum}
\end{figure}
%

\begin{figure}[tbp]
\plotone{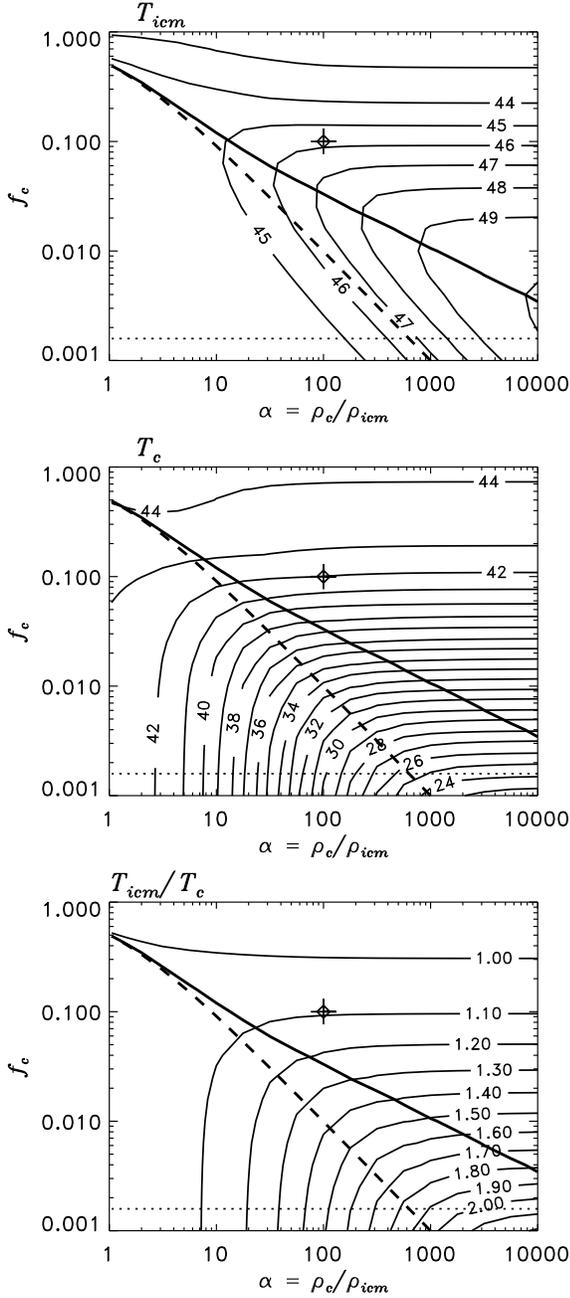}
\caption{ Temperatures of graphite dust heated by uniformly distributed sources,
as computed by
the MGEP model over a wide range of clumpy medium
parameters $(\alpha,f_{c})$. The top, middle, bottom panels show temperatures
in the ICM, the clumps, and the ratio, respectively.
The diamond and cross indicate results that were compared with MCRT. }
\label{MGEP_Tdust}
\end{figure}

Since the MGEP equations are computationally fast, we can very easily model
the escape and absorption of radiation over a wide range of parameters
characterizing the clumpy medium or the radiation source. Figures~\ref
{MGEP_Lum} and \ref{MGEP_Tdust} show the results of MGEP model calculations
over the range of clump filling factors, $0.001<f_{c}<1$, and density
ratios, $1<\alpha <10^{4}$, with clump radii $r_{c}=0.05$. The dust
composition and total mass is the same as that used in the above MCRT to
MGEP comparisons. The sources of radiation are isotropic and uniformly
distributed within the sphere, again having a black-body spectrum with
temperature $T_{s}=15,000$K and total luminosity $L_{s}=33,000L_{\odot }$.
In all the panels the diamond and cross marks the point $(\alpha
,f_{c})=(100,\,0.1)$ for which we presented detailed comparisons with MCRT,
and the horizontal dotted lines indicate the limiting value of $f_{c}$
(=0.0015) below which there are less than 10 clumps in the medium,
therefore, only filling factors above the dotted line are considered to be
statistically reliable.

Figure~\ref{MGEP_Lum} shows ({\sl upper panel}) the contours of the fraction
of the source luminosity that is absorbed by the clumpy medium ($%
L_{abs}/L_{s}$), and ({\sl lower panel}) the contours of the logarithm of
the ratio of luminosity absorbed by the ICM to that absorbed by clumps [$%
\log (L_{icm}/L_{c})$]. The dashed line (in all panels), extending from $%
(\alpha ,f_{c})=(1,\,0.5)$ to $(1000,\,0.001)$, is the curve 
\begin{equation}
f_{M}(\alpha )=\frac{1}{\alpha +1}\,\,,  \label{f_Mc=Micm}
\end{equation}
giving the locus parameters for which the ICM and clumps have equal mass,
that is, $M_{icm}=M_{c}$. The equation for $f_{M}(\alpha )$ is obtained from
eq.(\ref{Mc/Micm}), which we restate here as 
\begin{equation}
M_{c}\,=\,\left( \frac{\alpha f_{c}}{1-f_{c}}\right) M_{icm}\,\,.
\label{Mc_Micm}
\end{equation}
Note that if $f_{c}>$ $f_{M}(\alpha )$ then $M_{c}\,>M_{icm}$. The thick
solid line in all panels represents the locus of parameters $f_{L}(\alpha )$
for which the ICM and clumps absorb the same amount of energy, that is, $%
L_{icm}=L_{c}$. The dotted contours in the lower panel of Figure~\ref
{MGEP_Lum} indicate when the ICM absorbs less energy than the clumps. The
solid contours in the lower panel of Figure~\ref{MGEP_Lum} indicate that if $%
f_{c}<f_{L}(\alpha )$ then $L_{c}\,<L_{icm}$. Combining this with the fact
that $M_{c}\,>M_{icm}$ when $f_{c}>$ $f_{M}(\alpha )$ we have that for the
parameter region between the thick solid and dashed curves 
\begin{equation}
f_{M}(\alpha )<f_{c}<f_{L}(\alpha )\,\Rightarrow \left\{ 
\begin{array}{l}
M_{c}>M_{icm} \\ 
L_{c}<L_{icm}
\end{array}
\right\} \,,  \label{Tc<Ticm}
\end{equation}
which guarantees that the temperature of dust in clumps will be less than
that in the ICM since, compared to the ICM, a smaller amount of energy is
absorbed by a larger number of dust particles, which can radiate this energy
at a lower temperature.

Figure~\ref{MGEP_Lum} ({\sl upper panel})
also shows that for a given density contrast $\alpha _{0}$,
the minimum absorbed luminosity ($L_{abs}$) corresponds to the
filling factor $f_{c}=$ $f_{L}(\alpha _{0})$ for which $L_{c}=L_{icm}$. When 
$\alpha \gg 1$, the condition $L_{c}=L_{icm}$ is achieved when there is
approximate equality between the effective optical depth of the clumps and
that of the ICM, that is, $\tau _{mg}(\lambda _{p})\approx \tau
_{icm}(\lambda _{p})$, where $\lambda _{p}$ is the wavelength where most of
the source energy is absorbed. By equation (\ref{tau_eff_min_f}), the value
of $f_{c}$ for which $\tau _{mg}\approx \tau _{icm}$ is nearly the same
value for which $\tau _{eff}$ attains a minimum as a function of $f_{c}$
(see also Figures~\ref{mg_ff_min}, \ref{pabs_t2} and \ref{pabs_t10}).
Consequently, as $\alpha \rightarrow \infty $, the steepest descent of $%
L_{abs}$ is along the curve $f_{L}(\alpha )$. A detailed derivation of this
result and the following equations would require studying $\tau _{mg}$ and $%
\tau _{icm}$ as given by eqs.(\ref{tau_mg_scat}) and (\ref{tau_icm_scat}),
respectively, but here we assume that the effects of scattering are of
second order importance.

Let us study the curve $f_{L}(\alpha )$ in more detail. As $\alpha
\rightarrow 1$ the condition $L_{c}=L_{icm}$ is achieved by having $%
M_{c}=M_{icm}$, and so $f_{L}(\alpha )\approx f_{M}(\alpha )$ when $\alpha
<5,$ as shown by the merging of the dashed and thick curves in in Figure~\ref
{MGEP_Lum}. However, as $\alpha \rightarrow \infty $ we see that $%
f_{L}(\alpha )\gg f_{M}(\alpha )$. This behavior is due to the change in the
effective optical depth of the clumps, $\tau _{mg}$, since the clumps become
optically thick as $\alpha \rightarrow \infty $, so we derive an equation
for $f_{L}(\alpha )$ as follows. Starting with $\tau _{mg}(\lambda
_{p})=\tau _{icm}(\lambda _{p})$ and applying the mega-grains equations from
\S \ref{List_EAP} we obtain 
\begin{equation}
\frac{3f_{c}}{4r_{c}(1-f_{c})^{\gamma }}=\frac{\kappa (\lambda _{p})\rho
_{hom}}{(\alpha -1)\,f_{c}+1}\,\,,  \label{Lmg=Licm}
\end{equation}
assuming optically thick clumps. Since we propose that eq.(\ref{Lmg=Licm})
occurs when $L_{c}=L_{icm}$ we now use the symbol $f_{L}$ in place of $f_{c}$
for the solution. Then as $\alpha \rightarrow \infty $ we know that $%
f_{L}\ll 1$ but $\alpha f_{L}\gg 1$, so eq.(\ref{Lmg=Licm}) becomes 
\begin{equation}
\frac{3f_{L}}{4r_{c}}\approx \frac{\kappa (\lambda _{p})\rho _{hom}}{\alpha
f_{L}}\,\,.
\end{equation}
Solving for $f_{L}$ gives 
\begin{equation}
f_{L}(\alpha )\approx \sqrt{\frac{4r_{c}\kappa (\lambda _{p})\rho _{hom}}{%
3\alpha }}  \label{f_Lc=Licm}
\end{equation}
when $\alpha $ is large. Using $\kappa (\lambda _{p})\rho _{hom}=1.7$, which
occurs when $\lambda _{p}\sim 0.55\mu $m, gives $f_{L}(\alpha )\approx \frac{%
1}{3}\alpha ^{-\frac{1}{2}}$ which produces a perfect fit to the zero
contour of $\log (L_{icm}/L_{c})$ in Figure~\ref{MGEP_Lum} ({\sl lower panel})
when $\alpha>30 $ (not shown). The transition from the behavior
$f_{L}(\alpha)\approx f_{M}(\alpha )$ to the behavior
$f_{L}(\alpha )\propto \alpha ^{-\frac{1}{2}} $
occurs naturally around values of $\alpha $ for which eq.(\ref{f_Lc=Licm})
is greater than $f_{M}(\alpha )$, indicating that the clumps
are becoming optically thick. We find that the same value of $\kappa
(\lambda _{p})\rho _{hom}=1.7$ in eq.(\ref{f_Lc=Licm}) gives perfect fits to
the large $\alpha $ behavior of $f_{L}(\alpha )$ for all $0.1\leq r_{c}\leq
0.8$ (not shown).

%

Figure~\ref{MGEP_Tdust} shows contours of the temperature of graphite dust
in the ICM ($T_{icm}$; {\sl top panel}), in the clumps ($T_{c}$;{\sl \
middle panel}), and the temperature ratio, $T_{icm}/T_{c}$ ({\sl bottom panel%
}). The thick solid lines are the curves $f_{L}(\alpha )$ for which $%
L_{c}=L_{icm}$ and the dashed lines are the curves $f_{M}(\alpha )$ for
which $M_{c}=M_{icm}$. The contours of $T_{icm}$ wrap around $f_{L}(\alpha )$%
, whereas the contours of $T_{c}$ wrap around $f_{M}(\alpha )$. As $\alpha
\rightarrow \infty $ and $f_{c}\rightarrow 0$, we find that $T_{c}$
decreases whereas $T_{icm}$ increases slightly. From the relation $L\propto
MT^{4+\beta }$, where $\beta $ is the emissivity index, we get that 
\begin{equation}
\frac{T_{icm}}{T_{c}}=\left[ \left( \frac{M_{c}}{M_{icm}}\right) \left( 
\frac{L_{icm}}{L_{c}}\right) \right] ^{\frac{1}{4+\beta }}\,,
\label{Ticm/Tc}
\end{equation}
and with the conditions from eq.(\ref{Tc<Ticm}), this equation explains why
the situation of $f_{M}(\alpha )<f_{c}<f_{L}(\alpha )$ leads to $%
T_{c}<T_{icm}$. We can also explain why the temperature ratio must increase
as $\alpha \rightarrow \infty $. Substituting eq.(\ref{f_Lc=Licm}) into eq.(%
\ref{Mc_Micm}) gives 
\begin{equation}
\frac{M_{c}}{M_{icm}}\approx C\sqrt{\alpha }
\end{equation}
along the $f_{L}(\alpha )$ curve, where $L_{c}=L_{icm}$ and where $C$ is a
constant. Using this in eq.(\ref{Ticm/Tc}) we get that along the $%
f_{L}(\alpha )$ curve 
\begin{equation}
\frac{T_{icm}}{T_{c}}\approx \left( C\sqrt{\alpha }\right) ^{\frac{1}{%
4+\beta }}\,\,.
\end{equation}
Therefore $T_{icm}/T_{c}$ must increase as $\alpha \rightarrow \infty $. The
net result is that $T_{c}$ decreases faster than $T_{icm}$ increases because 
$M=M_{c}+M_{icm}$ is a constant whereas $L_{abs}=L_{c}+L_{icm}=2L_{icm}$ is
decreasing as $\alpha \rightarrow \infty $ and $f_{c}=f_{L}(\alpha )$. Note
that the spectrum of the emitted radiation is dominated by dust at
temperature $T_{c}$ if $f_{c}>f_{L}(\alpha )$, since then $L_{c}>L_{icm}$,
or it is dominated by dust at temperature $T_{icm}$ if $f_{c}<f_{L}(\alpha )$%
.

The temperature of silicates also exhibits the same variations but with
lower values since silicates absorb only about 25\% of what graphites absorb
for the chosen composition. The same type of pattern, shown here for
specific values of $r_{c}$, $T_{s}$, $L_{s}$, and source type, is found for
other values of $r_{c}$, $T_{s}$, $L_{s}$, and the central point source or
uniformly illuminating external source types. The dust temperatures are
lower as $L_{s}$ is decreased, and there is less variation in $T_{icm}/T_{c}$
as $T_{s}$ or $r_{c}$ is decreased. Cases of $T_{s}>20,000$K would require
modeling the ionization of gas and the heating of dust by absorption of Ly$%
\alpha $ photons.

\section{SUMMARY AND CONCLUSIONS}

We have introduced an analytical model for the escape and absorption of
radiation in two-phase clumpy media, based on combining the mega-grains
approximation of HP93 with escape/absorption probability formulae for
homogeneous media, as summarized in \S \ref{List_EAP}, and referred to by
the acronym MGEP. Enhancements of the mega-grains approximation developed in
this paper include: (1) improved and extended the formula for effective
optical depth to all clump filling factors, (2) improved the formula for the
effective albedo, (3) developed a new approximation for the effective
scattering asymmetry parameter, and (4) developed new approximations for the
fractions absorbed by each phase of the medium. We also developed a new
approximation for the fraction of photons from a uniformly illuminating
external source that are absorbed in a sphere of dust when scattering occurs.

The space of parameters describing a two-phase clumpy dusty environment is
six-dimensional: the clump filling factor, the clump to interclump medium
(ICM) density ratio, and the clump radii, together characterize the
morphology of the medium, and the interaction (absorption plus scattering)
coefficient, scattering albedo and asymmetry parameter are the three optical
parameters of the dust that vary with wavelength. The analytic MGEP model
was compared with Monte Carlo simulations of radiative transfer (MCRT) over
a subspace of the six-dimensional parameter space, and we found good
agreement for most of the parameter values checked. More importantly, the
qualitative behavior of the MGEP model agrees very well with MCRT
simulations, giving us confidence in applying the MGEP model to parameter
ranges that have not yet been checked by MCRT simulations. Three types of
source distributions were studied: the central point source (C), uniformly
distributed internal sources (U), and uniform illumination by external
sources (X). Source type U is the least absorbed, whereas source type X is
the most absorbed at low optical depths and source type C is the most
absorbed at high optical depths.

The MGEP model was shown to predict very well the emerging SED in a
realistic simulation of starlike sources heating a clumpy dusty medium,
as compared to MCRT simulations. Furthermore, the MGEP method requires just
seconds to compute a full spectrum simulation, in comparison to hours of
computation when using the MCRT method.

From MGEP simulations over a wide range parameters characterizing the clumpy
medium we find that for constant clump to ICM density ratio, $\alpha $, the
total luminosity absorbed by the clumpy medium attains a minimum at the
filling factor, $f_{c}$, for which the luminosity absorbed by clumps ($L_{c}$%
) and the ICM ($L_{icm}$) are equal. The curve of $f_{c}$ versus $\alpha $
for which $L_{c}=L_{icm}$ is found to be proportional to $\alpha ^{-\frac{1}{%
2}}$, a consequence of the clumps becoming optically thick, whereas the
curve of $f_{c}$ for which the clumps and ICM have equal mass is
proportional to $\alpha ^{-1}$, and these diverging behaviors cause the
temperature of dust in clumps to decrease as $\alpha \rightarrow \infty $
and $f_{c}\rightarrow 0$. Physically, the dust in opaque clumps shields
itself from radiation, thus reaching a lower equilibrium temperature than
dust in the ICM. The extra parameters gained by introducing clumpiness
allows for modeling more unusual relationships between the luminosity
absorbed by dust and the resulting dust temperatures.

\acknowledgements

We would like to acknowledge useful comments and suggestions given by the
referee, Mike Hobson, and by Michel Fioc. This research was supported by the
NASA Astrophysical Theory Program NRA97-12-OSS-098.

\appendix 

\section{FINAL FLUX APPORTIONMENT IN MONTE CARLO\label{App FinF}}

Let $N_{0}$ be the number of photons emitted as a group at the start of the
Monte Carlo simulation and suppose after $k$ iterations of traveling,
escape, absorption and scattering there are $N_{k}$ photons remaining in the
medium. Since after each scattering the flux weight of each photon is
reduced by the albedo $0<\omega <1$, the actual flux remaining after $k$
scatterings is $\omega ^{k}N_{k}$. For clarity in the following derivation
define the flux weight factor 
\begin{equation}
W_{k}\,\equiv \,\omega ^{k}\,\,.
\end{equation}
After one more iteration of traveling there are $N_{k+1}$ photons remaining,
each still having weight $W_{k}$, so that the flux escaping during the $k+1$
iteration is $E_{k+1}=W_{k}\,(N_{k}-\,N_{k+1})$. The remaining photons
interact with the medium and are apportioned into absorbed and scattered
fractions: the absorbed flux is $A_{k+1}=(1-\omega )\,W_{k}\,N_{k+1}$, and
the flux remaining after scattering is $\omega
\,W_{k}\,N_{k+1}=W_{k+1}N_{k+1}$. Carrying this analysis forward by
induction gives the general formula for the escaping flux after each
iteration: 
\begin{eqnarray}
E_{k+1} &=&W_{k}(\,N_{k}-\,N_{k+1})  \nonumber \\
E_{k+2} &=&W_{k+1}(N_{k+1}-N_{k+2})  \nonumber \\
&&\vdots  \nonumber \\
E_{k+n} &=&\omega ^{k+n-1}(N_{k+n-1}-N_{k+n})\,;  \label{MC_esc_flux}
\end{eqnarray}
and the absorbed flux after each iteration: 
\begin{eqnarray}
A_{k+1} &=&(1-\omega )\,W_{k}\,N_{k+1}  \nonumber \\
A_{k+2} &=&(1-\omega )\,W_{k+1}N_{k+2}  \nonumber \\
&&\vdots  \nonumber \\
A_{k+n} &=&(1-\omega )\omega ^{k+n-1}N_{k+n}\,.  \label{MC_abs_flux}
\end{eqnarray}
We seek expressions for the infinite sums of $E_{k+n}$ and $A_{k+n}$ over $%
n>0$ in order to terminate the Monte Carlo simulation at the $k$-th
iteration and still have an accurate estimate of the escaping flux. Assume
that for all $n>0$ 
\begin{equation}
\frac{N_{k+n}}{N_{k+n-1}}=\beta \quad ,  \label{remain_ratio}
\end{equation}
so that after $k$ iterations the fraction of photons remaining after each
additional iteration is assumed to have reached a steady state. Then we have 
\begin{eqnarray}
N_{k+n} &=&\beta ^{n}N_{k}\,\,, \\
N_{k+n-1}-N_{k+n} &=&(1-\beta )\beta ^{n-1}N_{k}\,\,.
\end{eqnarray}
The sum of the escaping flux is then: 
\begin{eqnarray}
\sum\limits_{n=1}^{\infty }E_{k+n} &=&\omega ^{k}\sum\limits_{n=1}^{\infty
}\omega ^{n-1}(N_{k+n-1}-N_{k+n})  \nonumber \\
&=&\omega ^{k}N_{k}(1-\beta )\sum\limits_{n=1}^{\infty }(\omega \beta )^{n-1}
\nonumber \\
&=&\omega ^{k}N_{k}(1-\beta )/(1-\omega \beta )\,.
\end{eqnarray}
Since $\omega ^{k}N_{k}$ is the known amount of flux remaining after $k$
iterations, the fraction that would eventually escape if the iterations were
continued is then 
\begin{equation}
f_{k}^{\,esc}=\frac{1-\beta }{1-\omega \beta }\quad ,  \label{fin_esc_frac}
\end{equation}
assuming that eq.(\ref{remain_ratio}) is true. Similarly for the sum of the
absorbed flux 
\begin{eqnarray}
\sum\limits_{n=1}^{\infty }A_{k+n} &=& (1-\omega )
\omega^{k}\sum\limits_{n=1}^{\infty}\omega ^{n-1} N_{k+n}  \nonumber \\
&=&\omega ^{k}N_{k}(1-\omega )\sum\limits_{n=1}^{\infty }\omega ^{n-1}\beta
^{n}  \nonumber \\
&=&\omega ^{k}N_{k}(1-\omega )\beta /(1-\omega \beta )\,,
\end{eqnarray}
and it is clear that 
\begin{equation}
f_{k}^{\,esc}\,+\,f_{k}^{\,abs}\,\,=\,\,\frac{1-\beta }{1-\omega \beta }\,+\,%
\frac{(1-\omega )\beta }{1-\omega \beta }\,\,=\,\,1\,,
\end{equation}
giving a check of the derivations. Note that the simpler assumption of $%
f_{k}^{\,esc}=\omega $ and $f_{k}^{\,abs}=1-\omega $ corresponds to the case
of $\beta =1/(1+\omega )$, which incorporates no knowledge of the particular
radiative transfer situation and could not possible work for all types of
sources and geometries. In contrast, the final iteration escaping fraction
given by eq.(\ref{fin_esc_frac}) is based on a partial history of absorption
and scattering and therefore can better estimate the true value.

To examine the effects of relaxing the assumption of eq.(\ref{remain_ratio}%
), define the remaining fraction at the $k$-th iteration: 
\begin{equation}
\beta _{k}\,\equiv \,\frac{N_{k}}{N_{k-1}}\quad ,
\end{equation}
the ratio of the number of photons remaining after the $k$-th iteration over
the number of photons in the medium before the $k$-th iteration. In our
experience with Monte Carlo simulations we find that the sequence $\beta _{k}
$ ($k=1,2,3,...)$ is monotonically either increasing or decreasing (does not
oscillate), and the direction of convergence depends on the source geometry
and the value of $g$, the phase function asymmetry parameter. In our Monte
Carlo simulations with uniformly distributed internal emitters, the sequence 
$\beta _{k}$ converges rapidly, and using $\beta _{k}$ in place of $\beta $
in eq.(\ref{fin_esc_frac}) gives an $f_{k}^{\,esc}$ that predicts the total
escaping flux quite well. The remaining fraction $\beta _{k}$ is found to be
an increasing sequence when $g<g^{*}(\tau )$, where $g^{*}(\tau )$ is
approximately given by eq.(\ref{g_tau}), since photons tend to get trapped
in the medium when scattering is more isotropic. When $g>g^{*}(\tau )$ then $%
\beta _{k}$ is a decreasing sequence, since photons tend to escape after
more scatterings if the angular scattering distribution is on average more
forward. When $g=g^{*}(\tau )$ we find that the sequence $\beta _{k}$ is
essentially constant. This dependence on the asymmetry parameter $g$ is
related to the validity of the escape probability formula for scattering
discussed in \S \ref{EPSS} and further analyzed in Appendix C2. The behavior
of the sequence $\beta _{k}$ is essentially the same for a uniformly
illuminating external source of photons. In the extreme case of a single
central source, it takes more scatterings for $\beta _{k}$ to converge and
it is always a decreasing function of $k.$ Thus $f_{k}^{\,esc}$ always
slightly overestimates the escaping fraction for the case of a central
source, but never underestimates it. When the scattering is more forward the
rate of decreasing $\beta _{k}$ is faster during the first few scatterings,
since in this case the likelihood that a photon will escape increases after
each scattering more than when the scattering is isotropic.

The discovery of the above equations and behavior was facilitated by the use
of dynamic arrays to represent a group of photons, a standard feature of the
Interactive Data Language (IDL), a product of Research Systems Incorporated
(RSI).

\section{RANDOM CLUMPS IN A TWO PHASE MEDIUM\label{App Nc}}

In section \ref{S.Clumpy Media} we introduced equation (\ref{Nclump}) for
the number of identical random clumps needed to produce a given volume
filling factor in a two-phase medium. Here we give a more detailed
derivation of that equation. Let $X$ be a randomly chosen point in the
medium. Then the probability that a randomly placed clump will contain the
point $X$ is simply 
\begin{equation}
p\equiv \frac{\upsilon _{c}}{V}
\end{equation}
where $\upsilon _{c}$ is the volume of a clump, and $V$ is the total volume
of the medium. Continue to randomly place more clumps in the medium, without
regard to overlaps, for a total of $N_{c}$ clumps. Then the probability, $%
P(n),$ that the point $X$ will be contained in $n$ clumps is given by the
binomial distribution: 
\begin{equation}
P(n)\,=\,\binom{N_{c}}{n}\,p^{n}\,(1-p)^{N_{c}-n}\quad .  \label{Binomial}
\end{equation}
In particular, the probability that $X$ is not in any clump is 
\begin{equation}
P(0)\,=\,\,(1-p)^{N_{c}}
\end{equation}
and so the fraction of the volume $V$ occupied by the clumps is 
\begin{equation}
f_{c}\,=\,1-P(0)\,=\,1\,-\,\,(1-p)^{N_{c}}\quad .
\end{equation}
Solving for $N_{c}$ gives the desired equation 
\begin{equation}
N_{c}\,=\,\,\frac{\ln (1-f_{c})}{\ln (1-p)}
\end{equation}
for the number of identical randomly placed clumps that will have a total
filling factor of $f_{c},$ when each clump has a filling factor $p.$

The expectation value for $n$ of the binomial distribution [eq.(\ref
{Binomial})] is 
\begin{equation}
E(n)=\sum_{n=0}^{N_{c}}nP(n)=N_{c}\,p=\frac{N_{c}\,\upsilon _{c}}{V}=Q_{c}\,,
\end{equation}
equaling the porosity of the clumps, $\,Q_{c},$ which was introduced for the
mega-grains approximation. As $p\rightarrow 0$ and $N_{c}\rightarrow \infty $
the binomial distribution is well approximated by the Poisson distribution
having the same expectation (e.g. \cite{bev92}): 
\begin{equation}
P(n)\,\,\simeq \,\,\frac{Q_{c}^{n}\,e^{-Q_{c}}}{n!}\quad .
\end{equation}
For $n=0$ this approximation gives 
\begin{equation}
f_{c}\,=\,1-P(0)\,\,\simeq \,\,1\,-\,e^{-Q_{c}}\quad ,
\end{equation}
providing another equation relating porosity and filling factor, which is
exact for our purposes, since $p<10^{-3}$ (e.g. when $r_{c}<0.1\,R_{S}$ )
and $N_{c}\gg 1$. When $f_{c}\ll 1$ then of course $f_{c}\,\simeq \,Q_{c}$.

Another quantity of interest is the average number of clumps encountered
along a randomly chosen line of sight, also called the covering factor, $%
F_{c}$. Consider a cylinder of radius $r_{c}$ and length $L$ centered on the
line of sight. The volume of the intersection of the cylinder with the
random clumps is $\pi r_{c}^{2}Lf_{c}$ and we can estimate the number of
clumps in the intersection by dividing by the volume of a clump: 
\begin{equation}
F_{c}\approx \frac{\pi r_{c}^{2}Lf_{c}}{\frac{4}{3}\pi r_{c}^{3}}=L\left( 
\frac{3f_{c}}{4r_{c}}\right) \,\,.
\end{equation}
This is actually more like a lower bound estimate since the clumps may
overlap. An upper bound is obtained by using the porosity in place of the
filling factor: 
\begin{equation}
F_{c}<L\left( \frac{3Q_{c}}{4r_{c}}\right) \,\,.
\end{equation}

\section{ESCAPE AND INTERACTION PROBABILITIES FOR A SPHERE\label{App Eprob}}

\subsection{No Scattering}

Here we derive equations (\ref{Pinteract}) and (\ref{Oster_EP}), which are
exact formulas for the interaction and escape probabilities of external and
internal sources, respectively, in a homogeneous sphere of dust, ignoring
the effects of scattering. Thus, scattering and absorption are considered
together as interactions causing extinction of photons along a line of
sight, and by extinction escape probability we mean the probability of
escaping without being scattered or absorbed. Consider a ray intersecting
the sphere at an angle $\theta $ with respect to the surface normal vector,
which we call the impact angle. The length of the chord created by the
intersection is $2R\cos \theta ,$ where $R$ is the radius of the sphere.
Defining $\tau \equiv \rho \kappa R$, the optical radius, where $\kappa $ is
the dust interaction coefficient and $\rho $ is the density of dust, the
extinction optical depth of the chord is $2\tau \cos \theta$, and this will
be used in the derivations.

To derive the interaction probability, eq.(\ref{Pinteract}), for the case of
a uniformly illuminating external source, consider a beam of photons
traveling in parallel rays impacting a hemisphere. Upon computing the
transmission of the beam we can get the interacting fraction of photons, and
then by symmetry this gives the interaction probability for all possible
beams impacting the sphere, i.e. a uniformly illuminating external source.
Let $I_{0}$ be the intensity of each parallel ray in the impacting beam.
Then the intensity emerging from the other side is reduced by the
extinction, which depends on the impact angle as follows: 
\begin{equation}
I_{out}(\tau ,\theta )\,=\,I_{0}\,\exp (-2\tau \cos \theta )\quad .
\end{equation}
The total flux emerging without interaction is computed by integrating over
all impact angles with respect to solid angle: 
\begin{eqnarray}
F_{out}(\tau ) &=&2\pi I_{0}\int_{0}^{\pi /2}e^{-2\tau \cos \theta }\cos
\theta \sin \theta \,d\theta   \nonumber \\
&=&2\pi I_{0}\int_{0}^{1}\mu e^{-2\tau \mu }\,d\mu   \label{F_out_x} \\
&=&\pi I_{0}\left[ \frac{1}{2\tau ^{2}}-\left( \frac{1}{\tau }+\frac{1}{%
2\tau ^{2}}\right) e^{-2\tau }\right] \,,  \nonumber
\end{eqnarray}
where we have used the substitution of variables $\mu =\cos \theta $ and
integration by parts. The flux that would emerge if the sphere was empty is
computed by integrating with $\tau =0$ in eq.(\ref{F_out_x}), obtaining 
\begin{equation}
F_{0}\,\equiv \,F_{out}(0)\,=\,2\pi I_{0}\,\int_{0}^{1}\mu \,d\mu \,=\,\pi
I_{0}\quad .
\end{equation}
So the fraction of photons that interact is 
\begin{eqnarray}
P_{i}(\tau ) &=&\frac{F_{0}\,-\,F_{out}(\tau )}{F_{0}}  \label{Pi_appendix}
\\
&=&1-\frac{1}{2\tau ^{2}}+\left( \frac{1}{\tau }+\frac{1}{2\tau ^{2}}\right)
e^{-2\tau }\,,  \nonumber
\end{eqnarray}
giving the interaction probability for a uniformly illuminating external
source. For the case of an optically thin sphere note that the extinction
behaves as 
\begin{equation}
e^{-2\tau \mu }\,\rightarrow \,1-2\tau \mu \text{\quad when }\tau \ll 1,
\end{equation}
and so the emerging flux is 
\begin{eqnarray}
F_{out}(\tau \sim 0)\, &\simeq &\,2\pi I_{0}\,\int_{0}^{1}\mu \left( 1-2\tau
\mu \right) \,d\mu   \nonumber \\
\, &=&\,\pi I_{0}\left[ 1-\frac{4\tau }{3}\right] \quad ,
\end{eqnarray}
and as expected this gives 
\begin{equation}
P_{i}(\tau \sim 0)\,\simeq \,\frac{4\tau }{3}\quad .  \label{Pi_tau->zero}
\end{equation}

For the case of uniformly distributed internal emission, the probability of
escaping without interactions, eq.(\ref{Oster_EP}), can be derived in a
similar fashion. Let $\epsilon $ be the emission per unit volume per second.
Then the non-interacting intensity emerging from a ray at an angle $\theta $
with respect to the surface normal is 
\begin{equation}
I_{out}(\tau ,\theta )\,=\,\frac{\epsilon }{\rho \kappa }\left( 1-e^{-2\tau
\cos \theta }\right) ,
\end{equation}
obtained in the standard fashion by integrating the transfer equation with
no scattering along the chord of optical length $2\tau \cos \theta $ through
the sphere. The total non-interacting flux emerging in any given direction
is computed by integrating with respect to solid angle, as in eq.(\ref
{F_out_x}), over all the parallel rays: 
\begin{eqnarray}
F_{out}(\tau ) &=&\frac{2\pi \epsilon }{\rho \kappa }\int_{0}^{\pi /2}\left(
1-e^{-2\tau \cos \theta }\right) \cos \theta \sin \theta \,d\theta  
\nonumber \\
&=&\frac{2\pi \epsilon }{\rho \kappa }\int_{0}^{1}\left( 1-e^{-2\tau \mu
}\right) \mu \,d\mu  \\
&=&\frac{\pi \epsilon }{\rho \kappa }\left[ 1-\frac{1}{2\tau ^{2}}+\left( 
\frac{1}{\tau }+\frac{1}{2\tau ^{2}}\right) e^{-2\tau }\right] ,  \nonumber
\end{eqnarray}
obtaining a result similar to the interaction probability above, due to the
same exponential term in the integral. If the medium had zero absorption and
scattering, the intensity emerging from a ray is 
\begin{equation}
I_{0}(R,\theta )\,=\,2\epsilon R\cos \theta \quad .
\end{equation}
The total flux emerging in any given direction is again computed by
integrating with respect to solid angle over all the parallel rays: 
\begin{eqnarray}
F_{0}(R) &=&\int_{0}^{2\pi }d\phi \int_{0}^{\pi /2}d\theta \cos \theta \sin
\theta \,I_{0}(R,\theta )  \nonumber \\
&=&4\pi \epsilon R\int_{0}^{1}\mu ^{2}d\mu \,=\,\frac{4\pi \epsilon R}{3}\,.
\end{eqnarray}
So the fraction of photons that escape the sphere is 
\begin{eqnarray}
P_{e}(\tau ) &=&\frac{F_{out}(\tau )}{F_{0}(R)}  \label{Pe_appendix} \\
&=&\frac{3}{4\tau }\left[ 1-\frac{1}{2\tau ^{2}}+\left( \frac{1}{\tau }+%
\frac{1}{2\tau ^{2}}\right) e^{-2\tau }\right] \,,  \nonumber
\end{eqnarray}
using the fact that $\tau =\rho \kappa R$, giving the extinction escape
probability for a uniformly distributed internal source.

\begin{deluxetable}{c|c|c|c|c}
\tablewidth{0pt}
\tablecaption{Multiple Scattering and Escape \label{EPS_tab}}
\tablehead{\colhead{\#} & \colhead{interacting} & \colhead{absorbed} &
		\colhead{scattered} & \colhead{escaping} }
\startdata
$1$ & $1-P$ & $(1-\omega )(1-P)$ & $\omega (1-P)$ & $P\omega (1-P)$ \\
$2$ & $\omega (1-P)^{2}$ & $(1-\omega )\,\omega \,(1-P)^{2}$ & $\omega
^{2}(1-P)^{2}$ & $P\omega ^{2}(1-P)^{2}$ \\
$3$ & $\omega ^{2}(1-P)^{3}$ & $(1-\omega )\,\omega ^{2}(1-P)^{3}$ & $\omega
^{3}(1-P)^{3}$ & $P\omega ^{3}(1-P)^{3}$ \\ 
$\vdots $ & $\vdots $ & $\vdots $ & $\vdots $ & $\vdots $ \\ 
$n$ & $\omega ^{n-1}(1-P)^{n}$ & $(1-\omega )\,\omega ^{n-1}(1-P)^{n}$ & $%
\omega ^{n}(1-P)^{n}$ & $P\omega ^{n}(1-P)^{n}$ \\ 
\enddata
\end{deluxetable}

\subsection{Including Scattering}

Equation (\ref{Lucy_EP}), which approximately includes the effects of
scattering into any extinction escape probability for the case of uniformly
distributed isotropic emitters in a bounded medium, was presented in Lucy et
al.\ (1991) but no derivation was given. In this section we derive the
equation and point out some interesting properties. The scattering albedo of
the medium is the fraction of the extinction optical depth that is due to
scattering: $\omega \,=\tau _{scat}/\tau _{ext}$, where $\tau
_{ext}\,=\,\tau _{abs}+\tau _{scat}$, is the sum of absorption and
scattering optical depths from the center to the boundary of the medium,
e.g. the radius of a sphere. Assume that we are given the escape
probability, $P_{e}(\tau )$, for absorption only, as a function of the
optical depth of the bounded medium. We can immediately use $P_{e}(\tau )$
to give the fraction of emitted photons that escape without interacting (not
absorbed {\em or} scattered), by using $\tau =\tau _{ext}$. For convenience
we shall identify $P\equiv P_{e}(\tau )$ in the following derivation. Thus a
fraction $P$ of photons escape with no interactions and a fraction $1-P$
interacts with the medium. Then by definition of the albedo, a fraction $%
(1-\omega )(1-P)$ is absorbed and the fraction $\omega (1-P)$ is scattered
for the first time. Assume that the scattered photons are distributed
uniformly in the medium and the directions of travel are isotropic. Then we
can apply the extinction escape probability again to find that a fraction $%
P\omega (1-P)$ escapes after one scattering. Repeating these steps, observe
that a fraction $\omega (1-P)^{2}$ interacts with the medium to result in a
fraction $\omega ^{2}(1-P)^{2}$ that scatters a second time. Continuing to
assume that the scattered photons are uniformly distributed and isotropic it
follows that a fraction $P\omega ^{2}(1-P)^{2}$ escapes after two
scatterings.

Table~\ref{EPS_tab} summarizes the analysis of the fractions in each state
and extends it by induction. The first column is the number of interactions
that have occurred. The column labeled ``scattered'' feeds back into the
``interacting'' column of the next row with another factor of $1-P$ to give
the next interaction. Note that for the first interaction, the photons that
are absorbed have experienced zero scatterings, so that the number of
scatterings for fractions in the ``interacting'' and ``absorbed'' columns is
one less than the interaction number. The induction presented in the table
shows that a fraction $P\omega ^{n}(1-P)^{n}$ escapes after $n$ scatterings.
Summing these escaping fractions over an infinite number of scatterings
gives the sought after formula for the total escaping fraction: 
\begin{eqnarray}
{\cal P}_{esc}(\tau ,\omega ) &=&P\sum_{n=0}^{\infty }\omega ^{n}(1-P)^{n} 
\nonumber \\
&=&\frac{P}{1-\omega (1-P)}\,\,.  \label{EPS}
\end{eqnarray}
Recall that the formula is valid (most accurate) at a single value of the
scattering asymmetry parameter, which in the case of spherical geometry is
approximately given by eq.(\ref{g_tau}). The absorbed fractions listed in
the third column of Table~\ref{EPS_tab} can also be summed to get the total
absorbed fraction: 
\begin{eqnarray}
{\cal P}_{abs}(\tau ,\omega ) &=&(1-\omega )(1-P)\sum_{n=0}^{\infty }\omega
^{n}(1-P)^{n}  \nonumber \\
&=&\frac{(1-\omega )(1-P)}{1-\omega (1-P)}\quad ,  \label{APS}
\end{eqnarray}
finding that ${\cal P}_{esc}(\tau ,\omega )+{\cal P}_{abs}(\tau ,\omega )=1$%
, which gives a check of the above derivations.

Let us define distributions, $p_{e}(n)$ and $p_{a}(n)$, for the probability
that a photon will escape or get absorbed after $n$ scatterings. To arrive
at a probability distribution, the fraction $P\omega ^{n}(1-P)^{n}$ that
escapes after $n$ scatterings must be normalized by dividing by the total
escaping fraction, obtaining 
\begin{eqnarray}
p_{e}(n) &=&\frac{P\omega ^{n}(1-P)^{n}}{{\cal P}_{esc}(\tau ,\omega )} 
\nonumber \\
&=&\left[ 1-\omega (1-P)\right] \omega ^{n}(1-P)^{n},
\end{eqnarray}
after substituting eq.(\ref{EPS}). As mentioned before, the number of
scatterings occurring for the absorbed fractions listed in the third column
of Table~\ref{EPS_tab} is actually $n-1$, where $n$ is the number of
interactions given in the first column. Therefore, the probability
distribution of absorption after $n$ scatterings is 
\begin{eqnarray}
p_{a}(n) &=&\frac{(1-\omega )\,\omega ^{n}(1-P)^{n+1}}{{\cal P}_{abs}(\tau
,\omega )}  \nonumber \\
&=&\left[ 1-\omega (1-P)\right] \omega ^{n}(1-P)^{n},
\end{eqnarray}
resulting in the fact that $p_{a}(n)=p_{e}(n)$ for all $n$. Of course this
can be true only when ${\cal P}_{esc}(\tau ,\omega )$ is valid. Monte Carlo
simulations verify that the escaping and absorbed probability distributions
are equal {\em only} when equation (\ref{EPS}) agrees with the Monte Carlo
results, which occurs only for a single value of the scattering asymmetry
parameter $g$, as discussed in section \ref{EPSS} for the case of spherical
geometry.

Defining $A\equiv 1-P$, the average number of scatterings that a photon
experiences before escape or absorption is calculated as 
\begin{eqnarray}
\left\langle n_{scat}\right\rangle _{*} &=&\sum_{n=0}^{\infty }n\,p_{e}(n) 
\nonumber \\
&=&\sum_{k=0}^{\infty }\left( \,\sum_{n=0}^{\infty
}p_{e}(n)\,-\,\sum_{n=0}^{k}p_{e}(n)\,\right)   \nonumber \\
&=&\sum_{k=0}^{\infty }\left( 1-\left[ 1-\omega A\right] \left[ \frac{%
1-\omega ^{k+1}A^{k+1}}{1-\omega A}\right] \right)   \nonumber \\
&=&\sum_{k=1}^{\infty }\omega ^{k}(1-P)^{k}  \nonumber \\
&=&\frac{1}{1-\omega (1-P)}\,-\,1  \nonumber \\
\, &=&\frac{{\cal P}_{esc}(\tau ,\omega )}{P_{e}(\tau )}\,-\,1\,\,\quad .
\label{E(nscat)}
\end{eqnarray}
Monte Carlo simulations for the case of uniform emission within a
homogeneous sphere of absorbers and scatterers again verifies that when the
asymmetry parameter takes on the particular value $g=g^{*}(\tau )$,
approximately given by eq.(\ref{g_tau}), then the average number of
scatterings experienced by photons is the same whether the final state is
escape or absorption: $\left\langle n_{scat}\right\rangle
_{esc}=\left\langle n_{scat}\right\rangle _{*}=\left\langle
n_{scat}\right\rangle _{abs}$. For other values of $g$ we find that $%
\left\langle n_{scat}\right\rangle _{esc}<\left\langle n_{scat}\right\rangle
_{*}<\left\langle n_{scat}\right\rangle _{abs}$ when $g<g^{*}(\tau )$, so
that absorption occurs after more scatterings than escape since there is
more absorption than predicted by eq.(\ref{EPS}), whereas $\left\langle
n_{scat}\right\rangle _{esc}>\left\langle n_{scat}\right\rangle
_{*}>\left\langle n_{scat}\right\rangle _{abs}$ when $g>g^{*}(\tau ),$ since
there are more escaping photons than predicted. We conjecture that the ratio
of $\left\langle n_{scat}\right\rangle _{esc}$ to $\left\langle
n_{scat}\right\rangle _{abs}$ can be affected only by $g$ and $\tau $
because the ratio is dependent only on geometry: the directions of
scattering and proximity to the boundary of the medium. The albedo $\omega $
gives probability of scattering relative to absorption and so affects the
magnitude of $\left\langle n_{scat}\right\rangle _{*}$, but it does not
enter into the geometry of photon paths, possibly explaining why $g^{*}(\tau
)$ is independent of $\omega $. Note that as the medium becomes optically
thick ($\tau \rightarrow \infty $) then $P\rightarrow 0$ and then eq.(\ref
{E(nscat)}) predicts that $\left\langle n_{scat}\right\rangle
_{*}\rightarrow \omega /(1-\omega )$ from below, and so for $g\leq
g^{*}(\tau )$ we have that $\left\langle n_{scat}\right\rangle _{esc}\leq
\omega /(1-\omega )$, which is effectively for all $g$ since $g^{*}(\tau
)\rightarrow 1$ as $\tau \rightarrow \infty .$

\section{ANALYSIS OF CLUMP OVERLAPS\label{App Cover}}

To extend the mega-grains approximation to high filling factors, consider
that as $f_{c}\rightarrow 1$ it is the increase of clump overlaps that makes
the mega-grains model become unrealistic. Since we ignore the extra density
occurring in overlaps, the effective radii of the clumps can be reduced to
eliminate most of the overlapping and the number of clumps then must be
increased to retain the same filling factor. The volume of clump overlaps is 
$V\,(Q_{c}-f_{c})$ and so the average overlap per clump is 
\begin{equation}
\frac{V\,(Q_{c}-f_{c})}{N_{c}}\,=\,\upsilon _{c}\,\left( \frac{Q_{c}-f_{c}}{%
Q_{c}}\right)   \label{clump_overlap_volume}
\end{equation}
where we have applied eq.(\ref{porosity}) and $\upsilon _{c}$ is the volume
of just one clump. Suppose there is a pair of clumps that overlap in volume
by the amount given in eq.(\ref{clump_overlap_volume}), then to find the
reduced radius at which the pair of clumps would not overlap we must solve 
\begin{equation}
\frac{\upsilon _{c}}{2}\left( \frac{Q_{c}-f_{c}}{Q_{c}}\right) \,=\,\frac{%
\pi }{3}\,h^{2}\,(3r_{c}-h)  \label{clump_overlap_depth}
\end{equation}
for $h$, the half-depth of the overlap, where $r_{c}$ is the clump radius.
The reduced radius that eliminates the overlap is then given by $r_{c}-h$.
We have solved eq.(\ref{clump_overlap_depth}) numerically for the full range
of filling factors $0<f_{c}<1$ and find that $r_{c}\,(1-f_{c})<r_{c}-h$ as $%
f_{c}\rightarrow 1$, thus eliminating the overlap. We propose to substitute $%
r_{c}(1-f_{c})^{\gamma }$ for all instances of $r_{c}$ in the mega-grains
equations, where $\gamma $ is an optional tuning parameter. In this new
model, when $\gamma =1$, the density of clumps is 
\begin{equation}
n_{c}\,=\,\frac{3f_{c}}{4\pi r_{c}^{3}(1-f_{c})^{3}}\,>\,\frac{3Q_{c}}{4\pi
r_{c}^{3}}
\end{equation}
thus greater than originally defined, and we assume that the radius
renormalization has effectively eliminated overlaps while preserving the
filling factor.

\section{DISTRIBUTION OF DUST TEMPERATURES AROUND A POINT SOURCE OF RADIATION%
\label{App Tdist}}

Given an isotropic point source of radiation in a spherically symmetric
medium, the resulting distribution of radiation and heating of the dust is
then a function of only the radial distance $r$ from the point source. We
will show that the probability distribution of dust temperatures can be
approximated by a power law function if the dust density and absorbed
luminosity versus $r$, and if the emitted luminosity versus dust temperature
can be approximated by power law functions. Letting $N(r)$ be the number of
dust grains in a sphere of radius $r$, our approach is to obtain
approximations for 
\begin{equation}
\frac{dN}{dT}\,=\,\frac{dN}{dr}\,\left( \frac{dT}{dr}\right) ^{-1}\,,
\label{dN/dT}
\end{equation}
which then gives the functional form for the distribution of temperatures.

Assume that the luminosity absorbed by the dust at distance $r$ varies like
an inverse power law with exponent $\eta $, 
\begin{equation}
L_{abs}(r)\,\sim m(r)\,\,r^{-\eta }\,\,,
\end{equation}
which in the optically thin case is nearly exact with $\eta =2$, where $m(r)$
is the mass of dust in a thin shell at radius $r$. When the dust is
optically thick we expect $\eta >2$, because the absorbed luminosity decays
more rapidly than any power law ($\eta $ is then also a function of $r$).
The luminosity emitted by the dust scales with the dust temperature, $T(r)$,
approximately as 
\begin{eqnarray}
L_{em}\left[ T(r)\right] &\equiv &4\pi m(r)\int_{0}^{\infty }B_{\nu
}[T(r)]\kappa _{\nu }d\nu  \nonumber \\
&\sim &m(r)\,T(r)^{4+\beta }\,\,,  \label{Lem[T(r)]}
\end{eqnarray}
if the emissivity per unit mass of the dust scales with frequency like $%
\,\kappa _{\nu }\sim \nu ^{\beta }$, and usually $0<\beta \leq 2$. Equating
the absorbed and emitted luminosities of the dust we get an approximate
relationship between dust temperature and radial distance: 
\begin{equation}
T(r)\sim \,\,r^{-\frac{\eta }{^{4+\beta }}}\,\,.  \label{T_vs_r}
\end{equation}
The rate of change in temperature with respect to radial distance then
varies as: 
\begin{equation}
\frac{dT}{dr}\,\sim \,r^{-\left( 1+\frac{\eta }{^{4+\beta }}\right) }\,\sim
\,T^{\left( 1+\frac{4+\beta }{\eta }\right) }\,\,,  \label{dT/dr}
\end{equation}
where we have also inverted eq.(\ref{T_vs_r}) and substituted for $r$.

Assume that the dust density is approximately a power law $\rho (r)\sim
r^{-\delta }$, so that the number of dust grains, $N(r)$, in a sphere of
radius $r$ also varies like a power law function: 
\begin{equation}
N(r)=4\pi \int_{0}^{r}\rho (s)\,s^{2}ds\,\,\sim \,r^{3-\delta }\,\,.
\end{equation}
Then 
\begin{equation}
\frac{dN}{dr}\,\sim \,r^{2-\delta }\,\sim \,T^{(\delta -2)\left( \frac{%
4+\beta }{\eta }\right) }\,\,,  \label{dN/dr}
\end{equation}
where we have again substituted for $r$ using eq.(\ref{T_vs_r}). Combining
eqs.(\ref{dT/dr}) and (\ref{dN/dr}) into (\ref{dN/dT}): 
\begin{eqnarray}
\frac{dN}{dT}\, &\sim &\,T^{(\delta -2)\left( \frac{4+\beta }{\eta }\right)
}\,T^{\left( -1-\frac{4+\beta }{\eta }\right) }  \nonumber \\
&\sim &\,T^{(\delta -3)\left( \frac{4+\beta }{\eta }\right) -1}\quad .
\end{eqnarray}
From this scaling approximation we presume that the distribution of dust
temperatures follows a power law 
\begin{equation}
p(T)\,=\,a\,T^{\mu }
\end{equation}
with exponent 
\begin{equation}
\mu \,=(\delta -3)\left( \frac{4+\beta }{\eta }\right) -1\,\,\,,
\end{equation}
which is certainly negative as long as $\delta \leq 3$. The normalizing
factor $a$ is determined by requiring 
\begin{equation}
\int_{T_{\min }}^{T_{\max }}p(T)dT\,\,=1\,\,,
\end{equation}
where $T_{\max }$ is usually the dust sublimation temperature ($\sim 1500%
{\sf K}$) and $T_{\min }$ is the minimum dust temperature. Actually, the
minimum dust temperature is the major free parameter in this theory, and is
determined by balancing the absorbed and emitted dust luminosities: 
\begin{equation}
\int_{0}^{R}L_{abs}(r)r^{2}dr=M\int_{T_{\min }}^{T_{\max }}p(T)K(T)dT\,,
\end{equation}
where $M$ is the total mass of dust contained in a sphere of radius $R$, and
where 
\begin{equation}
K(T)=\int_{0}^{\infty }B_{\nu }(T)\kappa _{\nu }\,d\nu 
\end{equation}
is the Planck averaged emissivity at temperature $T$. Most of the emission
comes from dust at temperatures near $T_{\min }$ since those dust grains
occupy most of the volume and the dust temperature decays rapidly with
distance from the point source. Once $T_{\min }$ is determined, the IR
emission spectrum from the dust grains is given by 
\begin{equation}
F_{\nu }=4\pi M\kappa _{\nu }\int_{T_{\min }}^{T_{\max }}B_{\nu
}(T)\,p(T)\,dT\,\,\,.
\end{equation}

The behavior of the emissivity of graphite or silicates changes at
wavelengths longer than about 10$\mu $m, therefore we find that the power
law approximation of the emissivity averaged with a Planck function at a
given temperature (proportional to the luminosity emitted by dust), 
\begin{equation}
K(T)\sim T^{4+\beta _{i}}\,\,,
\end{equation}
requires two exponents, $\beta _{i}$, with $i=1$ for $T>T_{b}$, and $i=2$
for $T\leq T_{b}$, where $T_{b}$ is called the emissivity break temperature,
corresponding to the wavelength at which the behavior of the dust emissivity
changes. From examination of the behavior of $K(T)$, the values of $\beta
_{i}$ and $T_{b}$ are found to be approximately 
\begin{equation}
\beta _{i}\,=\,\left\{ \, 
\begin{array}{l}
(1.0,\,\,2.0){\sf \quad }\text{(graphite)} \\ 
(0.5,\,\,2.0){\sf \quad }\text{(silicates)}
\end{array}
\right\}
\end{equation}
and 
\begin{equation}
T_{b}\,=\left\{ \, 
\begin{array}{l}
80{\sf K\quad }\text{(graphite)} \\ 
150{\sf K\quad }\text{(silicates)}
\end{array}
\right\} \quad .
\end{equation}
Consequently, the power law distribution of temperatures around a central
source will have an exponent depending on the dust temperature range: 
\begin{equation}
\mu _{i}\,=\,\,(\delta -3)\left( \frac{4+\beta _{i}}{\eta }\right) -1
\end{equation}
yielding power law probability distributions 
\begin{equation}
p_{i}(T)\,\equiv \,a_{i}T^{\,\mu _{i}}\quad ,
\end{equation}
where the constants $\,a_{i}$ are determined by requiring that $%
p_{1}(T_{b})=p_{2}(T_{b})$ and 
\begin{equation}
\int_{T_{b}}^{T_{\max }}p_{1}(T)dT\,+\int_{T_{\min
}}^{T_{b}}p_{2}(T)dT\,=1\,,
\end{equation}
yielding (defining $\gamma _{i}=\mu _{i}+1$) 
\begin{eqnarray}
a_{2} &=&a_{1}\,T_{b}^{\gamma _{1}-\gamma _{2}} \\
a_{1} &=&\frac{\gamma _{1}\gamma _{2}}{\gamma _{2}T_{\max }^{\gamma
_{1}}+(\gamma _{1}-\gamma _{2})T_{b}^{\gamma _{1}}+\gamma _{1}T_{\min
}^{\gamma _{2}}T_{b}^{\gamma _{1}-\gamma _{2}}}\,.  \nonumber
\end{eqnarray}
This approximation for the temperature distribution and resulting IR\
emission spectrum were found to be in reasonable agreement with Monte Carlo
simulations for the case of $\delta =0.$


\begin{thebibliography}{Bevington \& Robinson 1992, pp. 23-28}
\bibitem[Bazell \& Desert 1988]{Bazell88}  Bazell, D., Desert, F.X., 1988,
ApJ, 333, 353

\bibitem[Bevington \& Robinson 1992, pp. 23-28]{bev92}  Bevington, P.R. \&
Robinson, D.K. \thinspace 1992, Data Reduction and Error Analysis for the
Physical Sciences (McGraw Hill), Chapter 2

\bibitem[Boiss\'e (1990)]{boisse90}  Boiss\'{e}, P. \thinspace 1990, A\&A,
228, 483

\bibitem[Chan et al.\ 1997]{Chan97}  Chan, K.-W., Moseley, S.H., Casey, S.,
Harrington, J.P., Dwek, E., Loewenstein, R., V\'{a}rosi, F., and Glaccum, W.
1997, ApJ, 483, 798

\bibitem[Code \& Whitney (1995)]{codeblob95}  Code, A.D., and Whitney, B.A.
1995, ApJ, 441, 400

\bibitem[Cox 1995]{Cox95}  Cox, D.P. 1995, in The Interplay Between Massive
Star Formation, the ISM and Galaxy Evolution, eds. B. Kunth et al. (Cedex:
Editions Frontieres), 223

\bibitem[Dickey \& Garwood (1989)]{DG89}  Dickey, J.M., and Garwood, R.W.
\thinspace \thinspace 1989, ApJ, 341, 201

\bibitem[Draine (1985)]{Draine85}  Draine, B.T. 1985, ApJS, 57, 587


\bibitem[Dwek et al.\ 1997]{Dwek97}  Dwek, E. et al. 1997, ApJ, 475, 565

\bibitem[Elmegreen 1997]{elm97}  Elmegreen, B.G. \thinspace 1997, ApJ, 477,
196

\bibitem[Elmegreen \& Falgarone 1996]{elmfal96}  Elmegreen, B.G., and
Falgarone, E. \thinspace 1996, ApJ, 471, 816

\bibitem[Falgarone 1995]{Falg95}  Falgarone, E. 1995, in The Interplay
Between Massive Star Formation, the ISM and Galaxy Evolution, eds. B. Kunth
et al. (Cedex: Editions Frontieres), 95

\bibitem[Gaustad \& van Buren 1993]{GvB93}  Gaustad, J.E. and van Buren, D.
1993, PASP, 105, 1127

\bibitem[Gordon et al.\ 1994]{gordalb94}  Gordon, K.D., Witt, A.N.,
Carruthers, G.R., Christensen, S.A., and Dohne, B.C., \thinspace 1994, ApJ,
432, 641

\bibitem[Gordon, Calzetti, and Witt\ (1997)]{gordsb97}  Gordon, K.D.,
Calzetti, D., and Witt, A.N. \thinspace 1997, ApJ, 487, 625

\bibitem[Henyey \& Greenstein 1941]{H-G41}  Henyey, L.G., and Greenstein,
J.L. \thinspace 1941, ApJ, 93, 70

\bibitem[Hobson \& Scheuer (1993)]{hs93}  Hobson, M.P., and Scheuer, P.A.G.
\thinspace 1993, MNRAS, 264, 145

\bibitem[Hobson \& Padman (1993)]{hobpad93}  Hobson, M.P., and Padman, R.
\thinspace 1993, MNRAS, 264, 161 (HP93)

\bibitem[Knapp 1995]{knapp95}  Knapp, G. \thinspace 1995, Sky and Telescope,
May, p.20


\bibitem[(Lucy et al.\ 1991)]{Lucy91}  Lucy, L.B., Danziger, I.J., Gouiffes,
C., and Bouchet, P. \thinspace 1991, in Supernovae, ed. S.E. Woosley (New
York: Springer-Verlag), p.82

\bibitem[Lux \& Koblinger (1995, pp.40-41)]{LuxKob95}  Lux, I., and
Koblinger, L. \thinspace 1995, Monte Carlo Particle Transport Methods:
Neutron and Photon Calculations (Boca Raton, CRC Press), pp.40-41

\bibitem[Marscher et al.\ 1993]{Marscher93}  Marscher, A.P., Moore, E.M.,
and Bania, T.M. 1993, ApJ, 419, L101


\bibitem[McKee 1995]{McKee95}  McKee, C.F. 1995, in The Interplay Between
Massive Star Formation, the ISM and Galaxy Evolution, eds. B. Kunth et al.
(Cedex: Editions Frontieres), 223

\bibitem[Murthy et al.\ 1992]{MWH92}  Murthy, J., Walker, H.J., and Henry,
R.C. 1992, ApJ, 401, 574

\bibitem[Natta \& Panagia (1984)]{NatPan84}  Natta, A., and Panagia, N.
\thinspace \thinspace 1984, ApJ, 287, 228 (NP84)

\bibitem[Neufeld (1991)]{neufeld91}  Neufeld, D.A. \thinspace 1991, ApJ,
370, L85

\bibitem[Neumann 1951]{Neumann51}  Neumann, J. 1951, NBS Applied Math
Series, No. 12, (U.S. Gov. Printing Office), p.36

\bibitem[Norman \& Ferrara 1996]{norman96}  Norman, C.A., and Ferrara, A.
\thinspace 1996, ApJ, 467, 280

\bibitem[Osterbrock (1989, pp.385-386)]{oster89}  Osterbrock, D.E.
\thinspace 1989, Astrophysics of Gaseous Nebulae and Active Galactic Nuclei
(Mill Valley CA: University. Science Books), Appendix 2

\bibitem[Pfenniger \& Combes 1994]{PfenComb94}  Pfenniger, D., and Combes,
F.\thinspace \thinspace 1994, A\&A, 285, 94

\bibitem[(Press et al.\ 1992, pp.242-254)]{NumRecip}  Press, W.H.,
Teukolsky, S.A., Vetterling, W.T., and Flannery, B.P. 1992, Numerical
Recipes (Cambridge: Cambridge Univ. Press)

\bibitem[Rosen \& Bregman 1995]{RosBreg95}  Rosen, A., and Bregman,
J.N.\thinspace \thinspace 1995, ApJ, 440, 634

\bibitem[Rybicky \& Lightman 1979, pp.37-38]{Rybicky79}  Rybicky, G.B., and
Lightman, A.P. 1979, Radiative Processes in Astrophysics (John Wiley \&
Sons), pp.37-38

\bibitem[Sanders, Scoville, and Solomon 1985]{SanSco85}  Sanders, D.B.,
Scoville, N.Z., and Solomon, P.M. 1985, ApJ, 289, 387


\bibitem[Spitzer 1978]{Spitzer-ISM}  Spitzer, L. 1978, Physical Processes in
the Interstellar Medium (John Wiley \& Sons)

\bibitem[Stutzki \& Gusten 1990]{stutzki90}  Stutzki, J., and Gusten, R.
\thinspace 1990, ApJ, 356, 512

\bibitem[van Buren 1989]{vB89}  van Buren, D. 1989, ApJ, 338, 147

\bibitem[V\'{a}rosi \& Dwek 1997]{VD97}  V\'{a}rosi, F. and Dwek, E. 1997,
in The Ultraviolet Universe at Low and High Redshift, eds. W.H. Waller et
al. (New York: AIP), 370 (astro-ph/9905291)

\bibitem[V\'{a}rosi \& Dwek 2000]{VD00}  V\'{a}rosi, F. and Dwek, E. 2000,
in preparation

\bibitem[Waller et al.\ 1997]{Waller97}  Waller, W.H., V\'{a}rosi, F.,
Boulanger, F., and Digel, S.W., 1997, in New Horizons from Multi-wavelength
Sky Surveys, eds. B.J. McLean et al. (IAU Symp. 179: Kluwer Academic), 194

\bibitem[Witt (1977)]{witt77}  Witt, A.N. \thinspace 1977, ApJSup, 35, 1

\bibitem[Witt \& Gordon (1996)]{wittgor96}  Witt, A.N., and Gordon, K.D.
\thinspace 1996, ApJ, 463, 681


\bibitem[Wolf, Fischer \& Pfau (1998)]{Wolf98}  Wolf, S., Fischer, O., and
Pfau, W. 1998, A\&A, 340, 103
\end{thebibliography}
\end{document}